\documentclass[reprint,aps,prb,english,superscriptaddress,amssymb,amsfonts,longbibliography]{revtex4-2}
\usepackage{graphicx}
\usepackage{epsfig}
\usepackage{color}
\usepackage{amsmath}
\usepackage{amsfonts}
\usepackage{mathrsfs}
\usepackage{txfonts}
\usepackage{color}
\usepackage{mathtools}

\usepackage[dvipsnames,svgnames,x11names,hyperref]{xcolor}

\usepackage[breaklinks]{hyperref} 
\hypersetup{colorlinks=true, linkcolor=blue, citecolor=blue, filecolor=blue, urlcolor=blue}

\begin{document}
	\title{Analytical theory of cat scars with discrete time crystalline dynamics in Floquet systems}
	\author{Biao Huang}
	\email{phys.huang.biao@gmail.com} 
	\affiliation{Kavli Institute for Theoretical Sciences, University of Chinese Academy of Sciences, Beijing 100190, China}
	\date{\today}
	
	\begin{abstract}
		We reconstruct the spectral pairing (SP) theories to enable analytical descriptions of eigenstate spatiotemporal orders in translation-invariant system without prethermal conditions.
		It is shown that the strong Ising interactions and drivings alone stabilize a class of ``cat scar" eigenstates with tunable patterns, which lead to {\em local} discrete time crystal (DTC) dynamics. They exhibit Fock space localization and long-range correlations robust against generic perturbations in a disorder-free scenario.
		In particular, we introduce a symmetry indicator method to enumerate cat scars, with which a set of unexpected inhomogeneous scar patterns are identified  in addition to the ferromagnetic scars found before. These scars enforce DTC dynamics with rigid inhomogeneous patterns, offering a viable way to verify underlying eigenstate properties experimentally. Further, we prove rigorously that the strong Ising interactions enforces a selection rule for perturbations of different orders, which imposes an exponential suppression of spin fluctuations for Floquet eigenstates. Based on this property, three analytical scaling relations are proved to characterize the amplitudes, Fock space localization, and lifetime for DTC dynamics associated with cat scars.
		We further provide two practical methods to check whether certain DTC phenomena are dominated by single-spin dynamics or due to genuine interaction effects.
	\end{abstract}

	\maketitle
	
	%\tableofcontents

	\section{Introduction}
	
	Discrete time crystals (DTC) have emerged as an intriguing phase living far from equilibrium~\cite{Khemani2016,Keyserlingk2016,Else2016,Yao2017,Ho2017,Sacha2015,Zhang2017,Choi2017,Mi2022,Frey2022,Randall2021,Kyprianidis2021,Sacha2017,Sacha2020,Khemani2019b,Else2020,Zaletel2023,Kshetrimayum2020,Liu2022}. Phenomenologically, it features a reduced period $ nT $ ($ 1<n\in \mathbb{Z} $) for observables compared with system driving periods $ T $, thereby giving rise to the concept of spontaneous breaking of discrete time translation symmetry. 
	Up to now, numerous systems with different underlying mechanisms have been proposed to render such a phenomenon, both theoretically and experimentally~\cite{Sacha2017, Sacha2020, Khemani2019b, Else2020, Zaletel2023}. 
	Here, we focus on a particular mechanism dubbed spectral pairing (SP)~\cite{Else2016,Keyserlingk2016}, which remarkably fixes the spectral gap $ \Delta = 2\pi/nT $ between pairwise localized eigenstates, although individual levels shift considerably under perturbations. That locks the oscillation period to $ 2\pi/\Delta $ without fine-tuning. SP transcends Landau's paradigm in handling spontaneous breaking of time translation symmetry, and exemplifies unique principles of highly nonequilibrium nature.

	Currently, one central topic for DTC systems is to distinguish period-doubled oscillations attributable to different reasons. In doing so, a useful {\em method} of checking different initial states has been proposed recently~\cite{Luitz2020,Khemani2019b}.  For instance, in many-body localized (MBL) systems, SP is expected to occur for all eigenstates, and therefore initial states of arbitrary pattern exhibit {\em local} DTC oscillations with original patterns unchanged~\cite{Khemani2019b}. Also, for prethermal systems with Landau's spontaneous symmetry breaking~\cite{Else2017}, eigenstates close to low energy density sectors (i.e. for a ferromagnetic ground state) would exhibit SP. Correspondingly, initial states belonging to the same energy sector (i.e. starting from a ferromagnetic product state and include very few spin flips) would generate DTC dynamics, with the spins diffusing into the same patterns as the ground state for prethermal effective Hamiltonians (i.e. a ferromagnetic pattern). In contrast, initial states residing far away from the low-energy-density sector (i.e. an antiferromagnetic pattern) would show quick thermalization without dynamics. Their distinctions are emphasized by a new set of experiments recently~\cite{Mi2022,Randall2021,Frey2022,Kyprianidis2021}.
	
	Meanwhile, there is a third category of translation-invariant clean systems violating both MBL and prethermal conditions, where the mechanism for possible DTC oscillations~\cite{Rovny2018,Pal2018,Huang2018,Russomanno2017,Zeng2017,Lyu2020,Yu2019,Mizuta2018,Barfknecht2019} may be more controversial. By a similar analysis of checking various initial states, pioneering explorations have found several  rather different characters for DTC type of dynamics therein. For instance, numerics indicates that up to intermediate scales, homogeneous ferromagnetic initial states rendering period-doubled oscillations may be associated with SP of rare non-ergodic eigenstates~\cite{Yarloo2020,Pizzi2020}, dubbed ``scars"~\cite{Turner2018,Chandran2022}. These oscillations are shown to survive for exponentially long-time $e^{L}$ with the increase of system size. Meanwhile, initial states deviating from ferromagnetic ones may show diffusive oscillations that are attributable to other reasons. They include prethermal systems with approximate global U(1) symmetries~\cite{Luitz2020,Ho2020,Beatrez2023,Stasiuk2023}, and domain wall confinement~\cite{Collura2022} induced by extended Ising interaction ranges. A common feature in this class of systems is that the DTC lifetime is typically a fixed power-law of certain parameters, insensitive to the change of system sizes. To provide a more definitive understanding of the aforementioned phenomena, as suggested by several previous works~\cite{Mizuta2018,Pizzi2020}, it is desirable to formulate an independent analytical theory of eigenstate SP beyond the schemes for MBL and prethermal cases. The new theory is expected to yield a more comprehensive enumeration of possible scars with rigid SP. Also, it may offer more rigorous predictions to quantify the robustness of eigenstate SP in generic clean Floquet systems, and help distinguish certain similar phenomena that are attributable to different reasons.

	In this work, we explore such a possibility by presenting a reformulated SP theory. It generalizes our previous analysis~\cite{Huang2022} beyond few-body systems to achieve an analytical description of DTC type of dynamics in intermediate-scale spin systems.
	Our major results include two aspects.	
	
	First, we offer a practical way to quickly enumerate eigenstates with SP out of an exponentially large Hilbert space. Specifically, for translation-invariant systems, a symmetry indicator method is proposed, which predicts a {\em coexistence} of ferromagnetic (FM) {\em and} antiferromagnetic (AFM) scar patterns, with the latter largely unappreciated previously. In a more generic setting where sublattices are included, symmetry indicators further allow for identifying a richer variety of spin patterns for {\em local} DTC oscillations. We further show that an arbitrary scar pattern can be precisely engineered by weakly breaking translation symmetry. Specifically, the amplitudes of interactions are still homogeneous, while the signs of interaction on different bonds are allowed to change in this situation. As such, our theory may shed light on achieving on-demand engineering of SP for eigenstates, without constraints in dimensionality, interaction ranges, and the requirements of fully randomized interactions in contrast to the MBL cases.
	
	Second, to quantify the robustness of these scar enforced DTCs, we obtain for the first time {\em analytical} scaling relations showing pairwise Fock space localization and fixed spectral gaps for scars, which are robust against {\em generic} perturbation of strength $ \lambda $ up to system size $ L \lesssim 1/\lambda^2 $, i.e. $ L\lesssim10^2 $ for typical $ \lambda\sim0.1 $. Specifically, the amplitudes of DTC oscillation are shown to be dominantly rescaled by the first-order spin fluctuations. Such an amplitude rescaling, for separable perturbations, can be computed in a purely analytical fashion without fitting. Meanwhile, the Fock space pattern localization length, as well as the DTC lifetime scaling, are shown to be bounded by counting how many spins are interacting in each term of the perturbation Hamiltonian --- namely, by just checking the general form of perturbations. These scalings quantitatively characterize the eigenstate orders of the long-range-correlated ``cat scars", named in analogy to the Schr\"{o}dinger's cat eigenstates in MBL DTCs~\cite{Else2016}. Symmetry indicators and scaling rules provide a generic way long-sought-after to analytically characterize clean DTCs, especially to {\em quantitatively compute} their scaling behaviors with explicit analytical formulae.  We have compared the results above against numerics for extensive number of examples, and find good agreements in all cases.
	
	This work belongs to a series of works initiated in Ref.~\cite{Huang2022} aiming at an independent analytical description of possible SP and eigenstate localization in systems with or without disorders. We briefly discuss below the connection, distinctions, and relative progresses achieved in this work compared with the previous one.
	
	Generically, SP is a property for pairs of eigenstates being cat-like states. Such a state involves very small amounts (i.e. in our case, two) macroscopic Fock states, and therefore is intrinsically associated with localization physics in many-body Fock space. It was emphasized in Ref.~\cite{Else2016} that pairs of long-range correlated cat states could serve as independent origins for discrete time translation symmetry breaking without requiring the system to host any unitary symmetry~\cite{Zaletel2023}. The robustness of SP for long-range correlated cat states was further confirmed in Ref.~\cite{Keyserlingk2016} where anti-unitary symmetries were also broken. As such, although the original discussions in Ref.~\cite{Else2016} describe the situation where all eigenstates are cat states --- which necessarily means MBL by definition --- it is surely tantalizing to consider whether localized cat eigenstates, as independent and essential mechanisms enforcing SP, can be realized in other scenarios as well. Soon, it was found that Floquet prethermal systems with Landau's spontaneous symmetry breaking~\cite{Else2017,Machado2020} offers another platform to achieve localized cat eigenstates. Compared with MBL DTCs, the number of cat states in prethermal DTCs are relatively rare, but these cat states could be achieved with more flexible conditions regarding spatial dimensionality and interaction ranges. 
	
	Following this direction of research, it is of interst to further seek for alternative origins of cat eigenstates other than those induced by MBL or Landau's symmetry breaking. This effort may be especially useful in view of recent progresses in a re-examination of localization physics. In particular, a so-called avalanche mechanism~\cite{Roeck2017,Luitz2017,Thiery2018,Morningstar2020,Crowley2020} implies that the previously proposed phase transition~\cite{Basko2006,Oganesyan2007,Nandkishore2015} between thermal and MBL regimes may turn out to be a crossover, due to proliferation of many-body resonances in thermodynamic limits~\cite{Suntajs2020,Sels2021,KieferEmmanouilidis2020,Morningstar2022a,Long2022,Ha2023}. (See, however, different viewpoints~\cite{Crowley2022}). That means what one would encounter practically in experiments and numerics is a ``prethermal MBL" within finite size and/or time scales. In constrast, a true MBL {\em phase} may be either absent~\cite{Suntajs2020,Sels2021,KieferEmmanouilidis2021}, or occupy a much smaller parameter regime~\cite{Morningstar2022a,Long2022}. As an alternative approach towards a more definite understanding of localization physics, it may be worthwhile to explore whether an independent eigenstate SP and localization mechanism can be quantified rigorously from scratch, without assuming other background mechanisms {\em a priori}. 
	
	Constructing a generic analytical theory of localization physics once and for all in strongly interacting and strongly driven systems could be rather challenging. Indeed, up to now, a rigorous mathematical proof for Floquet MBL is still pending discovery~\cite{Zaletel2023}. However, it may be more viable to make progress by gradually generalizing the analytical framework so as to include more generic features. As a continued effort, this work qualitatively extends our previous constructions for SP of Floquet-Bloch scars (FBS)~\cite{Huang2022} in few-body systems. Cat scars reported here show different properties, and requires new techniques to analyze SP in an intermediate-scale setting.

	First, there is a crucial distinction between the two types of scars regarding in what space localization takes place. FBS features the localization in many-body {\em momentum} space, which naturally means that each FBS eigenstate is delocalized in the real-space Fock basis. Therefore, FBS's exhibit short-range correlations, and require spatial translation symmetry to {\em protect} their stability of SP --- henceforth ``Bloch" in the terminology FBS. In contrast, cat scars in this work exhibit pairwise localization in real-space Fock basis, and therefore demonstrate long-range correlations. As such, cat scars could endure perturbations breaking all crystalline symmetries. 
	
	Second, on the technical aspect, one major advancement in this work is to derive a selection rule of spin-flip effects for perturbations of different orders, which in turn legitimizes a perturbative treatment of certain Floquet systems with essentially {\em gapless} quasienergy spectrum.  This is indispensable to generalize our previous analysis for gapped few-body systems~\cite{Huang2022} into intermediate scales. 
	
	Specifically, for strongly interacting systems where interaction strength is {\em comparable} with Floquet driving frequency, the difference between the interaction energy of a certain eigenstate $E_1$ from that of scar eigenstate  $E_0$, i.e. $\Delta E = E_1 - E_0$, may be close to resonant frequency of the Floquet driving, i.e. $\Delta E - m\times (2\pi\hbar/T) \rightarrow 0, m\in\mathbb{Z}$. Such a many-body Floquet resonance may proliferate quickly with the increase of system size $L$, where density of states ramps up exponentially. At first glance, it may then appear inapplicable to consider a perturbative treatment in order to prove the localization and stability of a certain scar eigenstate. 
	
	However, a selection rule is derived in Sec.~\ref{sec:selection} of this paper, stating that the hybridization amplitude is suppressed exponentially $ \sim \lambda^{\delta s}$ for two Fock states differing by $\delta s$ spin flips, as enforced by the strong Ising interaction. Here $\lambda\ll1$ is the perturbation strength. Meanwhile, eigenstates are grouped into different domain wall sectors, mutually showing rather small hybridizations. Then, unless the interaction is fine-tuned to the exact resonance point at $\lambda = 0$ for nearby domain walls sectors, such resonances for hybridizing different Fock states are exponentially suppressed. We have demonstrated the mathematically proved selection rules using more intuitive illustrations, where eigenstate structures are shown in terms of averaged domain wall number and quasienergy for each Floquet eigenstate. By examining whether different domain wall sectors are strongly mixed or not, we clarify and confirm conditions for Floquet resonances to occur or vanish.

%	\begin{widetext} 
		
		\begin{figure*}[t]
			\parbox{18cm}{
				\includegraphics[width=15cm]{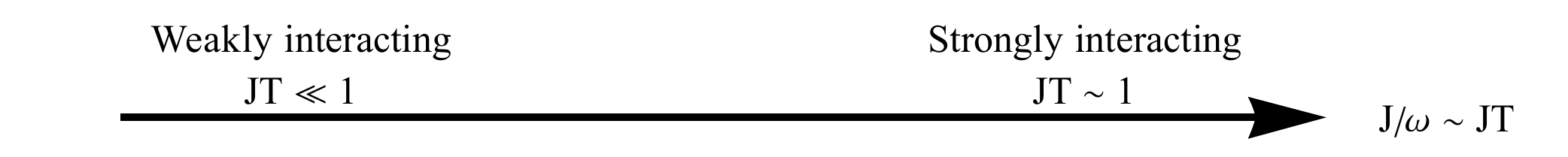}
				\\
				\parbox[t]{8cm}{
					\begin{flushleft}
						Dominant energy scale: Floquet driving frequency $\omega \Rightarrow$ prethermal time scale $\tau_* \sim e^{\omega/J} \sim e^{1/JT}$
					\end{flushleft}
					\begin{enumerate}
						\item 
						Prethermal DTC induced by Landau symmetry breaking (``at low-temperature")~\cite{Else2017,Machado2020,Kyprianidis2021}. 
						
						\item 
						Prethermal DTC induced by approximate U(1) conservation (``without temperature")~\cite{Luitz2020,Ho2020,Beatrez2023,Stasiuk2023}.
					\end{enumerate}
				}
				\qquad\qquad
				\parbox[t]{8cm}{
					\begin{flushleft}
						Dominant energy scale: interaction {\em and} driving frequency $J\sim\omega \Rightarrow$ Fock space localization {\em enforced} by  interactions
					\end{flushleft}
					\begin{enumerate}\setcounter{enumi}{2}
						\item 
						MBL DTC~\cite{Khemani2016,Else2016,Yao2017,Mi2022,Randall2021,Frey2022} with disordered interactions $J_j \in [J-W, J+W]$, where both $ J, W \sim \omega $.
						
						\item 
						Cat scar DTC ({\bf this work}) with either uniform or disordered $J_j \in [J-W, J+W]$, where only $J\sim \omega$ is required.
					\end{enumerate}
				}
			}
			\begin{picture}(0,0)
				\put(0,10){\line(0,1){130}}
			\end{picture}
			
			\begin{tabular}{ l c c c }
				\hline \hline 
				\parbox{3cm}{Types} & \parbox{6.5cm}{Conditions and crucial factors} & Lifetime scaling & DTC spin patterns
				\\ \hline 
				1. \parbox{3cm}{Prethermal Landau DTC} & 			
				\parbox{6cm}{
					{\scriptsize\quad}\\
					Mermin-Wagner theorem restricts spontaneous breaking of Ising symmetry at finite temperature to systems of dimensions $d\geqslant 2$ or those with long-range interaction}
				& 
				\parbox{4cm}{Grows exponentially until the prethermal time scale is reached, i.e. $\min(\tau_*, \sim e^L)$}
				&
				\parbox{3cm}{FM or AFM depending on energetics of effective Hamiltonian }
				\\  \hline 
				2. \parbox{3cm}{Prethermal U(1) DTC} & 
				\parbox{6cm}{
					{\scriptsize\quad}\\
					Global spin oscillations whose lifetime may be prolonged if interactions respect U(1) symmetries, i.e. for XXZ models }
				& 
				\parbox{4cm}{
					{\scriptsize\quad}\\
					Fixed power-law of perturbation strength insensitive to system size changes, i.e. $\sim 1/\lambda^2$.
				}
				&
				\parbox{3cm}{ Quickly diffuse into homogeneous pattern, then exhibit total spin oscillations }
				\\ \hline 
				3. \parbox{3cm}{MBL DTC} & 
				\parbox{6cm}{
					{\scriptsize\quad}\\
					One spatial dimension, short-ranged and disordered Ising {\em  interaction} (up to size and/or time scale where Floquet MBL is stable). } & 
				$\sim e^L$ 
				&
				\parbox{3cm}{Arbitrary spin patterns}
				\\ \hline 
				4. \parbox{3cm}{Cat scar DTC} & 
				\parbox{6cm}{
					{\scriptsize\quad}\\
					Intermediate-scale systems $L\lesssim 1/\lambda^2 \sim 10^{2\sim3}$, in arbitrary dimension with arbitrary interaction range and/or disorder strengths. }
				& 
				$\sim e^L$
				&
				\parbox{3cm}{
					On-demand engineered patterns
				}
				\\ \hline
			\end{tabular}
			\caption{\label{fig:int_strength}Several known DTC phenomena surviving in different parameter regimes and contrasts of their characters. }
		\end{figure*}
%	\end{widetext}
	
	It may be helpful to compare the cat scar DTCs introduced in this work with several other known examples of DTC systems. We briefly outline their distinctions in Fig.~\ref{fig:int_strength}.

	On the one hand, connections and distinctions between cat scars and MBL DTCs are relatively straightforward. They both survive in the strongly interacting regime where interaction strength $J$ is comparable to Floquet driving frequency $\omega$. Localization in these cases are both enforced by strong interactions. For cat scar DTCs, the relevant eigenstates showing SP is much more rare, typically of the order O(1), compared with MBL DTCs where majority eigenstates are expected to be cat states. Nevertheless, the requirement of such a system is significantly more flexible in terms of dimensionality, interaction range, and disorder strengths. Further, the spin patterns for cat scars can be precisely controlled by changing the signs of interactions, thereby reducing the burden of having interaction terms with large numbers of different amplitudes. This could be particularly useful in current noisy-intermediate-scale-quantum (NISQ) devices, because engineering different two-qubit gates for fully randomized interactions~(i.e. in Ref.~\cite{Mi2022}) costs notably more resources than having uniform interaction~(i.e. in Ref.~\cite{Mi2022a}), which in turn may impact the number of qubits accessible in experiments. 
	While this work chiefly focuses on engineering two pairs of cat states, a tunable number of cat states, in additional to the tunability of their patterns, can be achieved by including a limited number of different interaction amplitudes (see the examples in Sec.~\ref{subsec:symm}). 
	Thus, cat scar DTCs could be viewed as a new scheme featuring an on-demand engineering of cat eigenstates using corresponding resources, which supplements the scheme in MBL DTCs where both the number of cat states and resource burdens are maximized.
	%Since one major goal in computation science is to maximize the performance efficiency using {\em a given amount} of resource, from the practical point of view, cat scar DTCs could be viewed as a way of precise Hilbert space architecture in order to achieve an on-demand engineering of cat states.

	On the other hand, it may be a bit more subtle to distinguish cat scar DTCs from prethermal ones with Landau's symmetry breaking. In particular, although Mermin-Wagner theorem forbids a {\em finite temperature} phase transition of Ising symmetry breaking in 1D with short-range interaction, it is, however, possible for {\em zero temperature} symmetry breaking to occur. Then, we may be tentative to ask: are the cat scars found here associated with the Landau's symmetry breaking at zero temperature for certain ground states?
	
	The answer, however, turns out to be negative. Here, it is vitally important to note that the strong interaction for cat scars violates prethermal conditions, such that a static prethermal Hamiltonian cannot be defined in the first place. That leaves Landau's symmetry breaking irrelevant, either at finite or zero temperature. Specifically, in order to involve Landau's theory, it is necessary to start from a static Hamiltonian so as to define a set of thermodynamic quantities, such as conserved energy and temperature. Such an effective Hamiltonian can only be obtained in DTC systems if interactions are much weaker than the Floquet driving frequency, namely, the prethermal conditions are satisfied. For instance, in Ref.~\cite{Else2017}, it is shown that up to a global spin flip, if all other parameters in the Hamiltonians are much smaller than driving frequency (i.e. interaction $JT\ll1$, longitudinal fields $h_zT\ll1$ etc), it is possible to obtain an approximated static Hamiltonian  $H_{\text{eff}}$ by factoring out the global spin flips $U_F = P e^{-iH_{\text{eff}}}$ via unitary transformations. In turn, this static Hamltonian $H_{\text{eff}}$ describes the ordering of systems up to the prethermal time scale $\sim e^{\omega/J} \sim e^{1/JT}, JT\ll1 $, and forms the basis to define Landau's symmetry breaking. Correspondingly, the prethermal Landau DTC dynamics decays at the time scale $t\sim e^{1/JT}$, as the approximated effective Hamiltonian description fails at this point. That means if the prethermal condition is violated, as in our case $JT\sim 1 $, one cannot arrive at the effective prethermal Hamiltonian in the first place, leaving further discussions of Landau's symmetry breaking groundless. Then, we are forced to go beyond the Landau's scheme, and introduce the cat scar enforced DTCs attributable to the suppression of spin flips due to strong Ising interactions, as quantitatively described by the selection rules.
	%Consequently, features in prethermal Landau DTCs are not expected to occur here. For instance, increasing the driving frequency has been shown to significantly prolong the lifetime of prethermal Landau DTCs, because the thermalization time $\tau_* \sim e^{\omega/T}$ is delayed. However, for the cat scar DTCs, such an action means the interactions are weakened (recall that in Eq.~\eqref{eq:model}, all energy units are taken to be driving frequency, so $J$ in Eq.~\eqref{eq:model} means $ J/\omega$), which may actually shorten the cat scar DTC lifetime and amplitudes.

	Having compared cat scar DTCs with the other cases enforced by SP, we next briefly discuss the comparisons with DTCs without SP. In principle, as emphasized in Ref.~\cite{Luitz2020}, period-doubled oscillations without localized cat states generally would involve a diffusive dynamics, unlike the local DTC oscillations with a fixed spin pattern for the systems enforced by SP of cat states. In terms of early time evolutions, for diffusive systems, any initial states would quickly relax to a homogeneous spin configuration on average, where extensive numbers of different Fock states are involved. Then, the system may undergo certain global spin oscillations with the homogeneous patterns, and approximate conservation laws such as U(1) symmetries may delay the decay of global oscillations. In terms of late time dynamics, unlike systems with SP where DTC oscillations would persist for exponentially long time $\sim e^L$, the diffusive cases typically host a lifetime being certain fixed power-law insensitive to the change of system size $L$.
	
	In addition to discussing the distinctions between cat scar DTCs with several known examples listed above, as an application of the analytical framework constructed here, we also aim at offering practical ways to distinguish the many-body versus single-spin nature of DTC-like oscillations within spatial-translation-invariant settings, especially based on early time dynamics accessible to experiments. Our construction of SP theory here emphasizes the importance of {\em strong} Ising interaction. That results in a stable many-body cat scar robust against generic perturbations and shows rigid DTC oscillations persisting for exponentially long time $\sim e^L$. In contrast, for certain weakly interacting system, a clean DTC-like oscillation may emerge, which is instead dominated by single-particle effects for early-time dynamics. In Sec.~\ref{sec:diffusive}, we would discuss such an issue in detail, and offer two specific perspectives to distinguish many-body versus single-spin effects. We sketch the results below.
	
	First, we suggest to check whether certain DTC phenomena rely on fine-tuned single-spin echos. Specifically, a strong longitudinal magnetic fields followed by slightly imperfect spin flips in DTC models may add up into a spin echo for individual spins, if the longitudinal field strength is close to certain values. Such single-particle echos, when assisted by weak interactions, may result in certain phenomena resembling ``dynamical freezing"~\cite{Das2010,Haldar2018,Haldar2021,Haldar2022} with prolonged DTC lifetime as discussed in Refs.~\cite{Luitz2020}. Further, without transverse interactions in Ref.~\cite{Luitz2020}, the spin echo may even produce {\em local} oscillations that very much resembles a stable many-body DTC with SP. However, if one changes the longitudinal field strength away from the fine-tuned echo limit, the DTC-like oscillation is immediately suppressed, showing a strong dependence on single-spin physics. In contrast, the cat scars surviving in strongly interacting regimes are insensitive to whether such single-particle echos occur or not.
	
	Second, when perturbations only involve single-spin terms, i.e. for transverse magnetic field perturbations, a weakly interacting system may host emergent integrability for its lowest-order effective Hamiltonian. The approximate integrability results in a slow relaxation in the system's early-time dynamics. In contrast, for strongly-interacting system, we have obtained a rigorous effective Hamiltonian showing its overall non-integrable nature even at the lowest-order. It serves as a double-check that for strongly interacting systems, DTC oscillations are enforced by cat scars robust against generic perturbations, rather than on model fine-tuned integrability. Based on such an understanding, we suggest the inclusion of {\em two-spin terms} into perturbations, which could sharply distinguish the two cases. For weakly interacting systems dominated by integrable effective Hamiltonians at early time, adding interacting two-spin terms for perturbations drastically accelerates relaxation and suppresses DTC-like oscillations. In contrast, cat scar DTCs already live in the regime dominated by many-body effects, and are insensitive to such changes.

	Finally, results in this work may help push forward ongoing experiments. For instance, inhomogeneous DTC patterns found here offer a valuable opportunity to prove eigenstate orders in clean systems based on early time data, in parallel to MBL cases~\cite{Mi2022,Randall2021,Frey2022}. This contrasts situations without Fock space localization, where all initial states quickly relax to a homogeneous pattern and only total spin oscillation exists. 
	Also, the coexisting FM and AFM patterns in scar-enforced DTCs here are sharply distinct from prethermal cases, which host only one of these patterns in low-temperature sectors~\cite{Kyprianidis2021}.
	Such predictions may find applicability in platforms of intermediate scales, including superconducting qubits~\cite{Mi2022,Frey2022,Zhang2022,Xu2021}, nitrogen-vacancy centers~\cite{Randall2021,Beatrez2023}, trapped ions~\cite{Zhang2017,Kyprianidis2021}, and Rydberg atoms~\cite{Bernien2017,Bluvstein2021,Maskara2021}, all of which allow for single-site manipulation and detection. 
	%Cat scars with their peculiar SP and clean DTC features enrich the plethora of unique mesoscopic physics, ranging from fragmentation of Bose-Einstein condensates~\cite{Nozieres1982,Mueller2006,Evrard2021} to  new universal classes in Floquet thermalization~\cite{Morningstar2022}.  
	
	The remainder of this work is organized as follows. In Sec.~\ref{sec:model} we introduce a main model and briefly illustrate numerically the signatures of cat scars and the DTC dynamics. Then, the major contents for analytical framework is presented in Sec.~\ref{sec:analytics}. Further, the distinctions between single-spin and interaction effects are elaborated in Sec.~\ref{sec:diffusive}. And we conclude in Sec.~\ref{sec:con}. For intuitiveness of discussions, algebras are presented chiefly in the Appendix, while the main text would focus instead on a more physical discussion.

	\section{Exemplary model and cat scar signatures\label{sec:model}}
	
	To be concrete, we consider a periodically kicked Ising chain $ H_0(t+T)=H_0(t) $ constantly perturbed by $ H' $, namely, $ H(t) = H_0(t) + \lambda H' $ with
	\begin{align}\nonumber
		&\frac{H_0(t)T}{2\hbar} 
		=
		\begin{cases}
			(\pi/2)\sum_{j=1}^L \tau^x_j ,
			&
			t\in[0,T/2)
			\\
			\sum_{j=1}^L J_j \tau^z_j \tau^z_{j+1} ,
			&
			t\in [T/2, T)
		\end{cases}
		, 
		\\
		\label{eq:model}
		& \frac{H'T}{2\hbar} = \sum_{j=1}^L \left( \phi\tau^x_j \tau^x_{j+1} + \sum_{\mu=x,y,z} \theta_\mu\tau^\mu_j\right), \quad
		\phi^2 + \sum_{\mu=x,y,z} \theta_\mu^2 = 1.
	\end{align}
	That corresponds to the Floquet operator characterizing evolutions to period ends as
	\begin{align} \nonumber
		U_F &= {\cal T} e^{-(i/\hbar)\int_0^T dtH(t)} 
		\\
		\nonumber
		&=
		e^{ 
			-i\sum_{j=1}^L \left(
			J_j \tau^z_j \tau^z_{j+1} + \lambda ( \phi \tau^x_j \tau^x_{j+1} + \theta_x \tau^x_j + \theta_y \tau^y_j + \theta_z \tau^z_j)
			\right)
		}
		\\
		&
		\qquad \times
		e^{ 
			-i\sum_{j=1}^L \left(
			\frac{\pi}{2}\tau^x_j + \lambda ( \phi \tau^x_j \tau^x_{j+1} + \theta_x \tau^x_j + \theta_y \tau^y_j + \theta_z \tau^z_j)
			\right)
		}
	\end{align}
	where ${\cal T}$ means time-ordering.
	Here $ \tau^{x,y,z}_j $ are Pauli matrices for spins at sites $ j=1,2,\dots,L $.
	Uniform interaction $ J_j =1 $ is taken unless  specified otherwise. To reduce the effects of model fine-tuning, numerical results are allowed to average over random numbers $ \phi, \theta_\mu \in (0,1) $ so as to simulate generic perturbations. Then, perturbation strength is captured by a single parameter $ \lambda $. In the unperturbed limit $ \lambda=0 $, spins are perfectly flipped periodically $ \tau^z_j(nT) = (U_F^{\dagger})^{n} \tau^z_j U_F^{n} = (-1)^{n}\tau^{z}_j  $. A natural choice of observable is then the spatiotemporal magnetization orders
	\begin{align}\label{eq:corr}
		M(nT) =  \frac{(-1)^{n}}{L} \sum_{j=1}^L \langle \psi_{\text{ini}} | \tau^z_j (nT) | \psi_{\text{ini}}\rangle 
		\langle \psi_{\text{ini}} | \tau^z_j |\psi_{\text{ini}} \rangle.
	\end{align}
	DTC features a restored oscillation for $ \lambda\ne 0  $, as perturbations are neutralized generically by interactions. That means $ M(t) $ will assume a constant value over time, representing persisting period-$ 2T $ spin flips. Crucially, $ M(t) $ describes  oscillations of {\em individual} spins with respect to initial configurations $ \langle\psi_{\text{ini}}|\tau^z_j|\psi_{\text{ini}}\rangle $ {\em at each site}.
	%, unlike the  magnetization $ {\mathscr M}(t) = (-1)^t \sum_{j=1}^L \langle \psi_{\text{ini}} |\tau^z(t) |\psi_{\text{ini}}\rangle/L  $ for {\em total} spin oscillations. 
	Local information in $ M(t) $ is indispensable to illustrate inhomogeneous DTC patterns later.
	
	We choose the model in Eq.~\eqref{eq:model} due to two considerations. 
	On the one hand, the major point of this work is to analytically discuss the underlying {\em mechanism} for DTCs in strongly-interacting clean Floquet systems. Then, adopting a model closely related to previous numerics and experiments (i.e. for the unperturbed $H_0$) would facilitate illustration and comparison of the phenomenon. 
	On the other hand, the perturbation $H'$ considered here is more generic than many previous studies involving only a global spin tilting $\lambda \sum_{j=1}^L \tau^x_j $, i.e. resulting in evolutions $e^{-i\left(\frac{\pi}{2}-\lambda\right)\sum_{j=1}^L \tau^x_j} $ for the first-half of the period. We would show explicitly in Sec.~\ref{sec:diffusive} that such generic perturbations, especially for the two-spin terms $\sim \tau^x_j \tau^x_{j+1}$, are vitally important to distinguish the following two cases: (1) a cat-scar enforced DTC robust against generic perturbations, which lives in strongly interacting regimes, and (2) some other period-doublings dominated by single-particle physics, which sensitively rely on model or parameter fine-tunings. 
	
	\begin{figure}[h]
		\parbox{4.25cm}{\includegraphics[width=4.2cm]{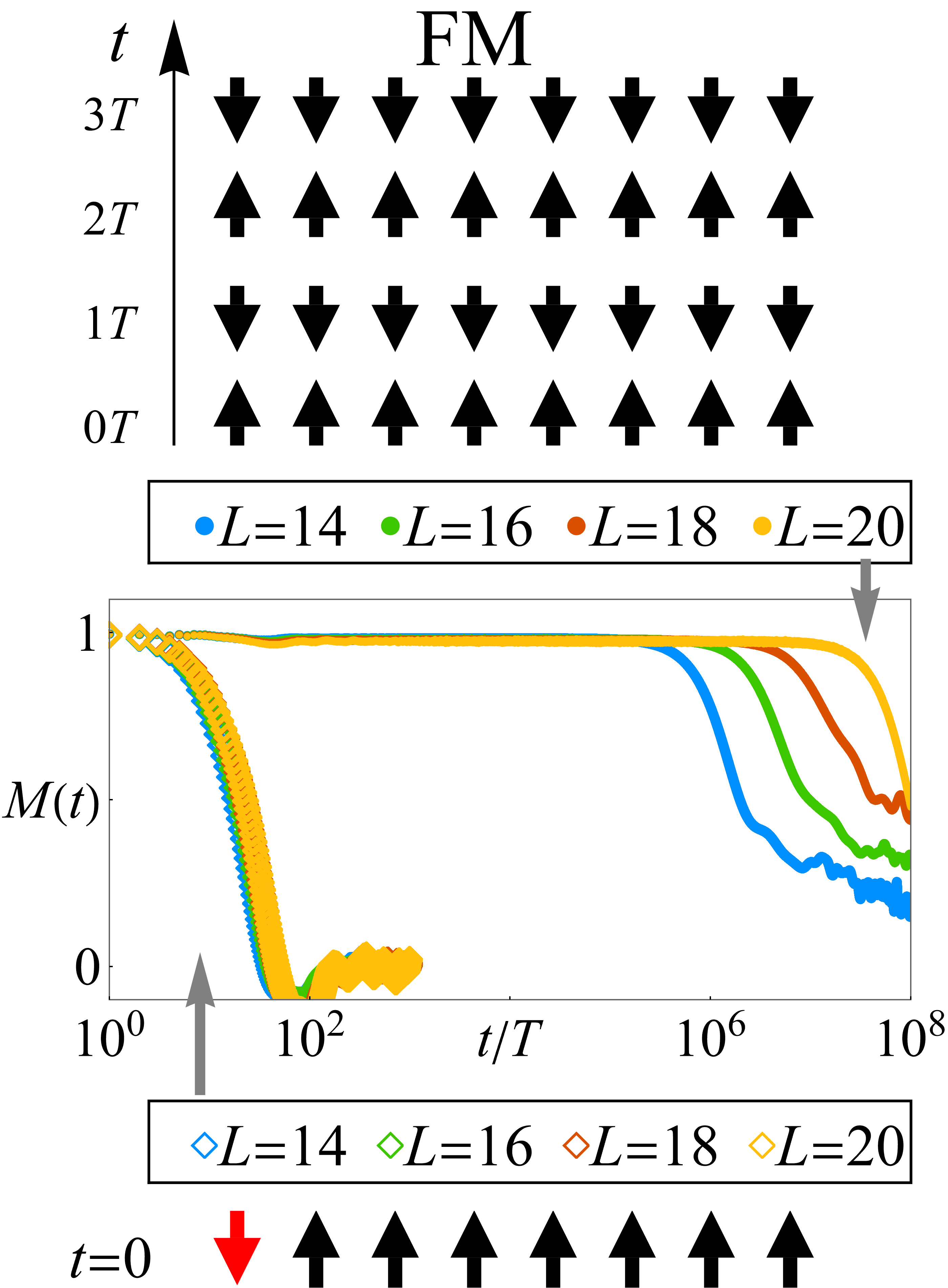}}
		\parbox{4.25cm}{\includegraphics[width=4.2cm]{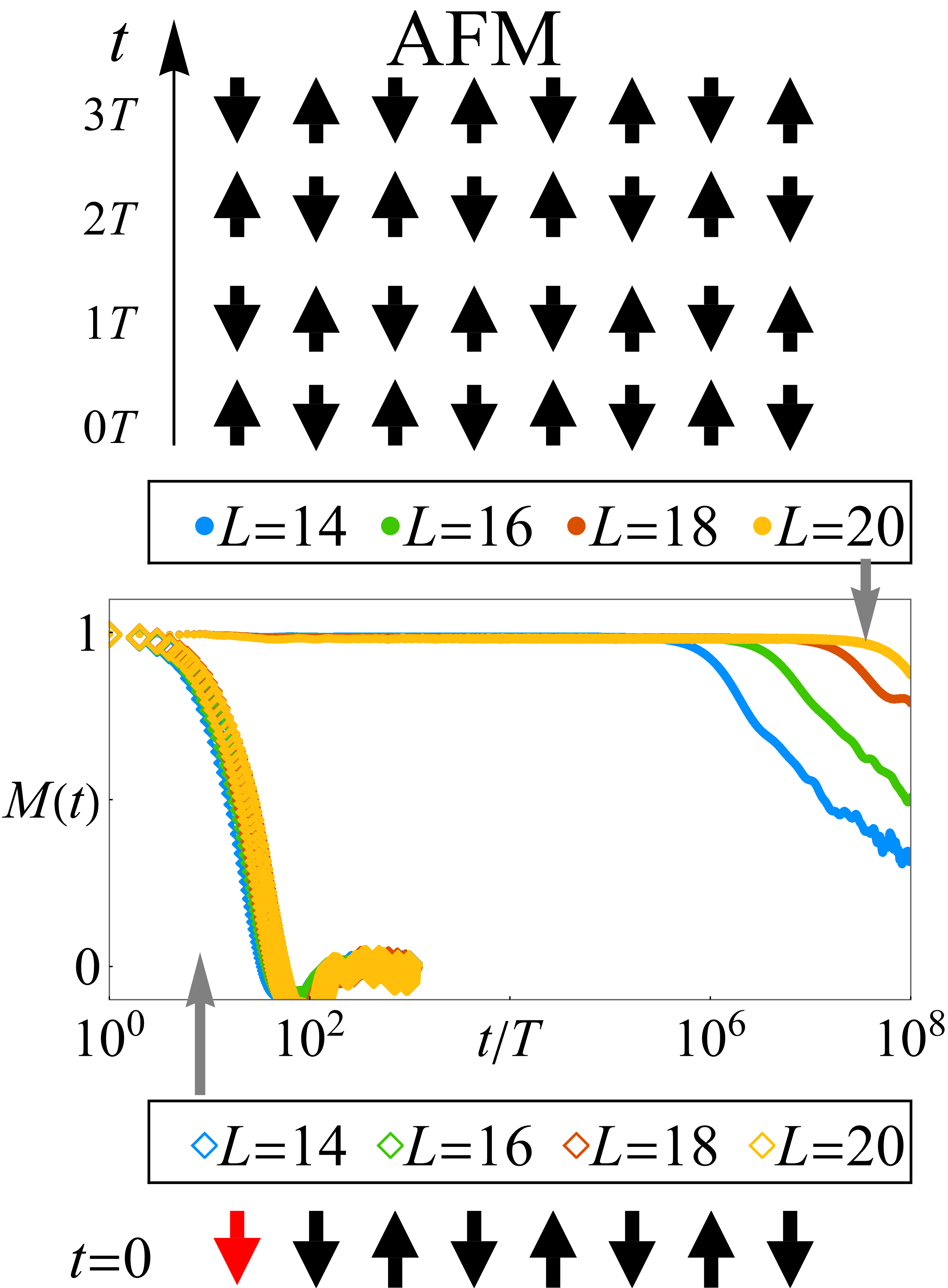}}
		\caption{\label{fig:dynamics} Two types of initial states (FM and AFM) leading to clean DTC oscillations, where the AFM type was unnoticed before. Flipping just one spin (denoted by red arrows) for the initial state drastically reduces the life time, indicating scar physics. For $ L=14, 16, 18, 20 $, we average data at each instant over $ 10^3, 10^3, 10^2, 10^1 $ samples of $ (\phi, \theta_{x,y,z}) $ respectively, and $ \lambda=0.05 $. Periodic boundary condition is taken throughout this work to eliminate edge effects.}
	\end{figure}
	
	Let us gain some intuitions through exact diagonalization of Eq.~\eqref{eq:model}. In Fig.~\ref{fig:dynamics}, we immediately see that both FM and AFM initial states lead to stable DTC oscillations persisting for exponentially long time. Contrarily, flipping just one spin (denoted by red arrows) for the initial states drastically shrinks the lifetime to $ t/T < 2\pi/\lambda\sim 10^2 $. That strongly indicates the coexistence of scars with FM and AFM configurations. 
	
	To characterize scars further, we state rigorously the concept of SP, which refers to Floquet eigen-solutions $ U_F|\omega_n\rangle = e^{i\omega_n}|\omega_n\rangle  $ satisfying two conditions. (i) {\em Pairwise Fock space localization}, where a pair of eigenstates $ |\omega_1\rangle, |\omega_2\rangle $ are dominated by different linear combinations of just two Fock product states $ |\{s_j\}_1\rangle, |\{s_j\}_2\rangle $. Here we denote $ |\{s_j\}\rangle\equiv |s_1\rangle \otimes |s_2\rangle \otimes \dots \otimes |s_L\rangle $, with $ \tau^z_j |s_j\rangle = s_j |s_j\rangle $, $ s_j=\pm1 $. Due to orthogonality, other eigenstates involve vanishing overlap with $ |\{s_j\}_{1,2}\rangle $. (ii) {\em Fixed spectral gap}, where the frequency difference $ \Delta\omega = \omega_{1} - \omega_{2} $ of the eigenstate pair remains unchanged under generic perturbations. With both conditions, a pertinent Fock initial state $ |\{s_j\}_1\rangle $ or $ |\{s_j\}_2\rangle $ overlaps chiefly with the spectral paired eigenstates, and results in oscillations of local observables, i.e. $ \langle \{s_j\}_1 | \tau_j^z(t) | \{s_j\}_1 \rangle \sim \langle\omega_1 |\tau^z_j|\omega_2\rangle e^{i\Delta\omega t} + c.c. $, with locked frequencies $ \Delta\omega $.

	\begin{figure}[h]
		\parbox[b]{4cm}{
			\includegraphics[width=4cm]{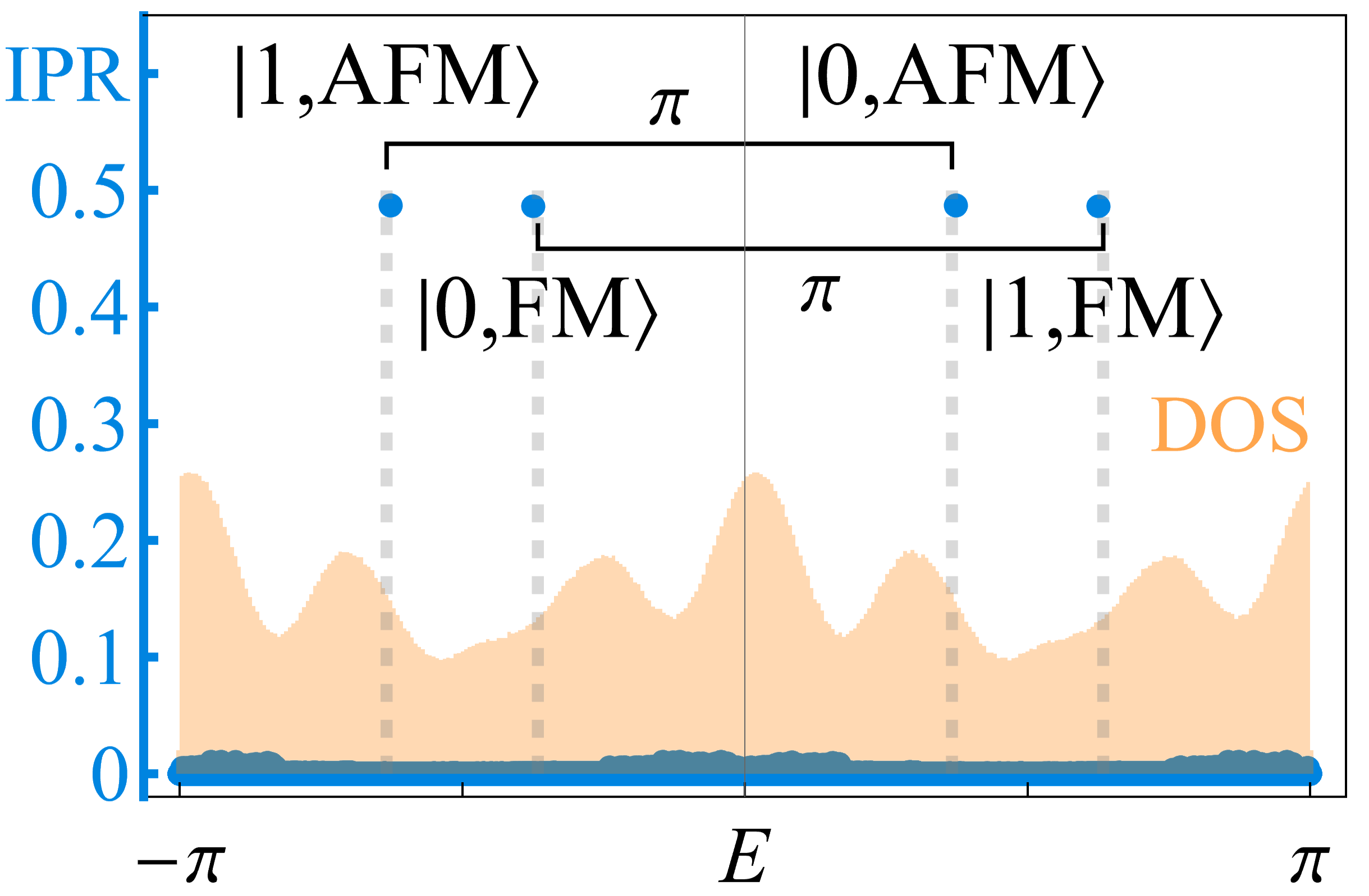}
			\\ (a) IPR and DOS for $ L=20 $}
		\parbox[b]{4cm}{
			\includegraphics[width=3.7cm]{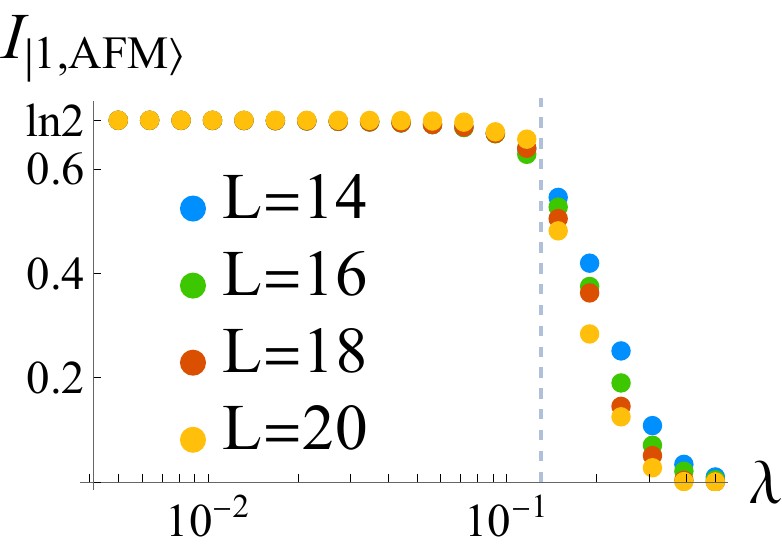}
			\\ (b) Scaling of MI for AFM scar}
		\caption{\label{fig:cats} Signature of cat scars. (a) IPR (blue dots) and DOS (yellow shadow). The scars are around quasienergy $ e^{iE(\ell,\text{FM})} = e^{-i[(\pi/2 + 1)L+\pi\ell]} $ and $ e^{iE(\ell,\text{AFM})} = \pm e^{-i[(\pi/2-1)L+\pi\ell]} $, where SP quantum numbers $ \ell=0,1 $. Scars feature exceptionally high IPRs signaling their Fock space localization. (b) Scaling of mutual information for the $ |1,\text{AFM}\rangle $ scar for sites $ j=1 $ and $ j=L/2+1 $. Data in (a) consists of 1 sample only, $ (\phi, \theta_x, \theta_y, \theta_z) = (0.3858, 0.7395, 0.3944, 0.3857) $ with $ \lambda=0.05 $, while (b) is averaged over many samples as in Fig.~\ref{fig:dynamics}.}
	\end{figure}
	Accordingly, SP in clean systems can be efficiently captured by eigenstates' inverse participation ratio IPR$(\omega_n) = \sum_{\{s_j\}} | \langle \{s_j\} | \omega_n\rangle |^4   $, as plotted in Fig.~\ref{fig:cats} (a). Larger value of IPR implies that an eigenstate $ |\omega_n\rangle $ is dominated by fewer configurations $ |\{s_j\}\rangle $, and therefore more Fock localized. While majority eigenstates do show vanishing IPRs typical of delocalized clean systems, there are four scars with exceptionally high IPRs hiding deeply inside the gapless Floquet spectrum (see density of states (DOS) in Fig.~\ref{fig:cats} (a)). These scars maintain a pairwise rigid quasienergy difference $ \Delta\omega \approx \pi $, satisfying SP condition (ii). Further, the value $ 0.5 $ for scar IPRs indicates that each scar is dominated by two Fock states. Such a Schr\"{o}dinger's cat type of eigenstates exhibit long-range correlations, which can be revealed by finite mutual information (MI) $ I = S_1 + S_{L/2} - S_{1,L/2} $ between distant sites~\cite{Else2016}. Here the entanglement entropy of a certain site $ S_j = -\text{tr}\left(\rho_j \ln \rho_j \right) $ is obtained from the reduced density matrix $ \rho_j = \text{tr}_{\{s_{k\ne j}\}} \rho $ for the chosen eigenstate $ \rho\equiv |\omega_n\rangle \langle \omega_n|  $. In Fig.~\ref{fig:cats} (b), $ I\rightarrow\ln 2 $ surviving finite $ \lambda $ is illustrated for the $ |1,\text{AFM}\rangle $ scar, as other scars behave similarly. Thus, SP condition (i) is also confirmed.
	
	Clean DTCs with unexpected AFM patterns call for a practical algorithm to enumerate underlying scars generically. Also, it is desirable to quantify the behaviors of these scar eigenstates in a generic setting. For these two purposes, we introduce an analytical framework below.
	
	%An uncorrelated state $ |\omega_n\rangle $ gives vanishing $ I $ as distant sites are mutually independent. However, as exemplified in Fig.~\ref{fig:cats} (b), these scars exhibit $ I\rightarrow \ln2 $ surviving finite perturbations. Following the terminology in Ref.~\cite{Else2017}, we would refer to $ |\text{(A)FM}\pm\rangle $ ``cat scars" here. The remainder of this work is then devoted to enumerating and analytically understanding them.

	\section{Quantitative predictions from analytical framework\label{sec:analytics}}
	
	\subsection{Symmetry indicators for cat scar patterns \label{subsec:symm}}
	
	We would start our analysis from the unperturbed limit $\lambda=0$, and observe the eigenstate structures for later applications. Under spatial translation invariance, a major property for the Floquet eigenstates is that there exists a large degree of degeneracies, which implies that these degenerate levels can be easily hybridized upon perturbations, leading to ergodic behaviors. Scars, contrarily, are the only non-degenerate ones defying such a fate. Specifically, at $ \lambda = 0 $ for Eq.~\eqref{eq:model},
	\begin{align}\label{eq:u0}
		U_0 \equiv U_F(\lambda=0) = (-i)^L \prod_j e^{-i \sum_j J_j \tau^z_j \tau^z_{j+1}} P,
	\end{align}
	where Ising symmetry $ P = \prod_{j=1}^L\tau^x_j $ flips all spins $ P|\{s_j\}\rangle = |-\{s_j\}\rangle $.
	Solutions to $ U_0|\ell,\{s_j\} \rangle = e^{iE(\ell, \{s_j\})} |\ell, \{s_j\}\rangle $ then read
	\begin{align}\nonumber
		&
		|\ell, \{s_j \} \rangle = \sum_{m=0,1} (-1)^{m\ell}  |(-1)^m\{s_j\} \rangle/\sqrt2,
		\qquad
		\ell=0,1 \text{ mod } 2,
		\\ 
		\label{eq:finetunesol}
		&
		E(\ell, \{s_j\}) =
		E_{\text{sp}}(\ell) + E_{\text{Ising}}(\{s_j\})\quad
		\text{mod} \quad 2\pi,
	\end{align}
	where the spectral pairing $ E_{\text{sp}}(\ell) = \pi \ell $ and Ising interaction energy $ E_{\text{Ising}}(\{s_j\}) = - \sum_j J_j s_j s_{j+1} $. Quasienergy shift $ -\pi L/2 $ due to the factor $ (-i)^L $ in $ U_0 $ is neglected.

	\begin{figure}
		[h]
		\includegraphics[width=8cm]{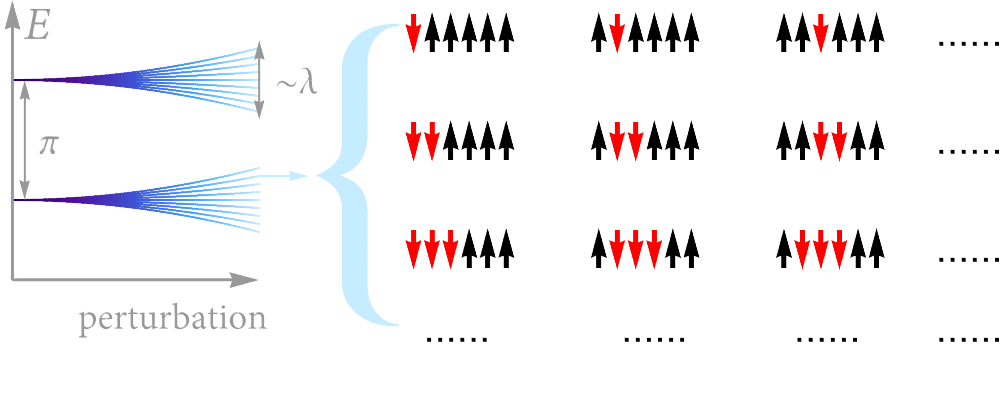}
		\caption{\label{fig:deg_schematic} $ L^2 $ configurations sharing the same total number of domain walls $w=2$, and therefore are degenerate at $\lambda=0$ with Ising energy $ E_{\text{Ising}} = J(L-4) $ in Eq.~\eqref{eq:dwsols}. That means under perturbations, those eigenstates will undergo a complete reconstruction, where {\em each} eigenstate involves a macroscopic number of configurations. In a reversed sense, if one chooses a single Fock state as the initial state, it will overlap with a macroscopic number of eigenstates with different quasienergy ranging over $ \sim\lambda $, and therefore any oscillation is expected to decay within the characteristic time scale $ 2\pi/\lambda $, as observed in Fig.~\ref{fig:dynamics} for the initial states with red arrows. }
	\end{figure}
	In the unperturbed limit $\lambda=0$, each eigenstate pair $ |0,\{s_j\}\rangle $ and $ |1,\{s_j\}\rangle $ in the solutions satisfy SP, as they consist of two Fock states $ |\pm\{s_j\}\rangle $ and differ in quasienergy by $ E(1,\{s_j\}) - E(0,\{s_j\}) = \pi $. However, the Ising energy for translation invariant systems ($ J_j=J $)
	\begin{align}\label{eq:dwsols}
		E_{\text{Ising}} = -J(L-2w), 
		\qquad 
		w=0,2,4,\dots,L
	\end{align}
 	only depends on the {\em total} number of domain walls (DW)
	\begin{align}
		w =\sum_{j=1}^L w_{j,j+1}, \qquad w_{j,j+1} =  (1-s_j s_{j+1})/2.
	\end{align}
	Here a DW denotes the bond connecting opposite spins which separates two FM domains. Then, majority eigenstates are grouped into degenerate subspace labeled by $ (\ell,w) $, each spanned by large numbers of configurations $ \{ |\ell,\{s_j\}_1\rangle, |\ell,\{s_j\}_2\rangle,\ldots \}$ with different allocations of $ w $ DWs. Under perturbation, each reconstructed eigenstate could involve all configurations within a subspace, and the degenerate levels are lifted into a continuous band of bandwidth $ \sim\lambda $, as schematically illustrated in Fig.~\ref{fig:deg_schematic} for the $w=2$ case. Both conditions of SP are therefore broken. Correspondingly, an initial Fock state generically overlaps with the whole delocalized band, so any oscillation is expected to dephase within the time scale $ 2\pi/\lambda $, as observed in Fig.~\ref{fig:dynamics} for the initial states with red arrows. 
	
	To identify non-degenerate scars efficiently, we introduce below a symmetry-based algorithm. Take spatial translation symmetry for instance, $ [\mathbb{T}_x, H(t)]=0 $, where $ \mathbb{T}_x |\{s_j\}\rangle = \mathbb{T}_x|s_1s_2\dots s_{L-1}s_L\rangle = |s_L s_1 s_2\dots s_{L-1}\rangle = |\{s_{j-1}\}\rangle $. Intuitively, scar configurations should exhibit identical DW numbers at all symmetry equivalent bonds, such that relocations of DWs cannot produce new degenerate configurations. Importantly, $ |\pm\{s_j\}\rangle $ host identical DW distributions. Then, scar patterns only need to satisfy a {\em projective} translation symmetry
	\begin{align}\label{eq:symm}
		\mathbb{T}_x|\{s_j\}\rangle = |\pm \{s_j\}\rangle,
	\end{align}
	where $ \pm $ signs precisely give the FM and AFM cat scars in Fig.~\ref{fig:cats} respectively, 
	\begin{align}\label{eq:scars}
		|\ell,\text{FM}\rangle \equiv |\ell,\{s_j=(+1)^j\}\rangle,
		\,\,
		|\ell, \text{AFM}\rangle \equiv |\ell, \{s_{j}=(-1)^j\}\rangle.
	\end{align}
	More rigorously, $ |\pm\{s_j\}\rangle $ are recombined into non-degenerate spectral pair satisfying $ P|\ell, \{s_j\}\rangle = (-1)^\ell |\ell, \{s_j\}\rangle $. Then, Eq.~\eqref{eq:symm} corresponds to the invariance of eigenstates $ \mathbb{T}_x |\ell,\{s_j\}\rangle = (\pm 1)^\ell |\ell, \{s_j\}\rangle $, so that DW operators 
	\begin{align}\label{eq:hatw}
		\hat{W} = \sum_{j=1}^L \hat{W}_{j, j+1}, 
		\qquad
		\hat{W}_{ij} = (1-\tau^z_i \tau^z_{j})/2,
	\end{align}
	act identically on all symmetry related bonds, i.e. $ W_{ij}|\ell, \{s_j\}\rangle = w_{ij}|\ell,\{s_j\}\rangle \, \Rightarrow \, W_{i+1,j+1} |\ell, \{s_j\}\rangle = \mathbb{T}_x B_{ij} \mathbb{T}_x^{-1} |\ell,\{s_j\}\rangle = w_{ij}|\ell,\{s_j\}\rangle $. As a crosscheck, note that the degenerate manifold with $w$ total domain walls would contain $C_L^w = L!/(L-w)! w!$ eigenstates, which denotes the $C_L^w$ ways to allocate the $w$ domain walls. Thus, we confirm that only for the FM ($w=0$) and AFM ($w=L$) subspaces, levels are non-degenerate in the unperturbed limit. Staying within the systems hosting crystalline spacegroup symmetries, Eq.~\eqref{eq:symm} constitutes a generic algorithm to quickly identify scars, as $ \mathbb{T}_x $ can be replaced by other symmetry operations.

	Clarifications for cat scar conditions are in order. First, translation symmetry $ \mathbb{T}_x $ is {\em not} required. Rather, removing $ \mathbb{T}_x $ means that degeneracy for eigenstates may be lifted and effects of spin fluctuations need not accumulate, which may even further stabilize and/or induce more cat scars. In other words, Eq.~\eqref{eq:symm} is to identify special patterns immune to destruction of SP by $ \mathbb{T}_x $, and scars survive all disorder strengths.
	%$ \mathbb{T}_x $ only serves to destroy SP by inducing degeneracy, and Eq.~\eqref{eq:symm} is to identify non-degenerate configurations immune to such destruction. In other words, cat scars survive all disorder strengths, while stronger disorders lifting degeneracies may stabilize SP for remaining eigenstates via MBL.
	Second, strong interaction of Ising type is {\em necessary} for SP, so as to validate the perturbative treatment including the vital selection rules, which we would elaborate in the next subsection. Note that strong random Ising {\em interaction} is also required in MBL DTCs to enforce localization ~\cite{Khemani2019b,Mi2022,Randall2021,Frey2022}.

	\begin{figure}
		[h]
		\parbox[b]{4.25cm}{\includegraphics[width=4.25cm]{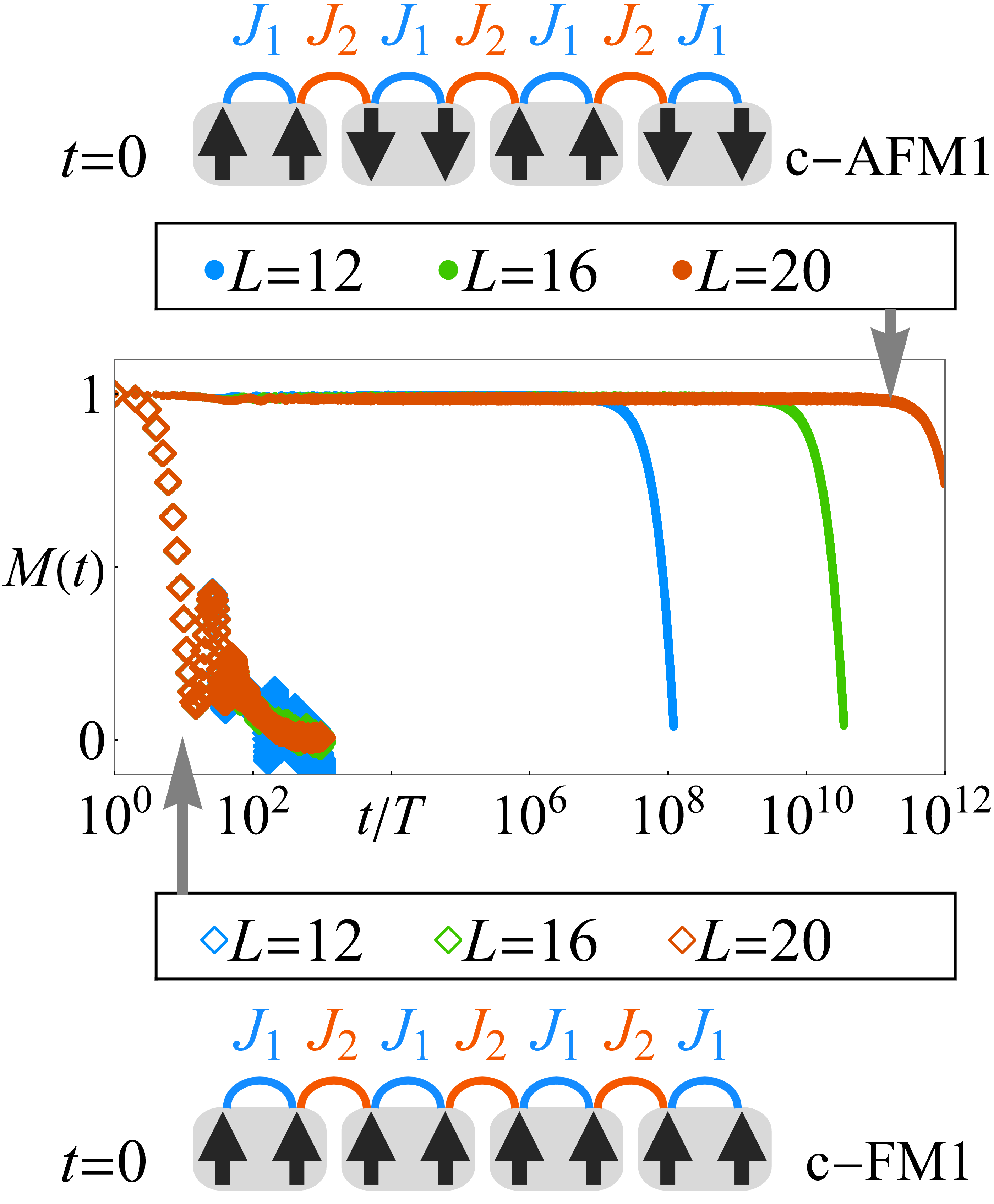} }
		\parbox[b]{4.25cm}{\includegraphics[width=4.25cm]{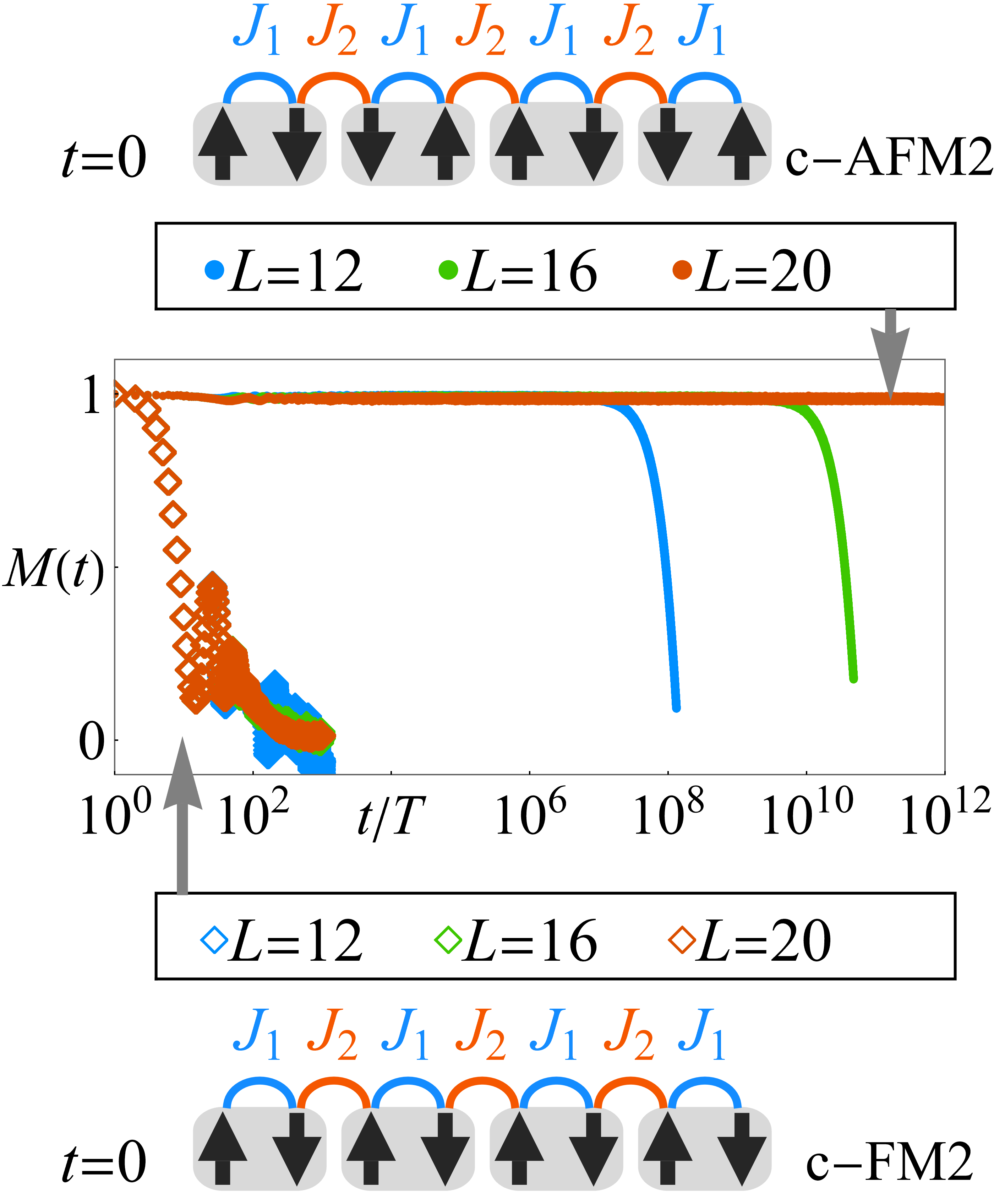} }
		\caption{\label{fig:cafm} Dynamics in systems with two sublattices, where $ J_1 = - J_2 = 1 $. The two non-degenerate c-AFM initial states lead to DTC phenomena persisting for exponentially long time, while the two degenerate c-FM configurations decay quickly upon perturbation. Here we take a single sample for $ (\phi,\theta_\mu) $ as in Fig.~\ref{fig:cats} (a), and $ \lambda=0.05 $. }
	\end{figure}
	With clarifications, we generalize $ \mathbb{T}_x $ in Eq.~\eqref{eq:symm} to generic spatial symmetry $ A $, where cat scar patterns $ \{s_j^{(\text{cat})}\} $ should satisfy
	(1)  projective  symmetry 
	\begin{align}\label{eq:symm_gen}
		A|\{s_j^{(\text{cat})}\}\rangle = |\pm \{s_j^{(\text{cat})}\}\rangle
	\end{align}
	and
	(2) no accidental degeneracy among unperturbed scars. 
	An example is given in Fig.~\ref{fig:cafm} exploiting Eq.~\eqref{eq:model} with $ J_j = (-1)^jJ $, which contains two sublattices hosting $ A = \mathbb{T}_x^2 $.  FM and AFM configurations in Fig.~\ref{fig:dynamics} are understood now as two composite-ferromagnetic (c-FM) patterns $ \mathbb{T}_x^2 |\{s_j^{(\text{c-FM})}\}\rangle = |+\{s_j^{(\text{c-FM})}\} \rangle $, but they are degenerate ($ E_{\text{Ising}} = 0 $) violating criterion (2). Contrarily, the two new composite-antiferromagnetic (c-AFM) patterns $ \mathbb{T}_x^2 |\{s_j^{(\text{c-AFM})}\}\rangle = |-\{s_j^{(\text{c-AFM})}\}\rangle $ are non-degenerate and yield expected DTC dynamics. Thus, we witness the counter-intuitive result that in certain clean systems, only inhomogeneous DTC patterns persist for exponentially long time, but not homogeneous total spin oscillations.

	\begin{figure}[h]
		\parbox[b]{4cm}{
			\includegraphics[width=4cm]{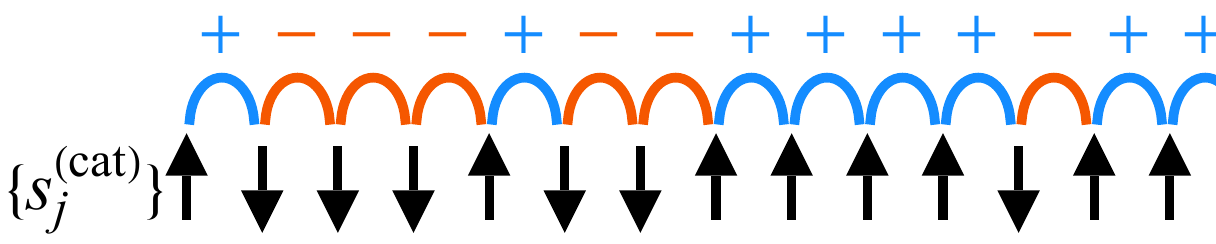}
			\includegraphics[width=4cm]{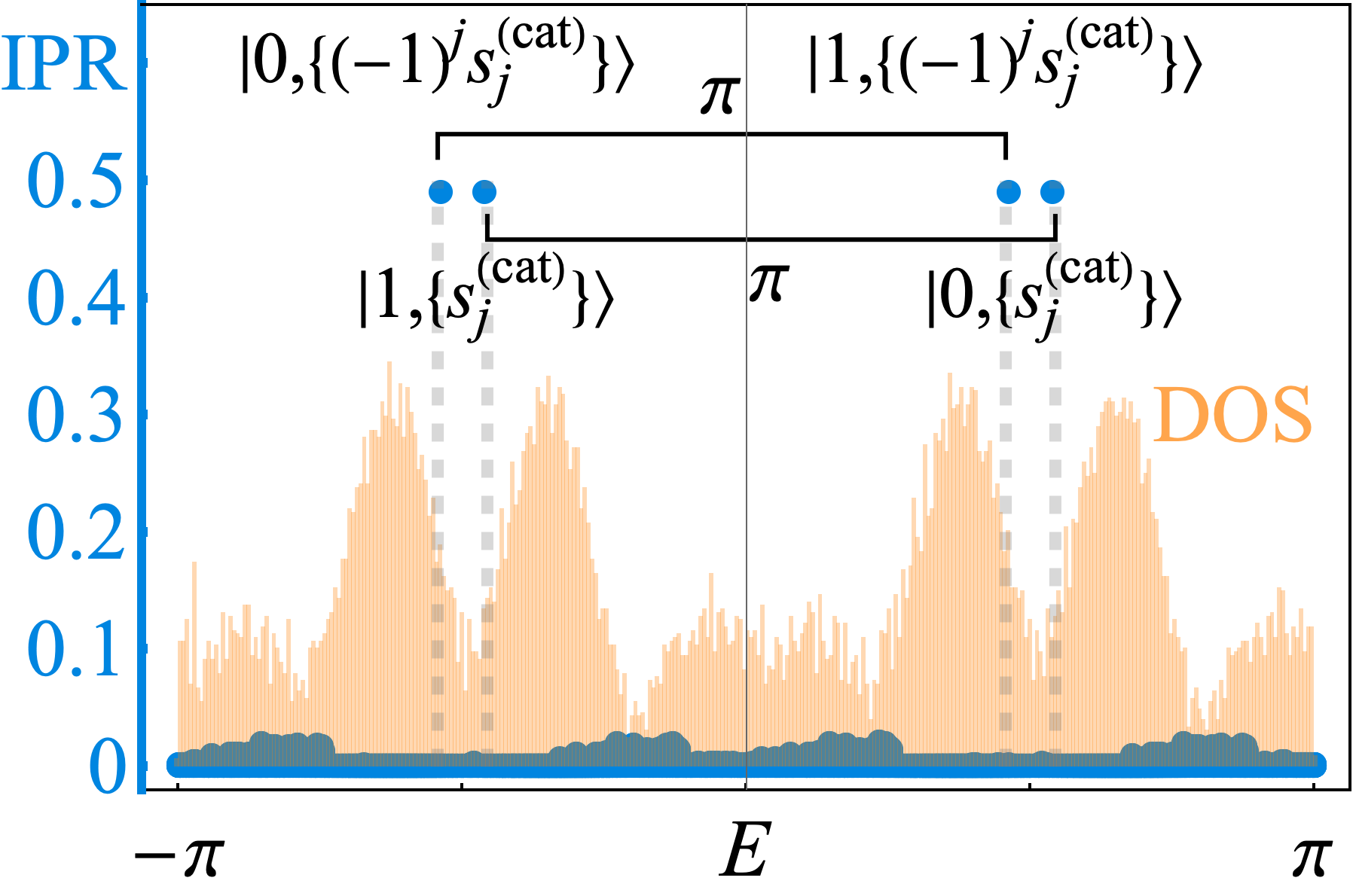}\\
			(a) IPR and DOS
		}
		\quad
		\parbox[b]{4cm}{
			\includegraphics[width=4cm]{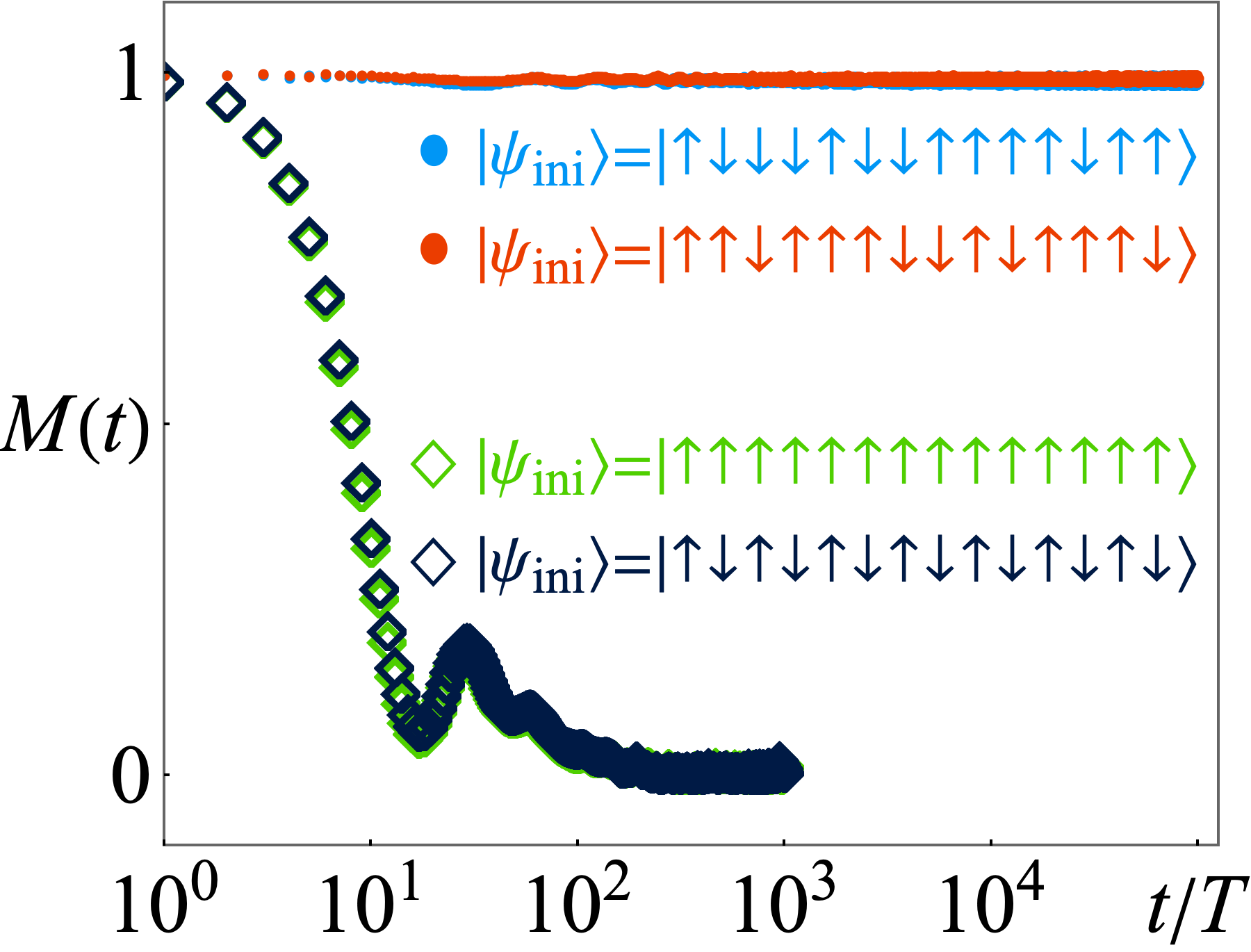}
			\\
			(b) Evolutions
		}
		\caption{\label{fig:arbitrary_pattern} Engineering of cat scars with arbitrary patterns, where interaction strength is uniform, while the signs are different as specified in Eq.~\eqref{eq:Jsign}. An example with $L=14$ is given here, where the cat scar patterns $\{s_j^{(\text{cat})}\}$ is given in the upper panel of (a). Correspondingly, we see persisting oscillations of such patterns in (b), while the previous FM and AFM ones become thermalizing patterns and decay quickly. Perturbation parameters are the same as in Fig.~\ref{fig:cats} (a).
		}
	\end{figure}
	As a further extension of the above analysis, we briefly mention the cases where an arbitrary pattern for cat scars can be achieved by weakly breaking the translation symmetry. Specifically, for any set of four desirable scar patterns,
	\begin{align}
		\pm \{s_j\}, \qquad
		\pm \{ (-1)^j s_j \},\qquad
		s_j = \pm 1.
	\end{align}
	one could engineer them with Ising interactions
	\begin{align}\label{eq:Jsign}
		J_j = (-1)^{s_js_{j+1}} J, 
		\qquad j=1,2,\dots, L.
	\end{align}
	Namely, the strength of interaction is still uniform, but their signs can change among different bonds. 
	An example is given in Fig.~\ref{fig:arbitrary_pattern}.
	
	\begin{figure}[h]
		\parbox[b]{4cm}{
			\includegraphics[width=4cm]{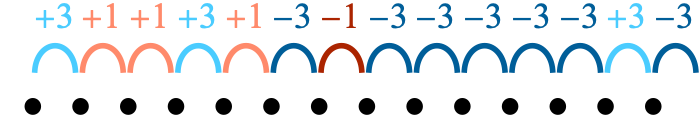}
			\includegraphics[width=4cm]{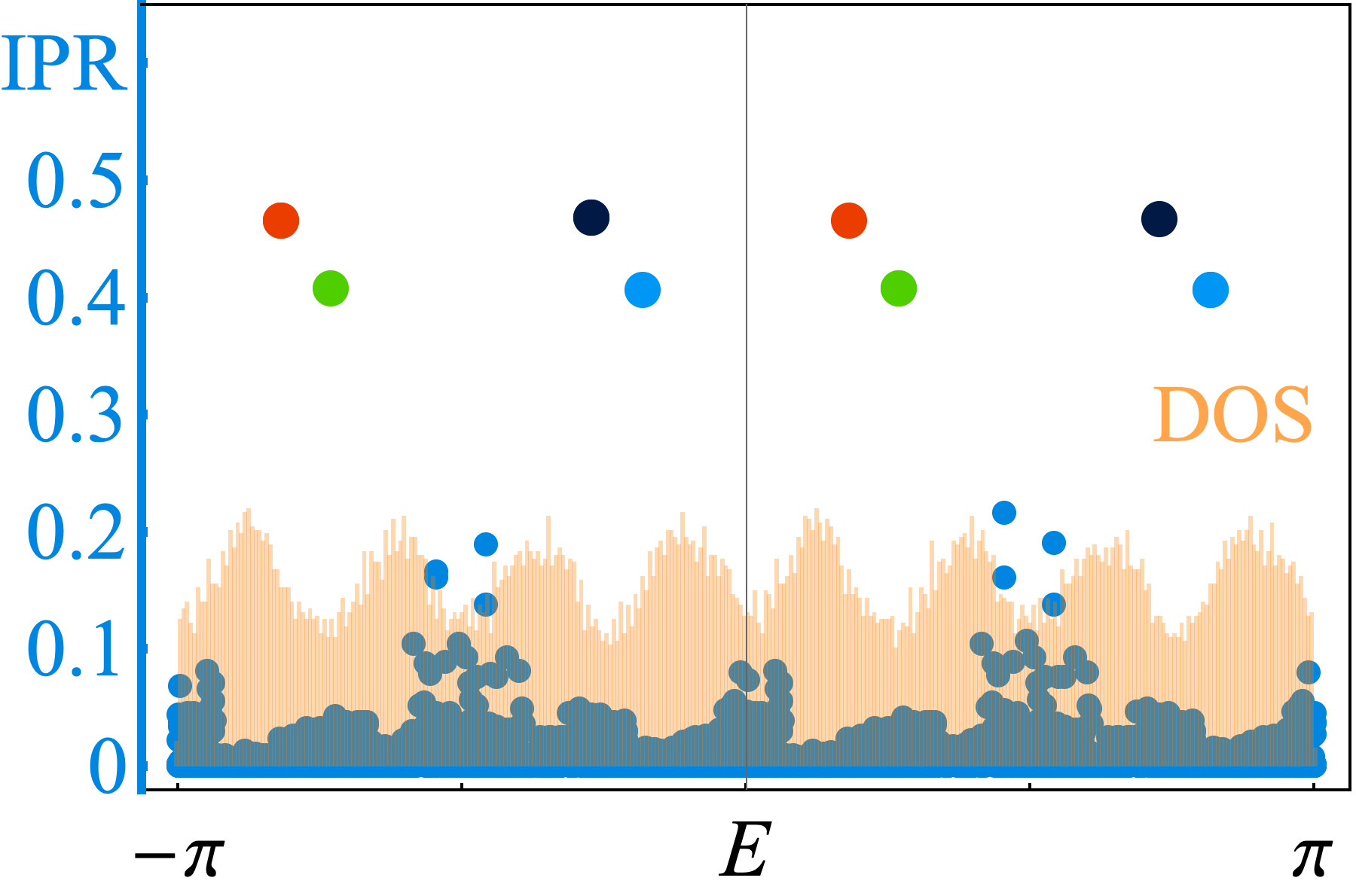}\\
			(a) IPR and DOS
		}
		\quad
		\parbox[b]{4cm}{
			\includegraphics[width=4cm]{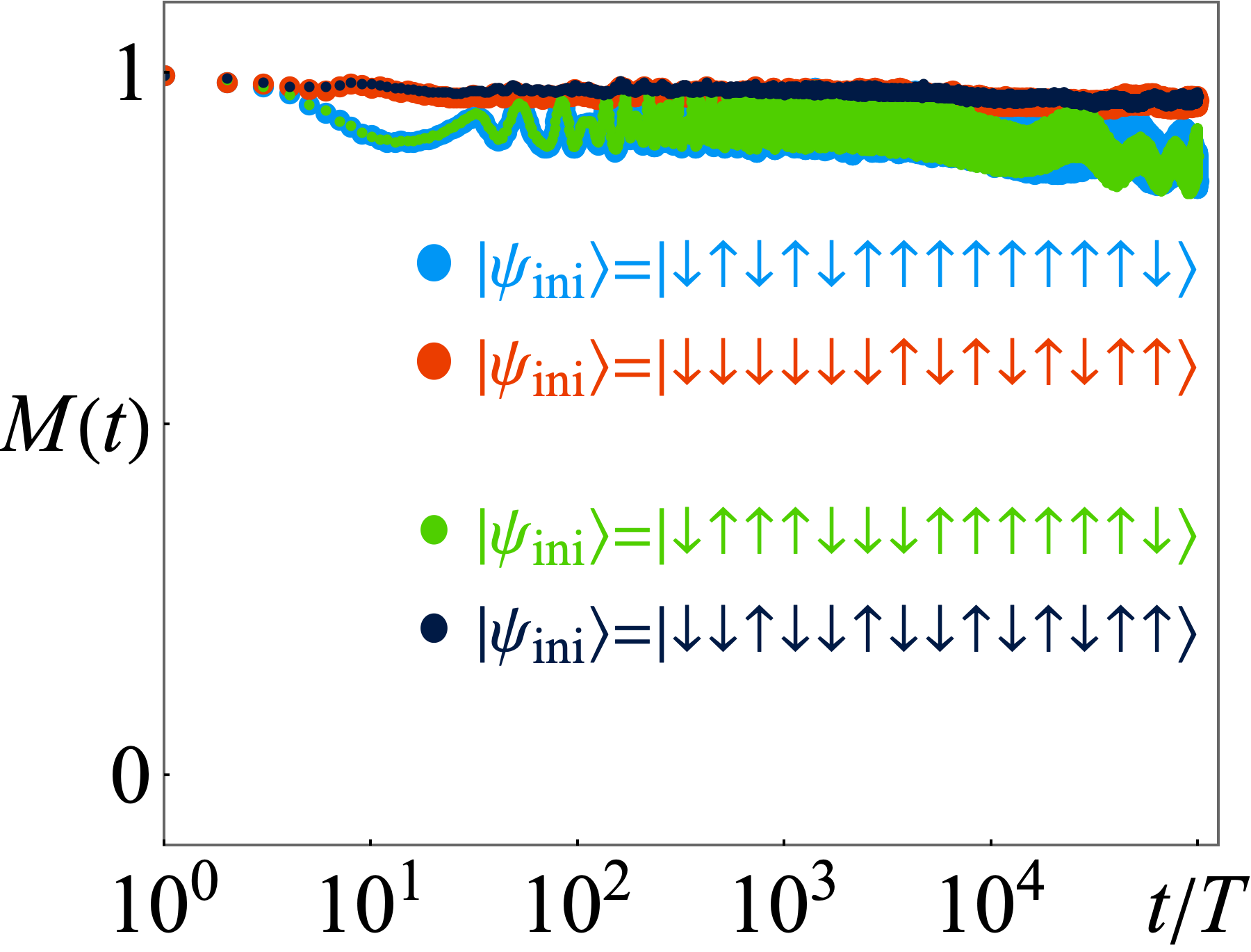}
			\\
			(b) Evolutions
		}
		\caption{\label{fig:morecats} Engineering of four pairs of cat scars with desirable patterns, in contrast to two pairs discussed previously. Here, $\{s_j\}_1, \{\tilde{s}_j\}_2$ in Eq.~\eqref{eq:morecats} are specified by the red and black configurations in (b) respectively. The associated interactions prescribed in Eq.~\eqref{eq:int_morecats} with $J_1 = 2, J_2 = 1$  result in the bond configuration in the upper panel of (a). System size reads $L=14$, and all other perturbations parameters are the same as in Fig.~\ref{fig:cats} (a). The four DTC patterns in (b) is caused by the four pairs of cat scars with large IPRs in (a) denoted by the same color.  Perturbation parameters are the same as in Fig.~\ref{fig:cats} (a).
		}
	\end{figure}
	In principle, the above scheme would allow for engineering more than two pairs of cat scars once the amplitudes of $J_j$ on different sites are allowed to change. For instance, if the desired four pairs of cat patterns are chosen,
	\begin{align}\label{eq:morecats}
		&\pm \{s_j\}_1,
		\quad \pm \{(-1)^j s_j\}_1,
		\qquad
		\pm \{\tilde{s}_j\}_2, \quad
		\pm \{(-1)^j \tilde{s}_j\}_2,
	\end{align}
	one could engineer them using the interaction configurations
	\begin{align}\label{eq:int_morecats}
		J_j = (-1)^{s_js_{j+1}} J_1 + (-1)^{\tilde{s}_j \tilde{s}_{j+1}} J_2,
	\end{align}
	namely, only two types of different amplitudes are involved. An example is given in Fig.~\ref{fig:morecats}.  Thus, an {\em on-demand} engineering of cat eigenstates can possibly be achieved. But further discussions on an extensive number of localized states would require a more definitive understanding of the avalanche effect, which is still under debate currently. Thus, we would postpone this topic to a separate future works. For the remaining part of this work, we would return to the translation invariant cases to rigorously establish the analytical theories of cat scar DTCs, so as to provide a solid anchor point for broader ranges of applications and extensions.

	\subsection{Emergent selection rule for generic perturbations \label{sec:selection}}
	
	In the previous subsection, we have been focusing on energetic degeneracy for the static Ising interaction energy $E_{\text{Ising}}$ in Eq.~\eqref{eq:finetunesol}. A non-degeneracy condition leads to the symmetry indicators for translation-invariant cases, as well as extensions to further engineering of cat scar patterns. These conditions point out the eigenstate configurations where a direct energy degeneracy is avoided. In this subsection, we would further consider the effect of non-degenerate perturbations so as to quantify the robustness of cat scars, and pay special attention to the possible Floquet resonance.
	
	%SP in DTCs should survive generic perturbations breaking all symmetries~\cite{Else2016,Keyserlingk2016}. Thus, we would exploit a strong-drive Floquet perturbation theory~\cite{Huang2022,supp} to quantify the robustness of cat scars in Eq.~\eqref{eq:scars}. 
	
	Specifically, to validate a perturbation treatment, one necessary condition is that the gap between two energy levels should be larger than the strength of perturbations trying to hyridize those levels. While a gapped level structure could be expected in a few-body setting as in Ref.~\cite{Huang2022}, for an intermediate-scale system as we have here, it is more complicated. As we can observe in Fig.~\ref{fig:cats}~(a), upon moderate perturbations, the spectrum is essentially gapless, where cat scars are deeply buried inside. More specifically, although the solutions Eq.~\eqref{eq:finetunesol} indicate a large energy separation between eigenstates with different total domain wall numbers, the energy separations may fairly be commensurate with Floquet driving frequency $\omega  = 2\pi$, which means they are energetically close-by upon absorbing multiple energy quanta $ m\omega $ ($m\in\mathbb{Z}$) from the driving. With the increase of system size $L$, the density of states ramps up exponentially within a finite Floquet quasienergy window $2\pi$. Thus, based purely on energetic considerations, it may appear rather unexpected that cat scars at $\lambda=0$ should maintain their stability against perturbations $\lambda\neq0$.

	Therefore, a more careful treatment of perturbation strength is needed before we proceed. To facilitate analysis, one could factor out perturbations $\lambda H'$ in the Floquet operator $U_F$ into
	\begin{align}\nonumber
		& 
		U_F(\lambda) = {\cal T} e^{-(i/\hbar) \int_0^T dt (H_0(t) + \lambda H'(t))} = U_0 U'(\lambda), 
		\\ \label{eq:factorVk}
		&
		U_0 = {\cal T} e^{-(i/\hbar) \int_0^T dt H_0(t)},
		\qquad
		U'(\lambda) = e^{i\sum_{k=1}^\infty \lambda^k V_k} ,
	\end{align}
	where perturbations of different orders are represented by $ V_k = V_k^\dagger $, and the unperturbed $ U_0 $ is solved in Eq.~\eqref{eq:finetunesol}. Algebras for factorization is shown in Appendix~\ref{smsec:selection}.
	
	Stability of SP for cat scars derives from a crucial configuration selection rule for $ V_k $. Specifically, the $ k $-th order perturbation with strength $ \lambda^k $ only relates Fock states differing by at most $ n_{\text{op}}k $ spins, namely, the {\em configuration selection rule} reads
	\begin{align}\label{eq:selection}
		&\lambda^k \langle \{s_j\} | V_k | \{\tilde{s}_j\}' \rangle \ne 0
		\quad 
		\Rightarrow
		\quad
		\frac{1}{2}\sum_j \left|s_j - \tilde{s}_j \right| \leqslant n_{\text{op}}k.
	\end{align}
	Here, the operator product order $ n_{\text{op}} $ counts the maximal number of operators being multiplied in individual terms of perturbation Hamiltonians, for instance,
	\begin{align}\nonumber
		&
		n_{\text{op}} = 1: \qquad H' \sim \tau^{\mu}_j,
		\\ \label{eq:nop}
		&
		n_{\text{op}} = 2: \qquad H'\sim \tau^{\mu}_j, \tau^\mu_i \tau^\nu_{j},
	\end{align}
	with $\mu, \nu = x, y, z$, and $ i,j=1,\dots,L$. So $ H' $ in Eq.~\eqref{eq:model} with both one-spin and two-spin terms gives $ n_{\text{op}}=2 $. We emphasize that the selection rule is an operator property for $U_F(\lambda)$, whose origin can be intuitively understood as follows. With perfect spin flips and Ising interactions, the zeroth order $ U_0 $ is highly localized, relating only pairwise Fock states $ \langle \{s_j\}| U_0 | \{\tilde{s}_j\}'\rangle \propto \delta_{\{s_j\} = - \{\tilde{s}_j\}'} $. Then, any matrix elements of $ U_F(\lambda) $ relating $ \{s_j\} $ to others $ \{\tilde{s}_j\}' \ne \pm \{s_j\} $ must entirely derive from perturbations $ \lambda H' $. For power counting, $ \lambda^kV_k $ involves multiplying $ k $ pieces of $ (\lambda H') $, which could flip at most $ n_{\text{op}}k $ spins. The selection rule implies that flipping more spins is suppressed exponentially by higher powers of $ \lambda^k $ in perturbation series. Rigorous algebraic proof using Baker-Campbell-Hausdorff-Dykin formula is presented in Appendix~\ref{smsec:selection}. We shall see below that this constitutes the mechanism beyond energetic reasoning to ensure scar localization.

\begin{figure}
		[h]
		\parbox{6cm}{\includegraphics[width=5.7cm]{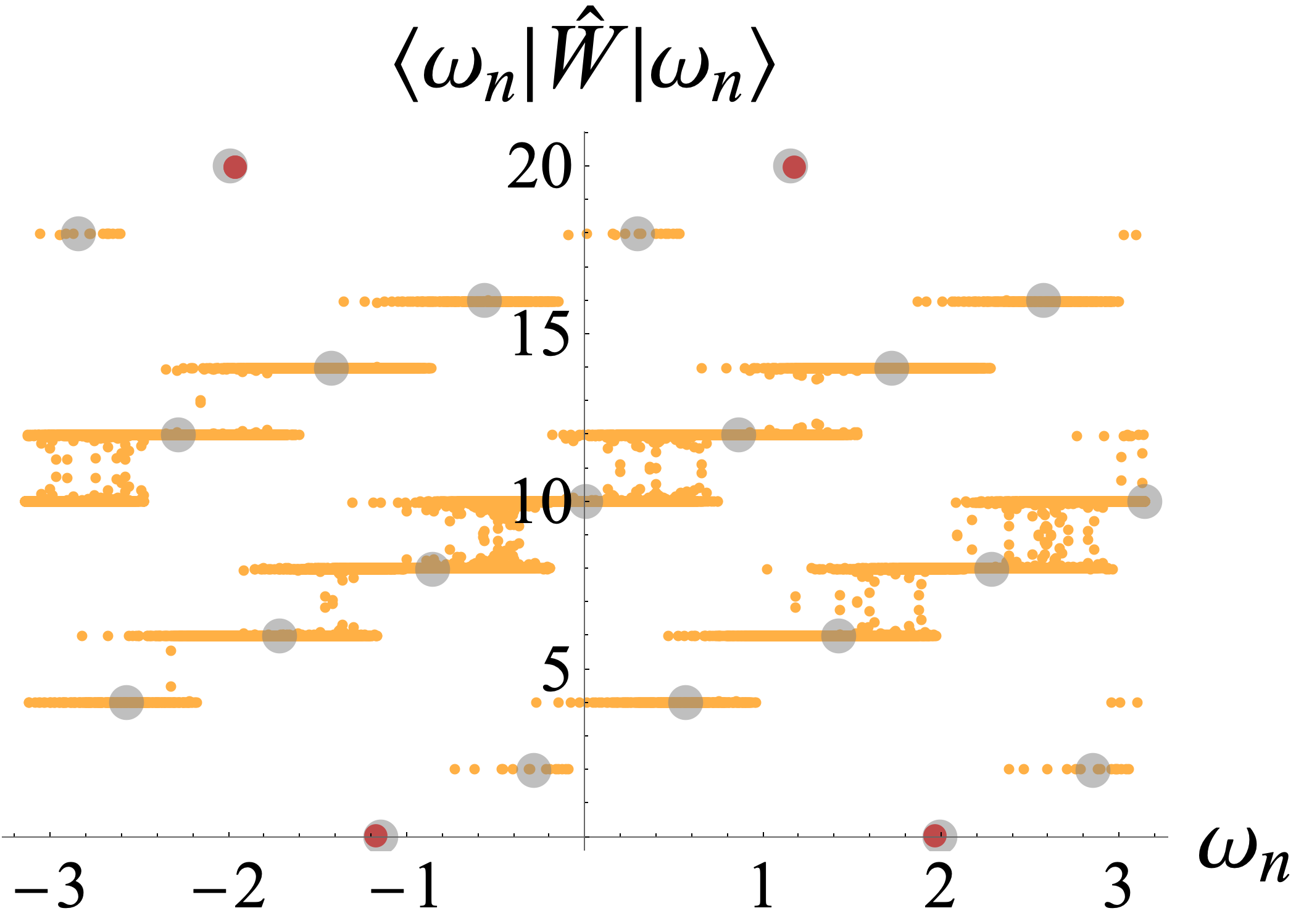} \\ (a) Domain wall resolved level distribution}
		\parbox{2.4cm}{
			\begin{tabular}{cc}
				$\langle W\rangle$ & $\langle r \rangle$
				\\
				18 & 0.5218
				\\
				16 & 0.5105
				\\
				14 & 0.5266
				\\
				12 & 0.5248
				\\
				10 & 0.5249
				\\
				8 & 0.5335
				\\
				6 & 0.5217
				\\
				4 & 0.5327
				\\
				2 & 0.5645
				\\
				\quad & \quad
				\\
				\quad & \quad
			\end{tabular}
		
		}
		
		\parbox{8.4cm}{\includegraphics[width=8.4cm]{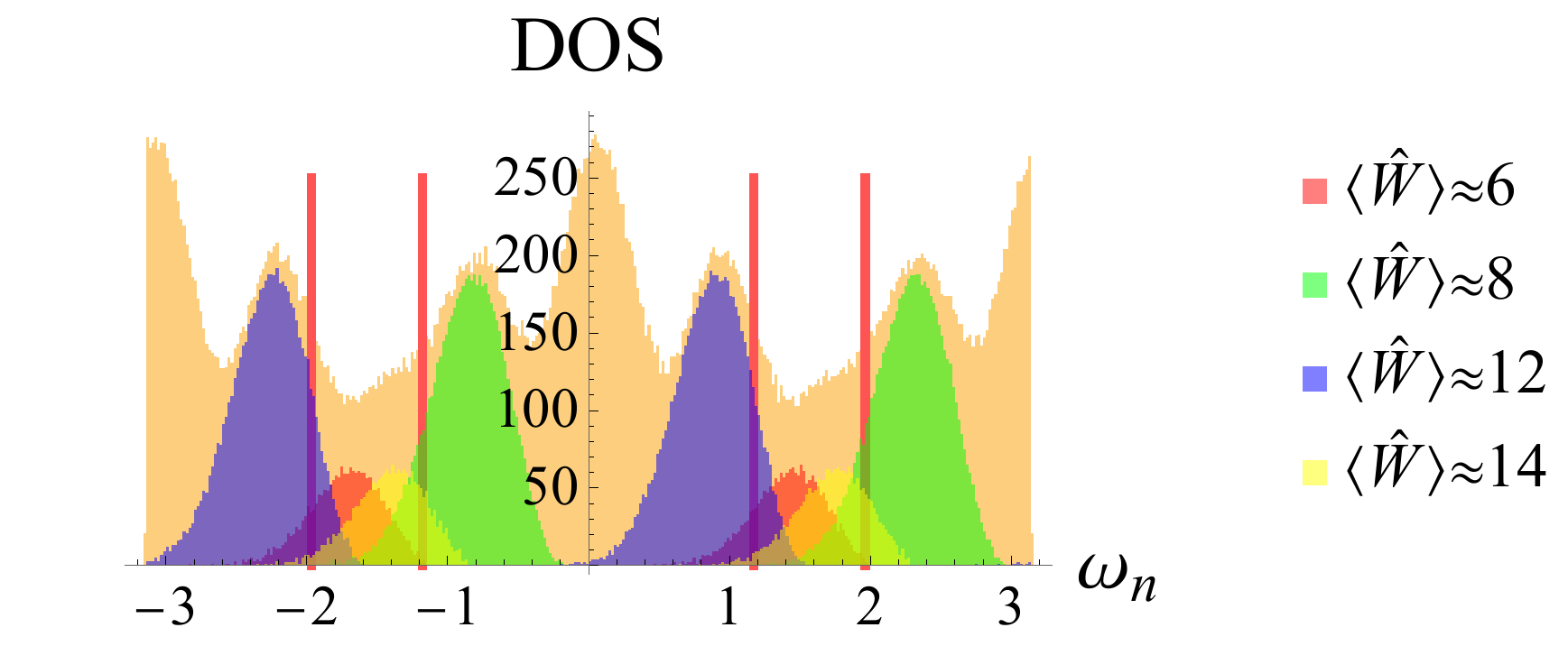} \\ (b) Density of states. Red lines highlight the scar quasienergy.}
		
		\caption{\label{fig:dw}Eigenstate structure with domain wall resolutions. 
		(a) The quasienergy $\omega_n$ for unperturbed ($\lambda=0$, gray dots) and perturbed ($\lambda=0.05$, yellow dots) Floquet operator corresponding to the model in Eq.~\eqref{eq:model}. Perturbed scars are highlighted by red dots. Vertical axis specifies the averaged total domain wall number for each eigenstate.
		(b) Density of states with total domain wall resolution, where we plot those quasienergies overlapping notably with the scar levels. Here we define, i.e. $\langle \hat{W} \rangle \approx 6$ by collecting all quasienergy levels whose averaged total domain wall number $\langle\omega_n |\hat{W} |\omega_n\rangle$ resides between $5$ and $7$, namely, $\langle\hat{W}\rangle \in [w-1, w+1] \approx w$. The system size and parameters are the same as those in Fig.~\ref{fig:cats} (a).
		}
	\end{figure}

	Let us check the structures of perturbed eigenstates more concretely, so as to understand how do the selection rules work. In Fig.~\ref{fig:dw} (a), we label the averaged domain wall number $\langle\omega_n | \hat{W} | \omega_n\rangle$ for each Floquet eigenstate $|\omega_n\rangle$ explicitly, where $\hat{W}$ is given in Eq.~\eqref{eq:hatw}, and $\omega_n$ is the quasienergy. Solutions in Eqs.~\eqref{eq:finetunesol} and \eqref{eq:dwsols} show that at $\lambda=0$, quasienergies are grouped into sectors of different total domain wall numbers $w$, as represented by gray dots in Fig.~\ref{fig:dw} (a). Each $w$ sector separates from nearby ones with $w\pm2$ by quasienergy difference $\Delta E = 4J \text{ mod } 2\pi$.  Under perturbation $\lambda\neq 0$, each degenerate set of levels (gray dot) are lifted into a band as denoted by yellow dots in Fig.~\ref{fig:dw} (a). 
	
	We observe that even after the perturbation, each eigenstate is still localized into a certain domain wall sector. This is surprising because there are apparently large numbers of eigenstates with different $\langle \hat{W} \rangle$ but the same quasienergy $\omega_n$, indicating possible Floquet resonance and strong hybridization of these levels. However, we see that only minor hybridization occurs for consecutive domain wall sectors, while in general, eigenstates seem to still preserve their original domain wall numbers on average. Further, we could verify the thermalization properties of these domain wall bands using level spacing statistics~\cite{Atas2013}. To do so, we assemble the levels according to their domain wall numbers as shown in Fig.~\ref{fig:dw} (b), i.e. levels with $\langle \hat{W} \rangle \in[w-1,w+1] \approx w$ are grouped into continuous bands. Then, calculations of level spacing ratios can be performed for each band, $ r_n = \min(\delta_n, \delta_{n+1})/ \max(\delta_n, \delta_{n+1})$, where $\delta_n = \omega_{n+1} - \omega_n$ denotes the consecutive gaps for the sorted quasienergy levels $\omega_{n} < \omega_{n+1}$. In the table of Fig.~\ref{fig:dw} (a), we list the averaged level spacing statistics $\langle r \rangle$ for each $\langle \hat{W} \rangle $ sector, and all such bands are in the strongly chaotic limit with Gaussian orthogonal ensemble $\langle r \rangle \rightarrow 0.54$, verifying the overall thermalizing nature of the system. Then, it is of interest to find out what suppresses Floquet many-body resonances that may hybridize these different $\langle \hat{W} \rangle$ sectors.

	The reason for domain wall separations can be understood by power counting of perturbations according to the selection rule in Eq.~\eqref{eq:selection}. Specifically, let us start from the unperturbed limit $\lambda=0$ and consider perturbations within the same domain wall sector and those among different $w$'s. Within the same domain wall sector, degenerate level hybridization can always start from the first-order which flips up to $n_{\text{op}}$ spins, i.e. by shrinking or expanding domain size without changing the total domain wall numbers. In contrast, two sectors $w_1, w_2 $ differing by $\delta w = |w_1 - w_2| $ domain walls necessarily requires flipping at least $\delta w/2 $ spins in order to create or annihilate $\delta w$ walls. That means the corresponding hybridization would not start until the perturbation order $k\geqslant \delta w/2n_{\text{op}}$ in witness of the selection rule, with corresponding maximal hybridization strength  $\lambda^{k\geqslant w/2n_{\text{op}}}$. In Fig.~\ref{fig:dw}, we observe that the scar levels ($w=0$ for FM and $w=L=20$ for AFM) are only close to thermalizing sectors of $w=6,8, 12, 14$, implying that $\delta w \geqslant 6$ at least. Thus, the hybridization between scars and other domain wall sectors involves a maximal strength $\lambda^{k> 6/4} $, which is subdominant compared with the intrasector hybridization strength $\lambda^1$ for a weak perturbation $\lambda\ll 1$. Once the degenerate levels are lifted for $\lambda\neq0$, according to the thermalizing behaviors for non-scar domain wall sectors, we would expect each level to be extended in the Fock space with total domain wall number $w$, whose Hilbert subspace dimension can be estimated as $C_L^w = L!/(L-w)!w!$. Each of such delocalized thermalizing eigenstate may involve all the $C_L^w$ configurations, among which only a small subset would deviate from scar configurations by the minimal spin flip number $\delta w/2n_{\text{op}}$ with hybridization strength $\sim \lambda^{\delta w/2n_{\text{op}}}$. Then, the localized scar would only hybridize with each delocalized level with an even weaker strength. 
	
	To summarize the above analysis, the condition to avoid Floquet resonance between scars and thermalizing levels is that in the {\em unperturbed} limit $\lambda=0$, scar quasienergy should not be identical to the nearby $\delta w/2 \leqslant n_{\text{op}}$ sectors. This way, even if scars are energetically resonant with other sectors $\delta w/2> n_{\text{op}}$, the hybridization strength is suppressed by $\sim \lambda^{k>\delta w/2n_{\text{op}}}/C_{L}^w$ due to both selection rules and the delocalization of thermalizing levels. 
	
	Since the absence of massive delocalizing Floquet resonance is of vial importance for later perturbative treatment, it is worth being more cautious, so we perform two additional tests to verify it.

	\begin{figure}[h]
		\parbox{2.7cm}{
			\includegraphics[width=2.7cm]{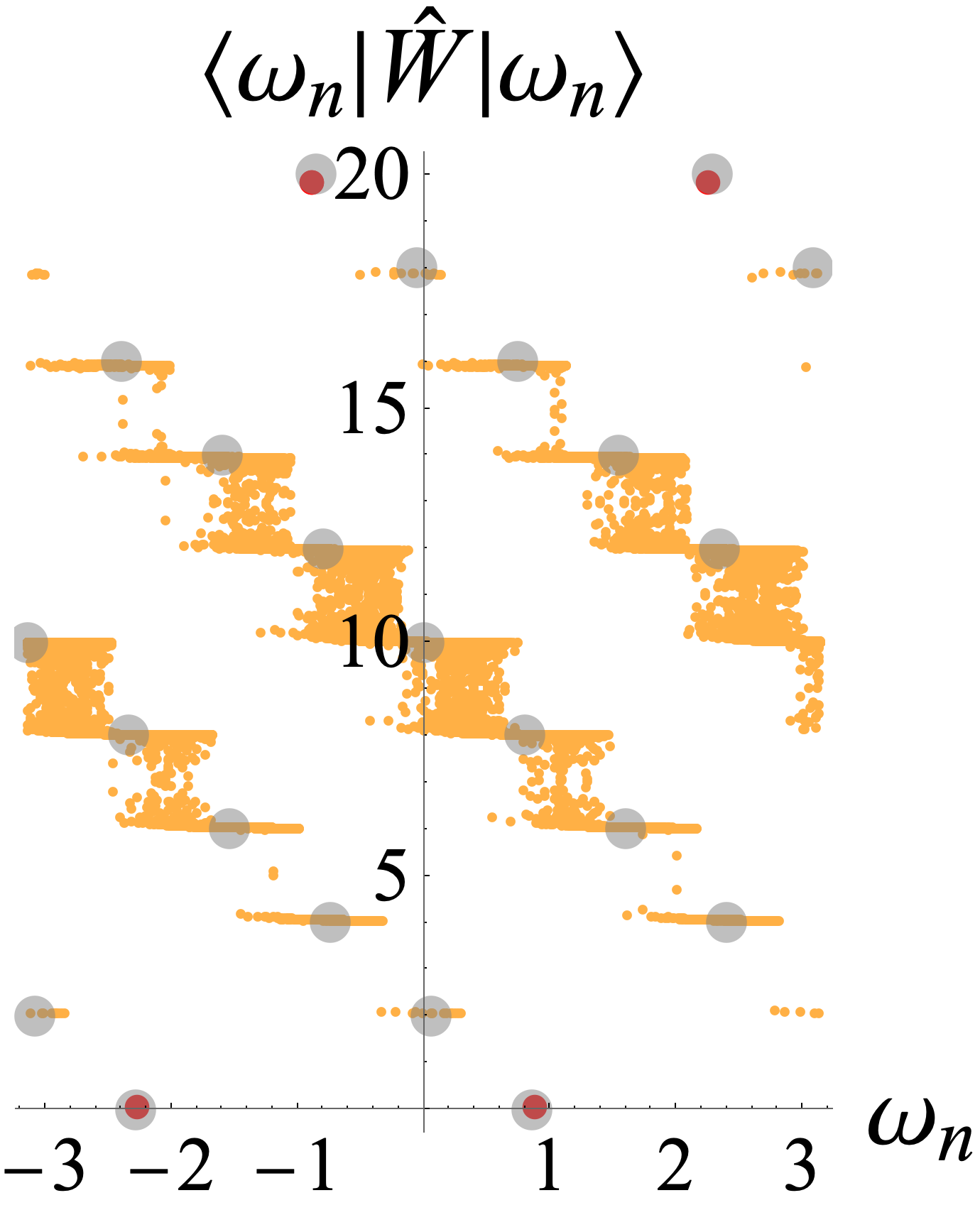}
			\\ (a) $J=\pi/4-0.2$
		}
		\parbox{2.7cm}{
			\includegraphics[width=2.7cm]{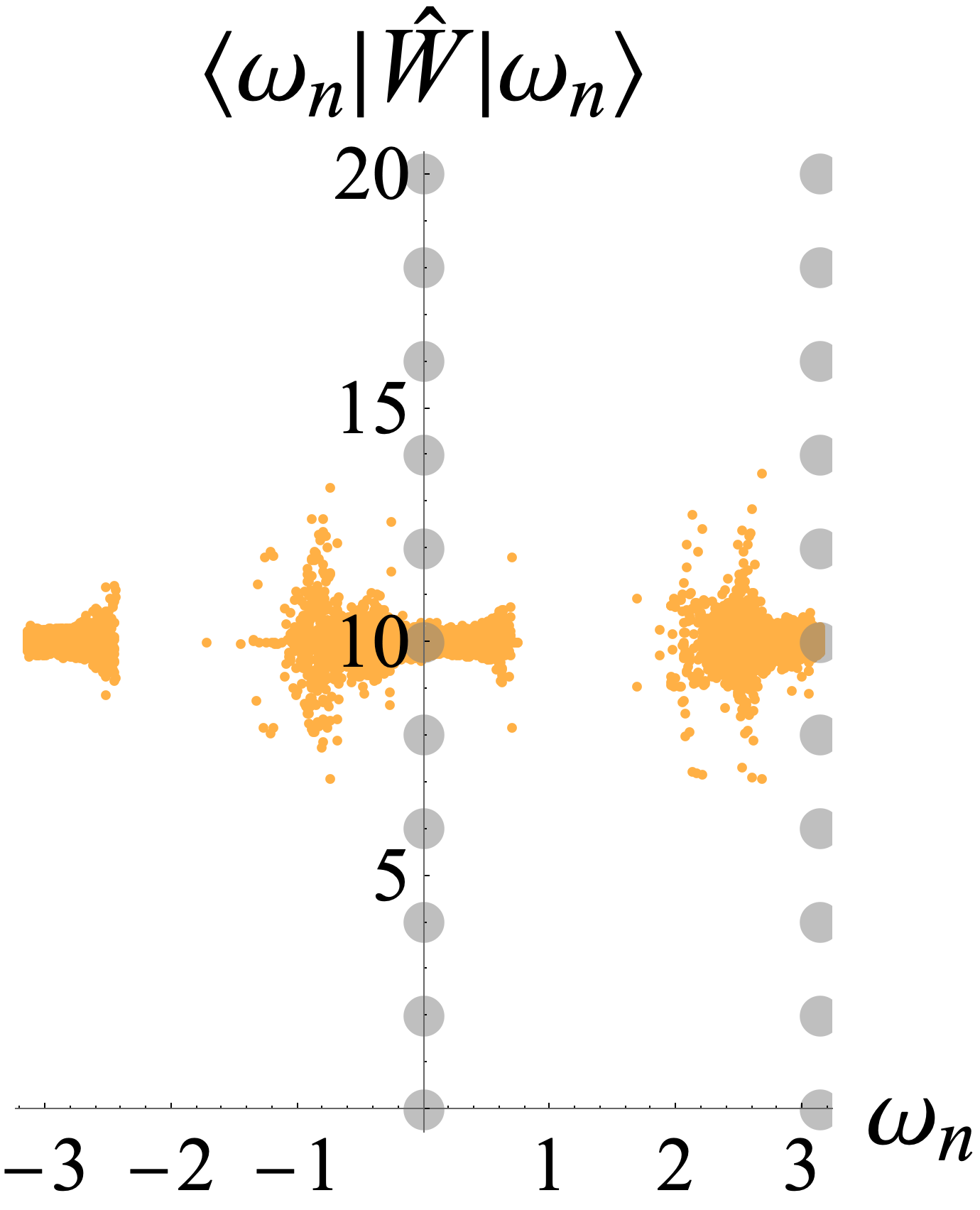}
			\\ (b) $J=\pi/4$
		}
		\parbox{2.7cm}{
			\includegraphics[width=2.7cm]{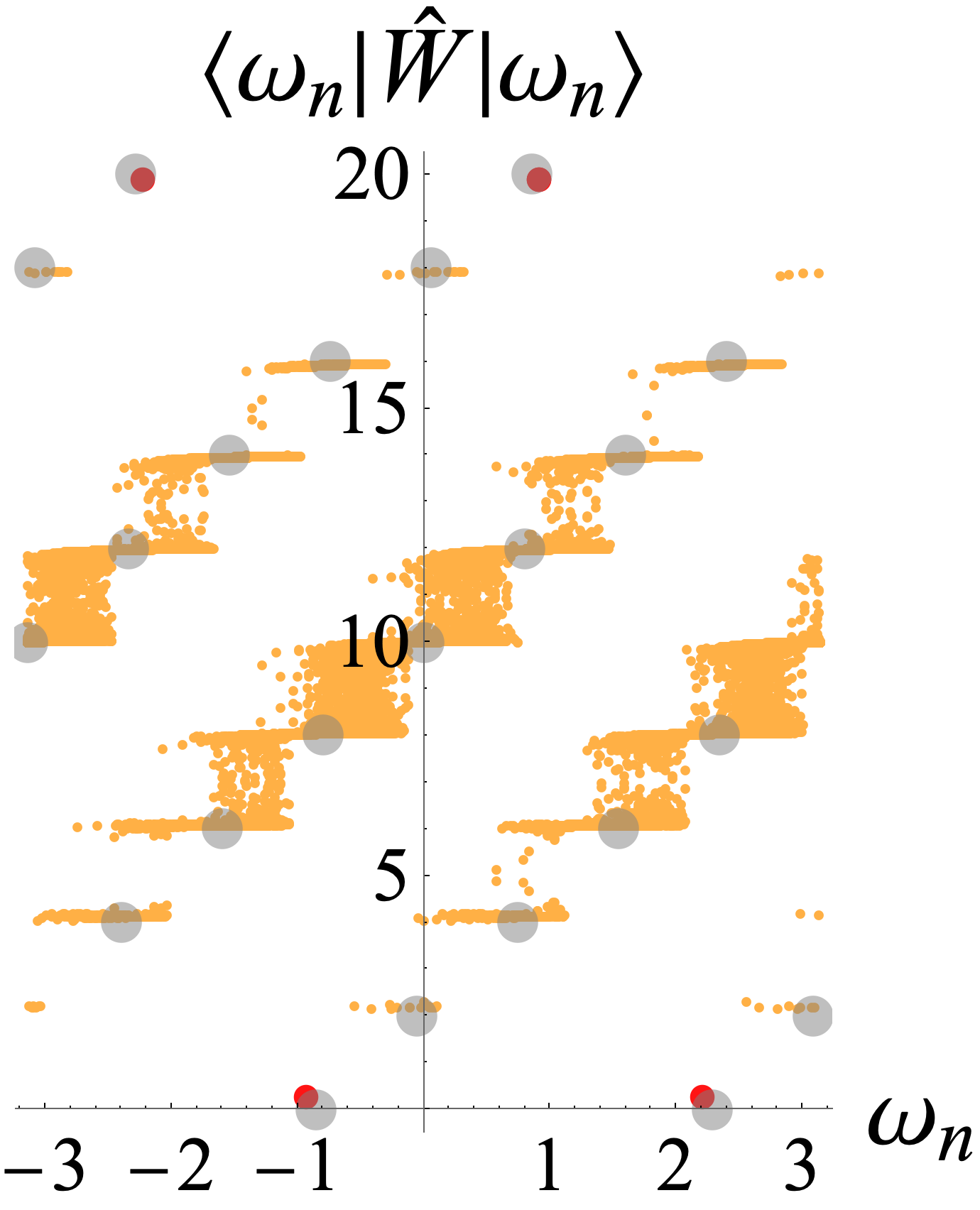}
			\\ (c) $J=\pi/4+0.2$
		}
		\caption{\label{fig:resonant} Domain wall resolved density of states, when level spacing between consecutive domain wall sectors $\Delta E = 2J, 2J+\pi $ are close to be in resonance with Floquet driving frequency $2m\pi$, namely, $J_{\text{res}} = m\pi/4, m\in\mathbb{Z}$. (a) (c) Off-resonant $|J-J_{\text{res}}|>\lambda$, where cat scars are preserved. (b) In exact resonance, where cat scars together with the domain wall structure are destroyed. System size and perturbation parameters are the same as in Fig.~\ref{fig:cats}. }
	\end{figure}

	First, let us observe how Floquet resonance between nearby domain wall sectors occur and vanish more quantitatively. To access the resonance, we could set the interaction exactly to be $J= \pi/4$, such that the energy separation between nearby domain wall sectors $\Delta E = 2J\delta w =  4J \approx \pi, \delta w = 2$. Then, all domain sectors $w$ would be in exact resonance with their consecutive ones $w\pm2$ (recall that for each domain wall sector, there are two spectral pairs differing by quasienergy $\pi$, so the total energy difference from Ising interaction $\Delta E \approx \pi$ and spectral pairing $\pi$ adds up to $ 2\pi = \omega$). As expected, from Fig.~\ref{fig:resonant} (b), we see that a complete resonance occurs for all levels in the system, as the intra-sector and inter-sector hybridization strengths are both $\sim \lambda^1$. Setting $J$ slightly away from the exact resonance point $\pi/4$ immediately suppresses the hybridization among different domain wall sectors and recovers the scar structure, as seen in Fig.~\ref{fig:resonant} (a) and (c).

	\begin{figure}[h]
		\parbox{4.2cm}{
			\includegraphics[width=4.2cm]{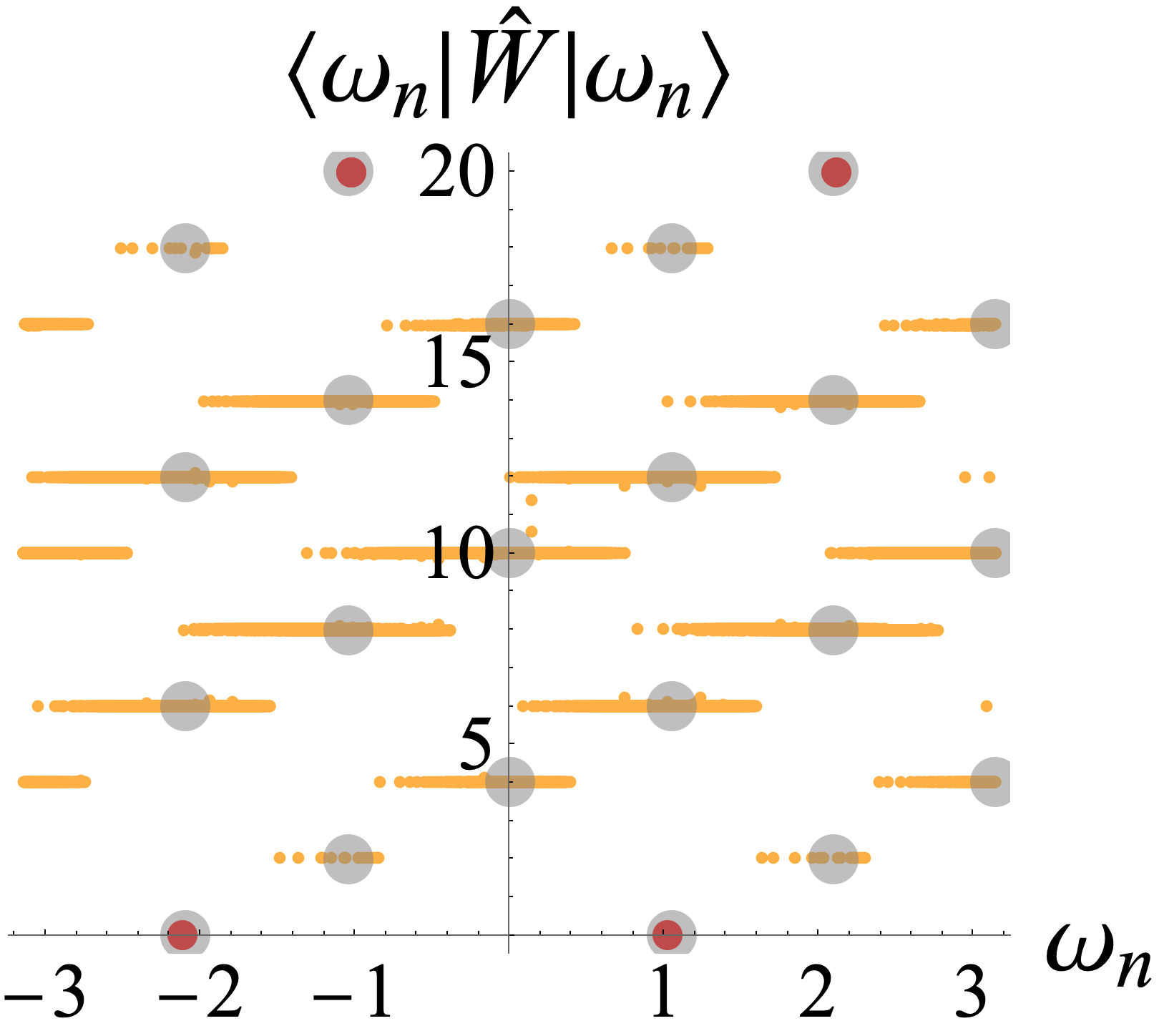}
			\\ (a) $J=\pi/3$
		}
		\parbox{4.2cm}{
			\includegraphics[width=4.2cm]{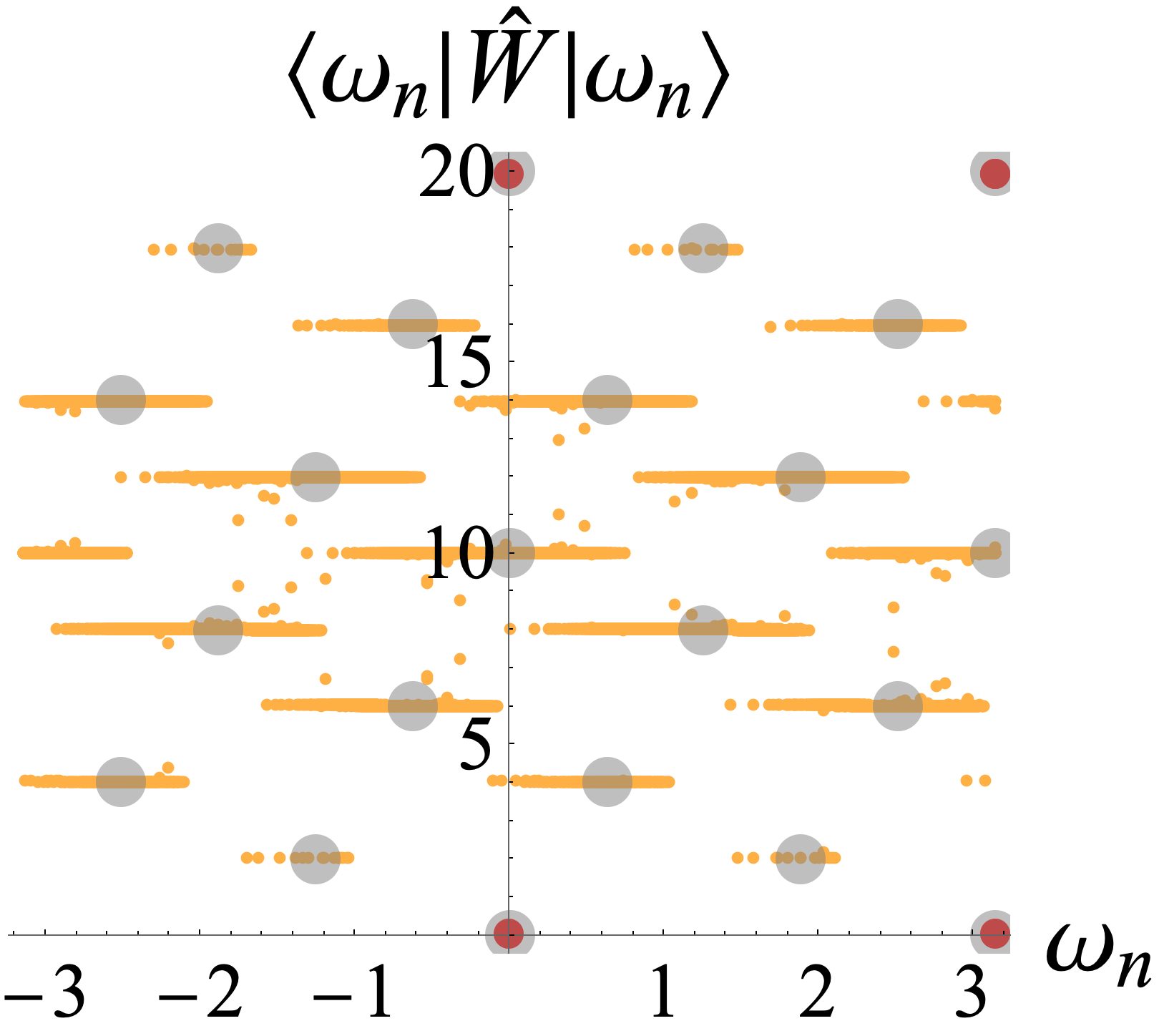}
			\\ (b) $J=2\pi/5$
		}
		\caption{\label{fig:off_res} Domain wall resolved density of states, where quasienergy degeneracy occurs between domain wall sectors $\Delta w > n_{\text{op}}$, where $n_{\text{op}}=2$ for the model in Eq.~\eqref{eq:model} whose perturbations are up to two-spin terms. Here, although the cat scars in $w=0,20$ sectors have {\em exactly} the same quasienergy with states in domain wall sectors $w=6$, due to exponential suppression of hybridization by selection rules, scars are still preserved. This result, together with Fig.~\ref{fig:resonant}, emphasize that a pure energetic consideration is insufficient. Rather, the stability of cat scars crucially relies on the configuration selection rules emergent from strongly interacting systems. Here, the system size and perturbation paramters are the same as those in Fig.~\ref{fig:cats} (a). }
	\end{figure}
	
	Second, let us also observe the effect where {\em exact} coincidence of quasienergy at the unperturbed point $\lambda=0$ occurs for two sectors separating by large numbers of domain wall differences $\delta w/2 > n_{\text{op}}$. In the previous analysis, we claim that due to the exponential suppression of inter-sector hybridization $\sim \lambda^{\delta w/2n_{\text{op}}}$ by selection rules, scars with exactly the same quasienergy as sectors separating by large numbers of domain walls would not lead to strong hybridization. Two such examples confirming the claim are presented in Fig.~\ref{fig:off_res}. For instance, in Fig.~\ref{fig:off_res} (a), the FM (or AFM) scars possess exactly the same quasienergy at $\lambda =  0$ with levels hosting $w=6, 12$ (or $w=8,14$) domain wall in Fig.~\ref{fig:off_res}, corresponding to minimal domain wall separation by $\delta w = 6$. As expected, we observe that although the scars (red dots) share the same quasienergy as several sets of levels (gray dots) in the unperturbed limit, scars would not notably hybridize with these levels due to the exponential suppression of spin flipping as given by the selection rule. Similarly, another example in Fig.~\ref{fig:off_res} (b) shows the coincidence of scar (red dots) quasienergies with those unperturbed levels hosting $w=10$ domain walls at $\lambda=0$ (gray dots). Once again, scars refuse to notably hybridize with these degenerate levels upon perturbation $\lambda\neq0$ (yellow dots).

	\subsection{Scaling exponents for cat scars}

	Based on the selection rule in Eq.~\eqref{eq:selection} and the eigenstate structure illuminated above, we are ready to analyze the stability and scaling behaviors of cat scars. To simplify discussions, FM and AFM patterns are both denoted ``cat" below, i.e. $ |\ell, \text{(A)FM}\rangle \equiv  |\ell, \{s_j^{\text{(cat)}}\} \rangle $, as analysis are identical for them. Perturbation series consists of iterative corrections $ |\tilde{\omega}_{\ell, \text{cat}}\rangle = e^{i\lambda^k S_k} \dots e^{i\lambda^2 S_2} e^{i\lambda S_1} |\ell, \{s_j^{(\text{cat})}\}\rangle + O(\lambda^{k+1}) $, where $ S_k $ diagonalizes the perturbed Floquet operator $ U_F(\lambda)|\tilde{\omega}_{\ell, \text{cat}}\rangle = e^{i\tilde{\omega}_{\ell, \text{cat}}} |\tilde{\omega}_{\ell,\text{cat}}\rangle $ at the order $ \lambda^k $, rendering corrected quasienergy 
	$ e^{i\tilde{\omega}_{\ell, \text{cat}}} = 
	e^{ i\left( E(\ell, \{s_j^{(\text{cat})}\}) + \sum_{k=1}^\infty \lambda^k \omega_{\ell, \text{cat}}^{(k)} \right) } $. Recall that the perturbed Floquet operator is also factored into the form $U_F(\lambda) = U'(\lambda)U_0$ previously, where $|\ell, \{s_j^{(\text{cat})}\}$ are eigenstates of $U_0$, and $U'(\lambda) = e^{i \sum_{k=1}^\infty \lambda V_k}$ are for perturbations of different orders.

	Three universal scaling relations can be obtained for perturbed cat scars $ |\tilde{\omega}_{\ell,\text{cat}}\rangle $. We first present the results below. These scaling formulae will be applied in later subsections to specific examples and compare with corresponding numerical investigations.
	
	{\bf Scaling (1)} Amplitudes for the original FM or AFM components in Eq.~\eqref{eq:scars} are rescaled by the dominant first-order fluctuations to
	\begin{align}\label{eq:scaling1}
		\alpha_0^2 \equiv |\langle \ell, \{s_j^{(\text{cat})}\}|\tilde{\omega}_{\ell,\text{cat}}\rangle|^2 = \frac{1}{1+\bar{V}_1^2 \lambda^2 L}  +O((\lambda^2L)^2)
	\end{align}
	Here, the first-order {\em local} perturbation strength 
	\begin{align}\nonumber
		\bar{V}_1^2 
		&= \frac{1}{8}\sum'_{\ell',\{s_j\}'} \left|
		\langle \ell', \{s_j\}'|V_{1,j=1} + V_{1,j=2}|\ell,\{s_j^{(\text{cat})}\}\rangle 
		\right|^2 \\ \label{eq:vbar}
		&\qquad 
		\times \csc^2 \left[ \left(E(\ell, \{s_j^{(\text{cat})}\})- E(\ell', \{s_j\}') \right)/2 \right]
	\end{align}
	characterizes spin fluctuations on top of FM or AFM patterns $ \{s_j^{(\text{cat})}\} $, where the factored perturbation in $U'(\lambda)$ reads $ V_1 = \sum_{j=1}^L V_{1,j} $, and summation $ \sum'_{\ell', \{s_j\}'} $ excludes the cat eigenstates under consideration. Unperturbed quasienergies are also shown in Eq.~\eqref{eq:finetunesol}.
	
	{\bf Scaling (2)} Overall spin configurations for perturbed cat scars $ |\tilde{\omega}_{\ell,\text{cat}}\rangle $ exhibit an exponential Fock space localization to the unperturbed patterns $ \{s_j^{\text{cat}}\} $: 
	\begin{align}\label{eq:scaling2}
		 |\langle  \{s_j \} |
		\tilde{\omega}_{\ell,\text{cat}} \rangle|^2 \propto \lambda^{ \Delta s_{\text{cat}} (\{s_j\})/\xi}
	\end{align}
	Here, the pairwise Fock space distance
	\begin{align}\label{eq:Ds}
		\Delta s_{\text{cat}}(\{s_j\}) = \frac{1}{2} \min\left(\sum_{j=1}^L|s_j - s_j^{(\text{cat})}|, \sum_{j=1}^L|s_j+s_j^{(\text{cat})}| \right)
	\end{align}
	counts how many spins are different between the configurations $ \pm\{s_j^{(\text{cat})}\} $ and another $ \{s_j\} $, serving as a measure of ``distance" in many-body Fock space. The corresponding localization length in Fock space is constrained by selection rules to
	\begin{align}\label{eq:exponent_Fock}
		\xi \leqslant n_{\text{op}}.
	\end{align}
	
	{\bf Scaling (3)} Spectral gap for pairwise perturbed scars approaches the unperturbed value $ \pi $ with exponential accuracy,
	\begin{align}\label{eq:scaling3}
		\Delta_{\pi} = |\tilde{\omega}_{1,\text{cat}} - \tilde{\omega}_{0,\text{cat}}| = \pi + O(\lambda^{L/\nu})
	\end{align}
	Similarly, due to selection rules, the spectral deviation exponent is constrained into
	\begin{align}\label{eq:exponent_lifetime}
		\nu \leqslant n_{\text{op}}.
	\end{align}

	Recall that the operator product order $n_{\text{op}}$ is defined in Eq.~\eqref{eq:nop}, which counts that in the bare perturbation Hamiltonian $H'$, up to $n_{\text{op}}$-spin terms are involved.

	Physically, scaling relations in Eqs.~\eqref{eq:scaling1}, \eqref{eq:scaling2} and \eqref{eq:scaling3} prescribe the universal behaviors of clean DTCs facing perturbations.
	Specifically, relations {\bf (1)} and {\bf (2)} render
	\begin{align}\label{eq:pertscar}
		|\tilde{\omega}_{\ell,\text{cat}} \rangle = \alpha_0  |\ell, \{s_j^{(\text{cat})}\} \rangle  + \sum_{\{s_j\}'} O(\lambda^{\Delta s_{\text{cat}} (\{s_j\}')/\xi}) |\{s_j\}'\rangle
	\end{align}
	so FM or AFM initial states overlapping chiefly with perturbed cat scars evolve as 
	\begin{align}\nonumber
		&
		|\psi(nT)\rangle = U_F^n |\{s_j^{(\text{cat})}\} \rangle 
		\\
		&
		\approx (\alpha_0/\sqrt{2}) \left(  
		e^{in\tilde{\omega}_{0, \text{cat}}} |\tilde{\omega}_{0,\text{cat}} \rangle  
		+ e^{in\tilde{\omega}_{1, \text{cat}}} |\tilde{\omega}_{1,\text{cat}} \rangle  
		\right) 
	\end{align}
	Minor overlaps of $ |\{s_j^{(\text{cat})}\} \rangle $ with other eigenstates contribute exponentially localized spin fluctuations.  Further using scaling relation {\bf (3)} and solutions of $ |\ell, \{s_j^{(\text{cat})}\} \rangle $ in Eq.~\eqref{eq:scars}, we obtain the dominant dynamics (recall that $M(nT)$ is defined in Eq.~\eqref{eq:corr})
	\begin{align}\nonumber
		 |\psi(nT)\rangle &\approx \alpha_0^2 \cos(O(\lambda^{L/\nu}) n)
		|(-1)^n\{s_j^{(\text{cat})}\} \rangle,
		\\ \label{eq:dtc}
		\Rightarrow\quad M(nT) &\approx \alpha_0^4 \cos^2\left( O(\lambda^{L/\nu} n) \right)
	\end{align}
	Thus, we observe period-$ 2T $ local spin flips $ |(-1)^n \{s_j^{(\text{cat})}\} \rangle $, with amplitudes for $ M(t) $ reduced to $ \alpha_0^4 \approx (1+ \bar{V}_1^2 \lambda^2 L)^{-2} $, thereby giving the estimation $ L\lesssim 1/\lambda^2 $. Within such intermediate scales, the DTC lifetime $ \sim (1/\lambda)^{L/\nu}T $ grows exponentially with the increase of system sizes, as indicated by the cosine modulation. 
	
	Proof for scaling relations is sketched below to illuminate their origins, while algebras are furnished in Appendix~\ref{smsec:scaling}. Recall that $ S_k $ in the perturbation series serves to cancel the off-diagonal terms of $ U_F $ proportional to $ \lambda^k $, which involves products of perturbations $ V_{k_1} V_{k_2} \dots V_{k_\alpha} $ with $ \sum_{p=1}^\alpha k_p = k $. Then, selection rules for $ V_{k_p} $ enforce that $ S_k $ similarly cannot flip more than $ n_{\text{op}}k $ spins.
	Consequently, the first-order correction $ |\tilde{\omega}_{\ell, \text{cat}} \rangle = \alpha_0 (1+i\lambda S_1) |\ell, \{s_j^{(\text{cat})}\} \rangle + O(\lambda^2) $ features fluctuations of  $ n_{\text{op}} $ nearby spins for local perturbation. Under translation invariance, amplitudes for spin flips are identical on different sites, and their accumulated effect renders the factor $ L $ in normalization constant $ \alpha_0 $ for scaling relation (1). 
	Further, it takes perturbation of orders $ \lambda^{ k \geqslant \Delta s_{\text{cat}} (\{s_j\}) /n_{\text{op}} } $ to flip $ \pm\{s_j^{(\text{cat})}\} $ by $ \Delta s_{\text{cat}} (\{s_j\}) $ spins into $ |\{s_j\}\rangle $ in $ |\tilde{\omega}_{\ell, \text{cat}} \rangle $, implying scaling relation (2) and the bound on $ \xi $. 
	Finally, in quasienergy corrections $ \lambda^k \omega_{\ell, \text{cat}}^{(k)} = \lambda^k \langle \ell, \{s_j^{(\text{cat})}\} | F_k | \ell, \{s_j^{(\text{cat})}\} \rangle = \frac{\lambda^k}{2} \sum_{m,m'=0}^1 (-1)^{(m-m')\ell} \langle (-1)^m \{s_j^{(\text{cat})}\} | F_k | (-1)^{m'} \{s_j^{(\text{cat})}\} \rangle  $, the spectral pair numbers $ \ell $ only appear in the cross terms $ m\ne m' $ for opposite patterns. Here $ F_k $ involves products of $ S_{k_1}\dots S_{k_\alpha} $ and $ V_{p_1}\dots V_{p_\beta} $ of total orders $ \sum_{j=1}^\alpha k_j + \sum_{j=1}^\beta p_j \leqslant k $, and therefore flips no more than $ n_{\text{op}}k $ spins. That means lower order $ F_{k< L/n_{\text{op}}} $ cannot flip all $ L $ spins and the cross terms for $ m\neq m' $ vanish. Then, $ \omega_{1, \text{cat}}^{(k)} = \omega_{0,\text{cat}}^{(k)}  $ up to $ k<L/n_{\text{op}} $, and the perturbed spectral gap maintains rigidity $ \tilde{\omega}_{1, \text{cat}} - \tilde{\omega}_{0, \text{cat}} = E(1, \{s_j^{(\text{cat})}\}) - E(0, \{s_j^{(\text{cat})}\}) + \sum_{k\geqslant L/n_{\text{op}}} \lambda^k (\omega_{1, \text{cat}}^{(k)} -\omega_{0, \text{cat}}^{(k)} ) = \pi + O(\lambda^{k\geqslant L/n_{\text{op}}}) $, proving scaling relation (3) and the bounds on $ \nu $. In the language more closely related to quantum computation~\cite{Else2016}, perturbations $ F_k $ represents a local circuit of depth $ n_{\text{op}} k $, so lower order $ F_{k< L/n_{\text{op}}} $ cannot disentangle the correlated cat scars into product states.
	
	In previous numerics~\cite{Huang2018,Zeng2017,Yarloo2020,Pizzi2020}, certain aspects of scalings have been speculated. New contributions in this work involve not only clarifying the underlying mechanism, but also proving the analytical scaling form, including the values or bounds of exponents. In the following subsections, we verify these scalings in Eqs.~\eqref{eq:scaling1}, \eqref{eq:scaling2} and \eqref{eq:scaling3} numerically, which shows {\em quantitative} agreements for both the model in Eq.~\eqref{eq:model} and alternative ones with different $ n_{\text{op}} $.

	\subsection{Applications and numerical verifications}
	
	In this subsection, we would compare the analytical scaling relations in Eqs.~\eqref{eq:scaling1}, \eqref{eq:scaling2} and \eqref{eq:scaling3}, as well as the bounds on exponents in Eqs.~\eqref{eq:exponent_Fock} and \eqref{eq:exponent_lifetime}, against numerical investigations of the models in Eq.~\eqref{eq:model} and Eq.~\eqref{eq:ufh}.
	
	First, let us benchmark the IPR scaling using the relations in Eq.~\eqref{eq:scaling1}, which is intimately related to the DTC oscillation amplitudes in Eq.~\eqref{eq:dtc}.  Recall that for all zeroth-order solutions in Eq.~\eqref{eq:finetunesol}, $ \sum_{\{s_j\}'} |\langle \{s_j\}' | \ell, \{s_j\}\rangle|^4 = \sum_{\{s_j\}'} |\langle \{s_j\}'| \frac{1}{\sqrt2} \sum_{m=0,1} (-1)^{m\ell} |(-1)^m\{s_j\}\rangle |^4 = 1/2 $, then from Eq.~\eqref{eq:pertscar} we have
	\begin{align}\nonumber
		\text{IPR}(\tilde{\omega}_{\ell, \text{cat}})
		=
		\sum_{\{s_j\}''} |\langle \{s_j\}''| \tilde{\omega}_{\ell, \text{cat}}\rangle |^4 \approx
		 \frac{	\alpha_0^4}{2}.
	\end{align}
	To leading orders, using Eq.~\eqref{eq:scaling1}, the IPR scaling corresponding to (half of) the DTC oscillation amplitude reads
	\begin{align}\label{eq:iprscaling}
		\text{IPR}(\tilde{\omega}_{\ell, \text{cat}})
		\approx \frac{1}{2}\frac{1}{(1+\bar{V}_1^2 \lambda^2  L)^2}
		\approx
		\frac{1}{2}-\bar{V}_1^2 \lambda^2 L.
	\end{align}
	Thus, the leading order deviation of IPR from the unperturbed value $ 1/2 $ takes the universal scaling form $ \bar{V}_1^2 \lambda^2 L $. 
	
	\begin{figure}
		[h]
		\includegraphics[width=6cm]{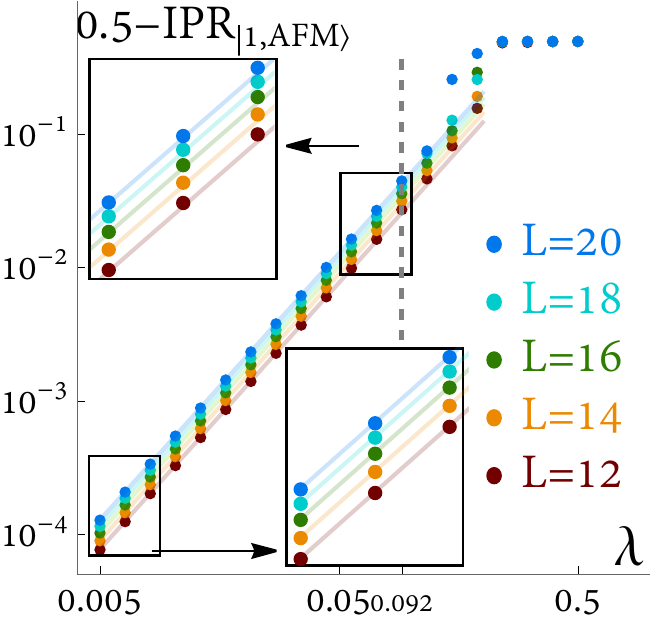}
		\caption{\label{fig:iprscaling1} Universal scaling behaviors for $ |1,\text{AFM}\rangle $ scar IPRs. Dots are numerically data of IPR for the $ |1,\text{AFM}\rangle $ scar eigenstates, and the lines are prescribed by the universal scaling relation $ \bar{V}_1^2 \lambda^2 L $ in Eq.~\eqref{eq:iprscaling}, with a single fitting parameter $ \bar{V}_1^2 \approx 0.2564 $. The value of $ \bar{V}^2 $ is obtained by fitting the data for small system size $ L=12 $ alone. In turn, we see that the analytical scaling matches numerical data for all $ L $ up to $ \lambda \lesssim 0.1 $. }
	\end{figure}

	To test the analytical results directly, we compute the IPR deviations for our main text model in Fig.~\ref{fig:iprscaling1}.  For generic perturbations $\lambda H'$ in Eq.~\eqref{eq:factorVk}, usually it is hard to obtain a closed form for the factored $V_k$. Nevertheless, it is worth emphasizing that since $ \bar{V}_1 $ is the same for {\em all} system sizes, one can conveniently obtain its value by fitting the scaling form to numerical data for a single small system size. In turn, the universal scalings for all larger system sizes can be predicted via Eq.~\eqref{eq:iprscaling}. An example is given in Fig.~\ref{fig:iprscaling1}. Indeed, the expected behaviors for $ 1/2 - \text{IPR}(\tilde{\omega}_{1,\text{AFM}}) \propto \lambda^2L $ shows up.

	\begin{figure}
		[h]
		\parbox{4.2cm}{\includegraphics[width=4cm]{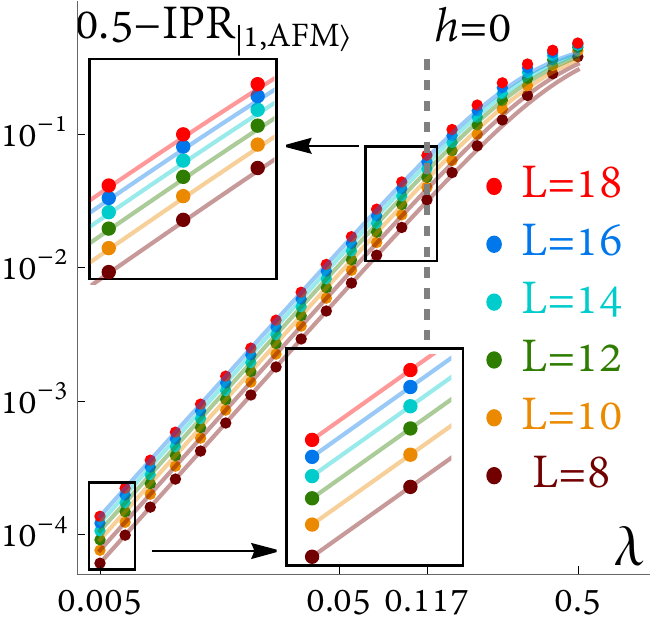} }
		\parbox{4.2cm}{\includegraphics[width=4cm]{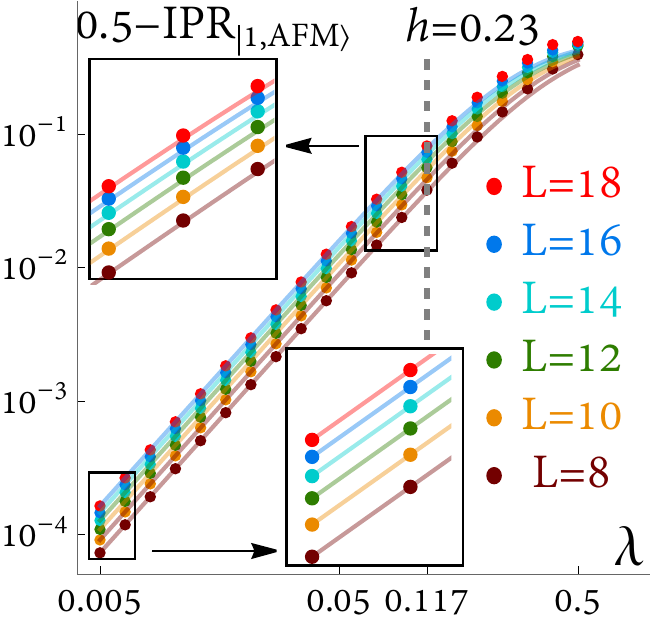} }
		\parbox{4.2cm}{\includegraphics[width=4cm]{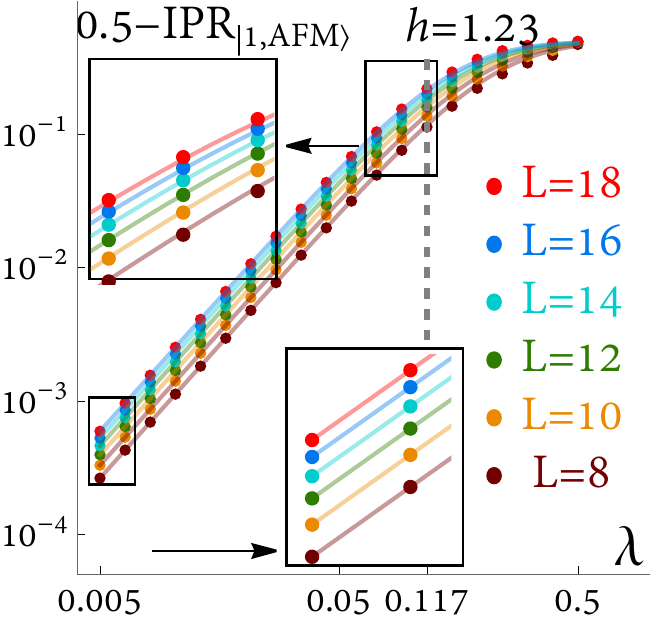} }
		\parbox{4.2cm}{\includegraphics[width=4cm]{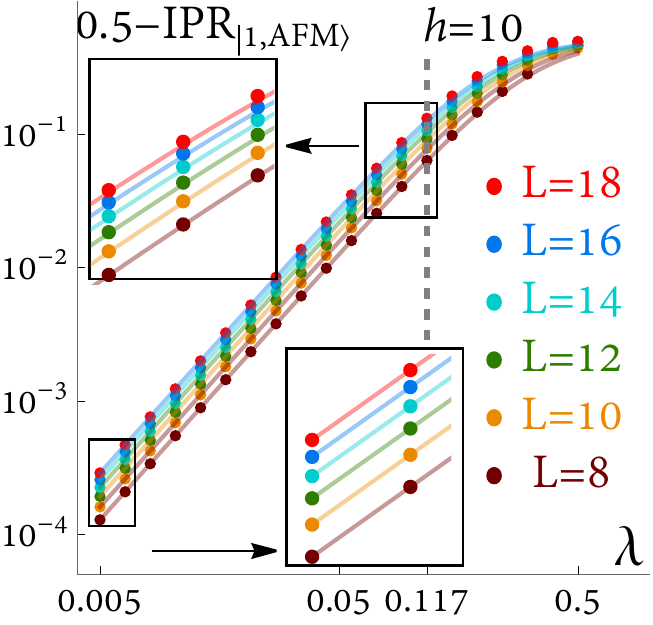} }
		\caption{\label{fig:ipr_simplemodel} The scaling of IPR deviation for the model in Eq.~\eqref{eq:ufh} with separable perturbation. Dots are numerical data for $ U_F $ in Eq.~\eqref{eq:ufh}, while lines are analytical scaling functions $ \frac{1}{2}- $~IPR~$\approx \frac{1}{2} \left( 1 - (1+\bar{V}_1^2 \lambda^2 L)^{-2}\right) $ in Eq.~\eqref{eq:iprscaling}, with $ \bar{V}_1^2 $ explicitly given in Eq.~\eqref{eq:v1barsimple}. Here $ J=1 $, and other parameters are denoted in the figures. We see that the analytical formula precisely predicts the scaling behaviors {\bf without any fitting parameters}. Also, we verify that the longitudinal fields $ h $ indeed only lead to perturbative effects. }
	\end{figure}

	\begin{figure*}[t]
		\parbox[b]{4.5cm}{\includegraphics[width=5cm]{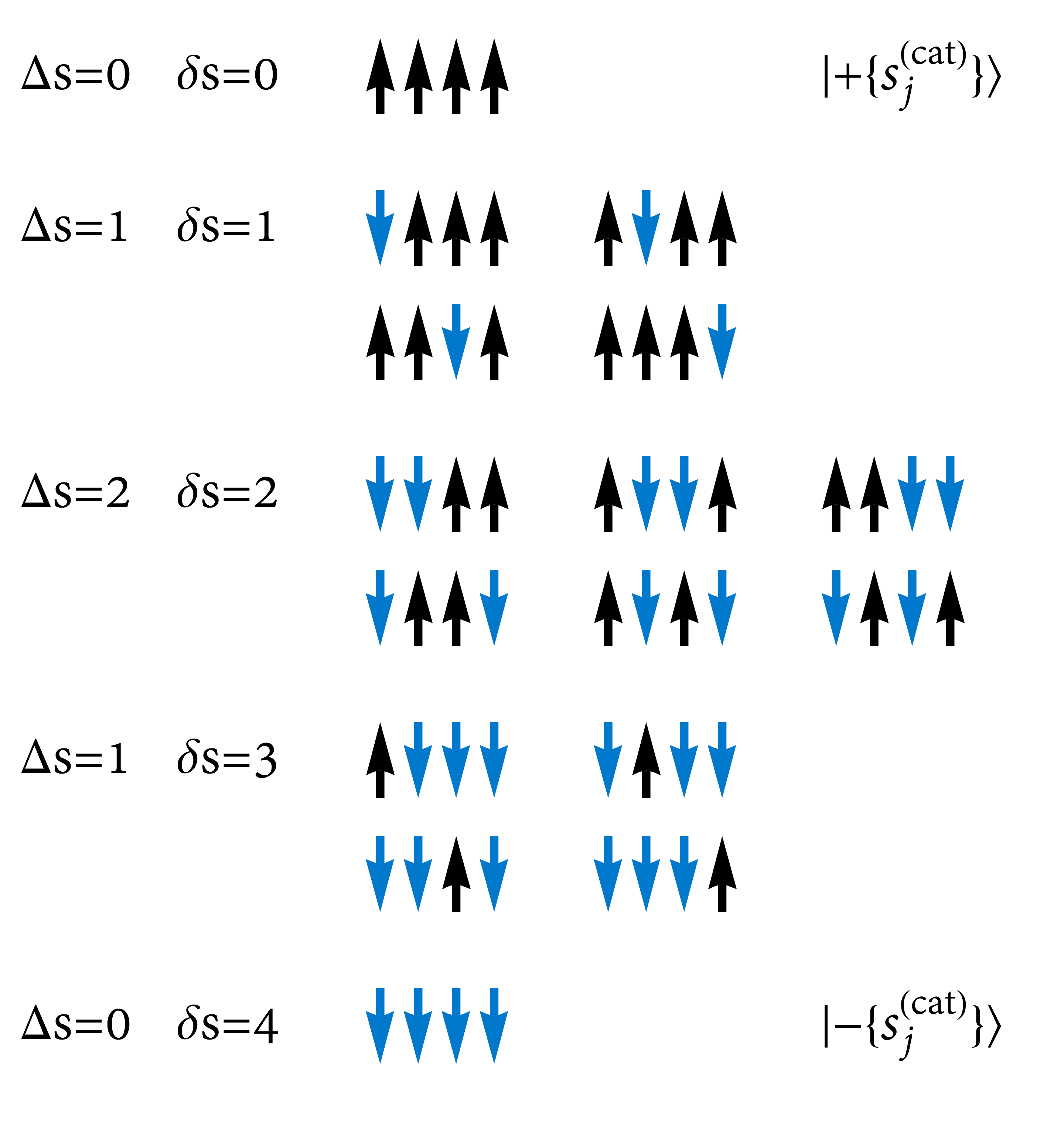} \\ \quad \\ \quad \\ (a) Schematic for $ |\ell, \text{FM}\rangle $ scars}
		\parbox[b]{8cm}{
			\parbox[b]{8cm}{
				\includegraphics[width=8cm]{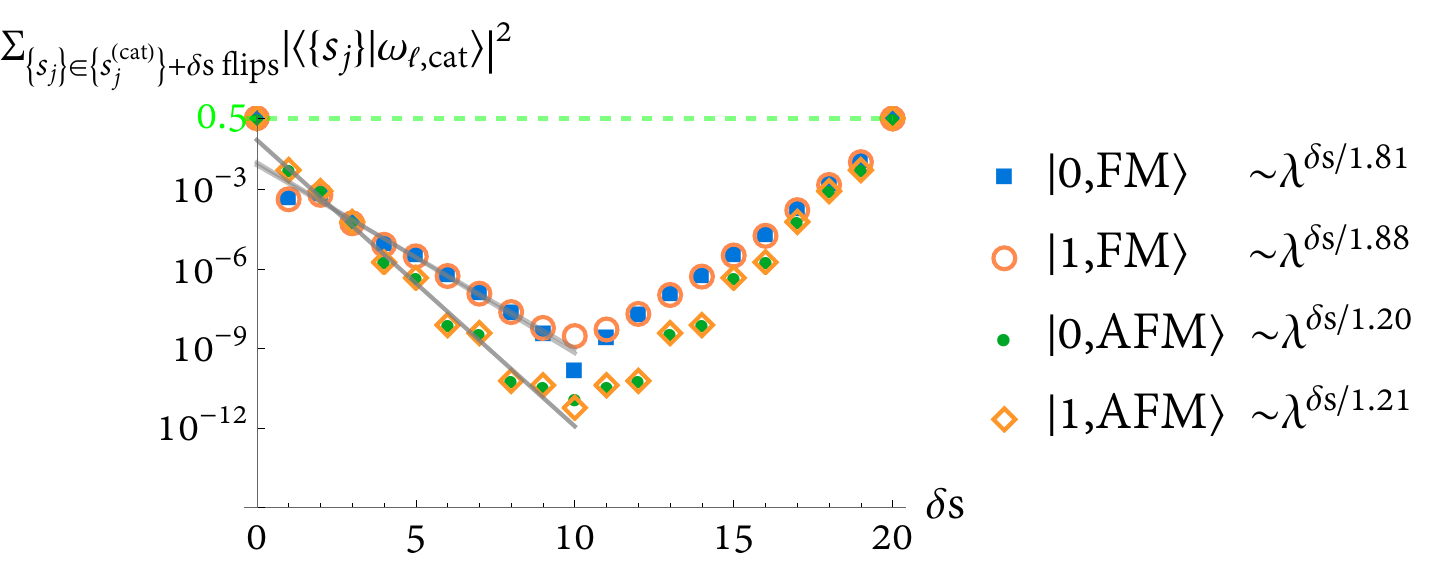}\\ (b) $ (\phi, \theta_x, \theta_y, \theta_z) \approx (0.3858, 0.7395, 0.3944, 0.3857) $ }
			\\
			\parbox[b]{8cm}{
				\includegraphics[width=8cm]{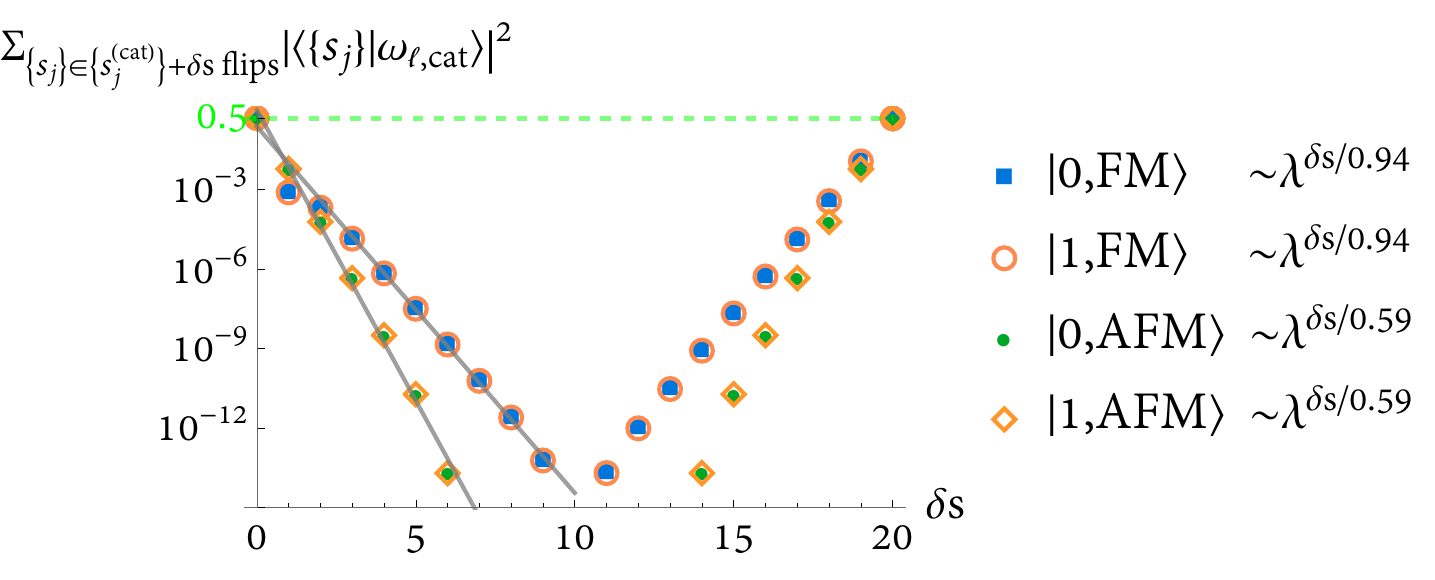}\\ (c) $ (\phi, \theta_x, \theta_y, \theta_z) \approx (0, 0.8016, 0.4275, 0.4180) $ } 
		}
		\parbox[b]{4.5cm}{\includegraphics[width=5cm]{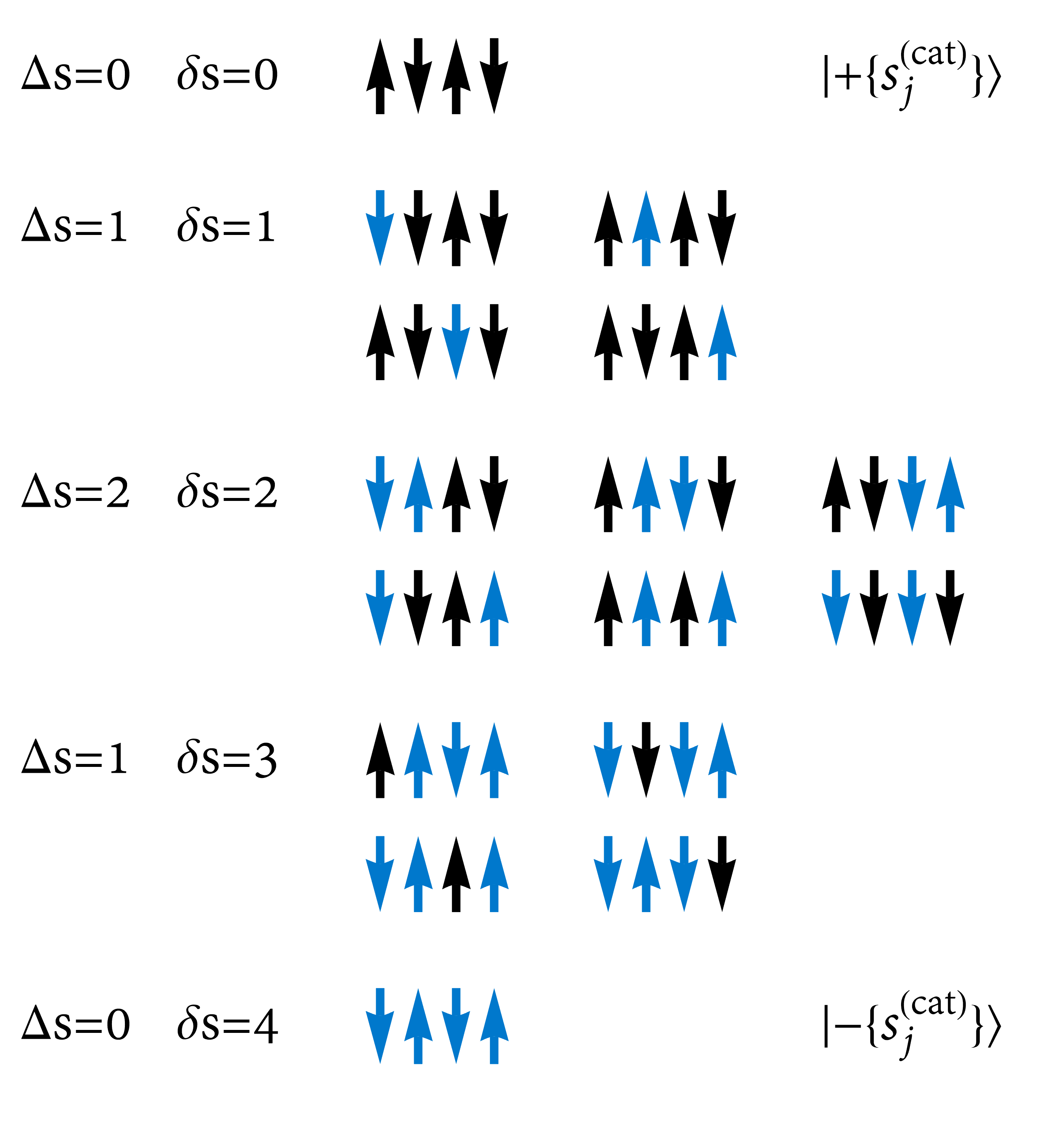} \\ \quad \\ \quad \\ (d) Schematic for $ |\ell, \text{AFM}\rangle $ scar}
		
		\caption{\label{fig:fock_loc} Numerical test of Fock space localization for cat scars. 
			(a) (d) Schematic illustration for counting spin flips in systems of size $ L=4 $, which quantify the Fock space distance between two configurations. 
			(b) Fock space localization when both one-spin ($ \theta_\mu \tau^\mu_j $) and two-spin ($ \phi \tau^x_j \tau^x_{j+1} $) perturbations are present. In this case, the Fock localization exponent $ \sim \lambda^{\delta s/\xi} $ is bounded by $ \xi\leqslant n_{\text{op}} = 2 $.
			(c) Fock space localization when only one-spin ($ \theta_\mu \tau^\mu_j $) perturbation is present. Here, due to a modified selection rule, the Fock localization exponent $ \sim \lambda^{\delta s/\xi} $ is bounded by $ \xi \leqslant n_{\text{op}} = 1 $. In (b) and (c), the dots are numerical data for the model in Eq.~\eqref{eq:model}. Gray lines are fittings with exponents specified in the legends for all scars. Perturbation strength reads $ \lambda = 0.05 $, system size $ L=20 $, and other parameters are specified in the figures. From the data, we do see a Fock space localization onto the cat scar configuration pairs $ |\pm \{ s_j^{(\text{cat})} \} \rangle $, as the wave function amplitudes for perturbed cat scars decay exponentially with the increase of spin flips with respect to $ \pm\{s_j^{(\text{cat})}\} $.  }
	\end{figure*}

	Moreover, for models with separable perturbations in $ U_F $, the constant $ \bar{V}_1 $ can indeed be {\em computed} exactly, such that the IPR scaling can be obtained without any fitting. To illustrate the calculations, and also to test our analytical theories further, we discuss the following Ising chain under both transverse and longitudinal fields,
	\begin{align}\label{eq:ufh}
		U_F = e^{-i(J\sum_{j=1}^L \tau^z_j \tau^z_{j+1} + h \sum_{j=1}^L \tau^z_j)} e^{-i\left( \frac{\pi}{2} - \lambda\right) \sum_j \tau^x_j}.
	\end{align}
	Such a model, including its disordered variants, have played important roles in recent numerics and experiments. We first note that a gauge transformation reduces it to
	\begin{align} \nonumber
		\tilde{U}_F 
		&= 
		e^{i(h/2)\sum_j \tau^z_j} U_F e^{-i(h/2) \sum_j \tau^z_j}
		= U_0 U',
		\\
		U' &= e^{i\lambda \sum_j (\tau^x_j \cos(h) - \tau^y_j \sin(h))}. 
	\end{align}
	Here, the zeroth order $ U_0 $ is the same as in Eq.~\eqref{eq:u0}. Compared with the generic form $ U' = e^{i\sum_{k=1}^\infty \lambda^k V_k} $, we see that here 
	\begin{align}
		V_1 = \sum_j (\tau^x_j \cos(h) - \tau^y_j \sin(h)),
	\end{align}
	while all others $ V_{k\geqslant2}=0 $. Thus, for both types of scars, a straightforward calculation using Eq.~\eqref{eq:vbar} gives the analytical form for perturbation strength
	\begin{align}\label{eq:v1barsimple}
		\bar{V}_1^2 = \frac{1}{4}\left( \frac{\cos^2h}{\sin^2 2J} + \frac{\sin^2h}{\cos^2 2J} \right).
	\end{align}
	We compare the analytical scaling relations for this model with numerics in Fig.~\ref{fig:ipr_simplemodel}. As expected, without {\em any} fitting parameters, Eqs.~\eqref{eq:iprscaling} and \eqref{eq:v1barsimple} agree well with the numerical data.

	The reduction of FM or AFM amplitudes in cat scars can be understood as the effect of domain-wall fluctuations. Specifically, in Eq.~\eqref{eq:vbar}, the first order perturbation introduces new configurations differing from FM or AFM ones by $ n_{\text{op}} $ spins as allowed by selection rules for $ V_{1,j} $, corresponding to domain wall creation or annihilation on top of the background FM or AFM configurations. That in turn rescales the normalization constant $ \alpha_0 $. In translation invariant systems, such effects simply accumulate, giving rise to the factor $ L $ in Eq.~\eqref{eq:scaling1}.

	Second, in Fig.~\ref{fig:fock_loc}, we consider the model in Eq.~\eqref{eq:model}, and numerically evaluate the scaling relation {\bf (2)} in Eq.~\eqref{eq:scaling2} of Fock space localization for cat scars. 
	Here, the Fock space distance $\delta s $ between a certain configuration $\{s_j\}$ and one of the cat scar configuration pair $\{s_j^{(\text{cat})}\}$ is defined as
	\begin{align}
		\delta s_{\text{cat}}(\{s_j\}) = \frac{1}{2} \sum_{j=1}^L \left|s_j^{(\text{cat})} - s_j \right|,
	\end{align}
	while the pairwise deviation $\Delta s$ is defined in Eq.~\eqref{eq:Ds}.
	A schematic illustration of counting Fock space distance, both in terms of $\delta s$ and $\Delta s$, is given in Fig.~\ref{fig:fock_loc} (a) and (d) for FM and AFM cat scars respectively.
	Note that the localization length $ \xi $ in Eq.~\eqref{eq:exponent_Fock} is bounded by the selection rule for $ V_k $'s. Namely, if the perturbation Hamiltonians involve up to two-spin terms $ n_{\text{op}} = 2 $, the selection rules for $ V_k $ is that it could flip up to $ n_{\text{op}}k = 2k $ spins. That subsequently gives the bound $ \xi\leqslant n_{\text{op}} = 2 $ in Eq.~\eqref{eq:exponent_Fock}. This is verified in Fig.~\ref{fig:fock_loc} (b) that for all the four scars (corresponding to the case in Fig.~\eqref{fig:cats} (a)), $ \xi $ is bounded by 2. We could further test the bounds by reducing the perturbations to involving only one-spin terms, namely, setting $ \phi=0 $ for $ \phi\tau^x_j \tau^x_{j+1} $ leaving only $ \theta_\mu \tau^\mu_j $ in the perturbations of Eq.~\eqref{eq:model}. Consequently, $ n_{\text{op}} =1 $ and the selection rules is modified to that $ V_k $ can only flip $ n_{\text{op}} k = 1\times k $ spins, such that localization length is bounded by $ \xi\leqslant 1 $. The analytical result is again confirmed in Fig.~\ref{fig:fock_loc} (c), with the FM scars appearing to saturate the new bounds.

	%Up to this point, we have proved the counter-intuitive Fock space localization for cat scars in clean systems. Its reasons can be summarized as follows. (1) The zeroth order Floquet operator $ U_0 $ is highly localized, relating only pairwise opposite Fock states $ U_0 |\{s_j\}\rangle \propto |- \{s_j\}\rangle $. Meanwhile, the four cat scars are the only non-degenerate eigenstates in the fine-tuned limit $ \lambda=0 $. (2) The strong Ising interaction validates the perturbative treatment, and especially the selection rules that the $ \lambda^k $-th order perturbation can only flip as many as $ n_{\text{op}} k $ spins. The operator product order $ n_{\text{op}} $ counts at most how many spins operators are multiplied, i.e. if the bare perturbations involve both one-spin ($ \theta_\mu \tau^\mu_{j} $) and two-spin ($ \phi \tau^x_j \tau^x_{j+1} $) terms, $ n_{\text{op}} = 2 $; and if $ \phi=0, \theta_\mu\ne 0 $, $ n_{\text{op}}=1 $. (3) The spin flipping in perturbative corrections can be thought of as a scattering process, where $ \lambda^k $-th order terms could scatter the cat configurations $ \pm\{s_j^{(\text{cat})}\} $ by $ n_{\text{op}}k $ spins. Thus, flipping more spins away from $ \pm\{s_j^{(\text{cat})}\} $ is suppressed by higher powers of $ \lambda $, giving rise to the exponential Fock space localization.

	\begin{figure*}
		[t]
		\parbox[b]{8.5cm}{
			\parbox{4cm}{
				\includegraphics[width=4.2cm]{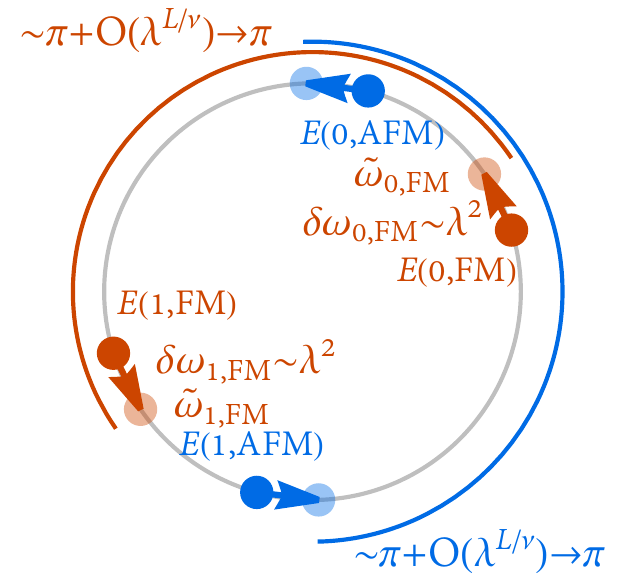} \\ (a) Schematics of SP}
			\\ \quad \\
		}
		\parbox[b]{8cm}{\includegraphics[width=6cm]{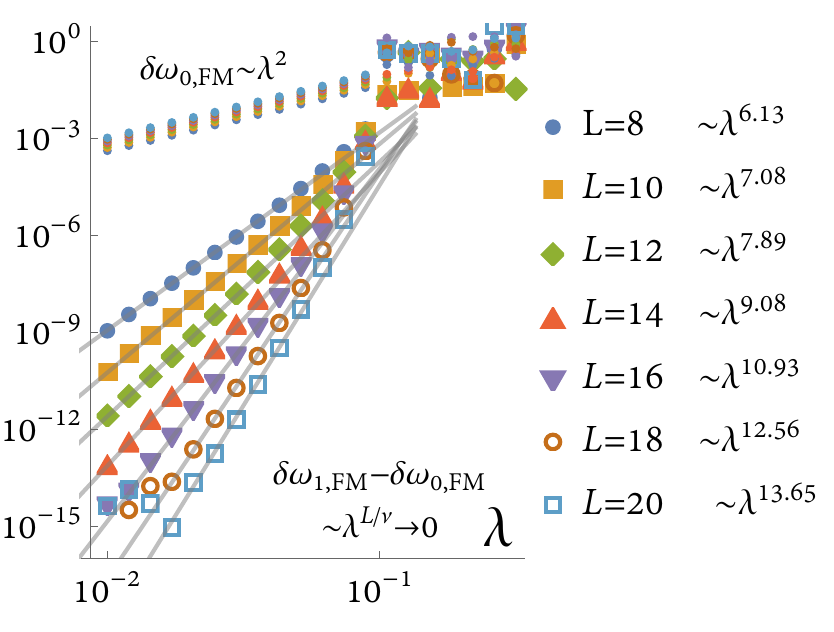} \\ (b) $ \tilde{\omega}_{\ell,\text{FM}} $ with $ (\phi, \theta_x, \theta_y, \theta_z) \approx (0.3858, 0.7395, 0.3944, 0.3857) $}
		\\
		\parbox[b]{8cm}{\includegraphics[width=6cm]{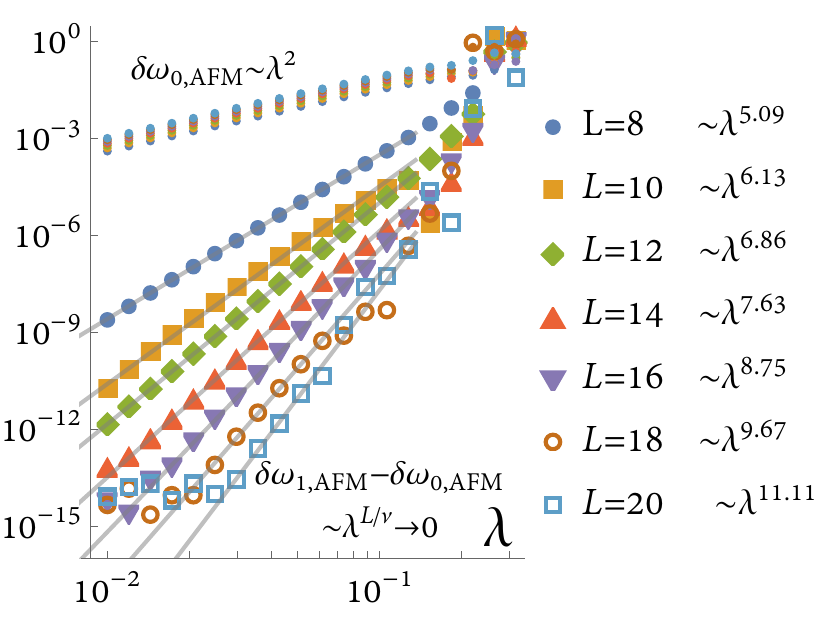} \\ (c) $ \tilde{\omega}_{\ell,\text{AFM}} $ with $ (\phi, \theta_x, \theta_y, \theta_z) \approx (0.3858, 0.7395, 0.3944, 0.3857) $}
		\parbox[b]{8cm}{\includegraphics[width=6cm]{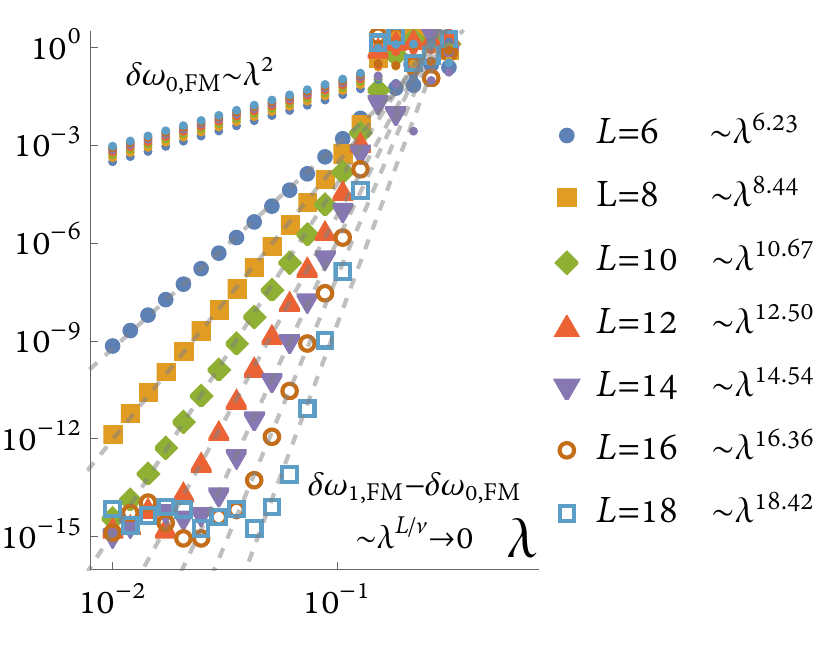} \\ (d) $ \tilde{\omega}_{\ell,\text{FM}} $  with $ (\phi, \theta_x, \theta_y, \theta_z) \approx (0, 0.8016, 0.4275, 0.4180) $}
		\caption{\label{fig:sp} Scaling of the exponentially small SP deviation from $ \pi $ for the perturbed cat scar pair. (a) Schematic illustration of the SP mechanism: each scar level could shift considerably $\sim\lambda^2$ under perturbation $\lambda$; but they shift by almost the same amount, so the quasienergy difference between a pair of cat scar approaches $\pi$. (b) -- (d) Scaling for the spectral gap deviation away from $ \pi $, which vanishes exponentially $ O(\lambda^{L/\nu}) $ with the increase of system size $ L $. Note that the exponent $ \nu\leqslant n_{\text{op}} $ is bounded by the operator product power for perturbation Hamiltonians. In (b) (c) there are both one-spin $ \theta_\mu \tau^\mu_j $ and two-spin terms $ \phi \tau^x_j \tau^x_{j+1} $, so $ \nu\leqslant n_{\text{op}}=2 $. Instead, we turn off the two-spin perturbations $ \phi=0 $ in (d), which reduces the exponent to $ \nu\leqslant 1 $. The spectral gap deviations for FM scars in this case saturates the bound $ \delta\omega_{1,\text{FM}} - \delta\omega_{0,\text{FM}} \sim O(\lambda^L) $. Parameters in (b) -- (c) are the same as in Fig.~\ref{fig:fock_loc} (b), while parameters in (d) are the same as those in Fig.~\ref{fig:fock_loc} (c). }
	\end{figure*}

	Finally, we compare the SP deviation scaling in Eq.~\eqref{eq:scaling3} with numerical investigations of Eq.~\eqref{eq:model}. 
	
	The schematic picture for the fixed spectral gap $ \pi $ is illustrated in Fig.~\ref{fig:sp} (a). At the fine-tuned point $ \lambda=0 $, the cat scar $ E(1,\{s_j^{(\text{cat})}\}) $ separates from its partner $ E(0, \{s_j^{(\text{cat})}\}) $ by quasienergy $ \pi $, where $\{s_j^{(\text{cat})}\}$ denotes FM or AFM. Under perturbation $ \lambda $, each level $ E(\ell,\{s_j^{(\text{cat})}\}) $ is shifted to $ \tilde{\omega}_{\ell, \text{cat}} $ by a significant amount $ \tilde{\omega}_{\ell, \text{cat}} - E(\ell, \text{cat}) = \delta\omega_{\ell, \text{cat}} \sim \lambda^2 $. However, both levels $ \ell=0,1 $ shift by almost identical amounts $ \delta\omega_{1,\text{cat}} - \delta\omega_{0,\text{cat}} = O(\lambda^{L/\nu}), \nu\leqslant n_{\text{op}} $, such that the perturbed scars still show a spectral gap $ \tilde{\omega}_{1, \text{cat}} - \tilde{\omega}_{0, \text{cat}} = \pi + O(\lambda^{L/\nu}) $, with deviations $ O(\lambda^{L/\nu}) = O(e^{-|\ln(1/\lambda)|L/\nu}) $ shrinking exponentially with the increase of system sizes $ L $.
	
	The spectral gap scaling is verified in Fig.~\ref{fig:sp} (b) -- (d). In all cases, as shown by the small dots, the quasienergy correction for {\em individual} levels $ \delta\omega_{\ell, \text{cat}} $ always scale as $ \lambda^2 $, which are relatively large. However, both scars $ \ell=0,1 $ shift by the same amount, such that pairwise deviations $ \delta\omega_{1,\text{cat}} - \delta\omega_{0, \text{cat}} $ vanish exponentially with the increase of system sizes $ L $. To further test the bounds for scaling exponent $ \nu $, note that when both one-spin ($ \theta_\mu \tau^\mu_j $) and two-spin ($ \phi \tau^x_j \tau^x_{j+1} $) perturbations are present, $ \nu\leqslant n_{\text{op}}= 2 $, as observed in (b) and (c) for both FM and AFM scars. We could further test the bounds similar to what was done in Fig.~\ref{fig:fock_loc} (d) by shutting down the two-spin terms in perturbing Hamiltonians, i.e. $ \phi=0 $, and only allow for one-spin perturbations $ \theta_\mu \tau^\mu_j $. Then, we see in Fig.~\ref{fig:sp} (d) that indeed the exponent $ \nu $ for spectral gap deviation $ O(\lambda^{L/\nu}) $ saturates the new bound $ \nu\leqslant n_{\text{op}} = 1 $.
	
	The exponent $ \nu $ in Eq.~\eqref{eq:exponent_Fock} satisfies the same bound as the pattern localization length $ \xi $ in Fock space as given by Eq.~\eqref{eq:exponent_lifetime}. However, it is worth clarifying that they are not the same quantity because they arise from different perturbation orders. The Fock space localization length $ \xi $ is chiefly contributed by lower order perturbations $ k< L/2n_{\text{op}} $, as the $ \lambda^{k} $-th order terms would involve spin configurations $ |\{s_j\}\rangle $ separating from FM or AFM ones $ \pm\{s_j^{(\text{cat})}\} $ by $ \leqslant n_{\text{op}}k $ spin flips, thereby giving the exponential scaling as in Fig.~\ref{fig:fock_loc}. In contrast, the exponent $ \nu $ for spectral gap deviations is determined by higher orders terms $ k\geqslant L/n_{\text{op}} $, because all lower-order quasienergy corrections $ \lambda^{k< L/n_{\text{op}}} $ are strictly the same for cat scar pairs  $ \omega_{1,\text{cat}}^{(k)} - \omega_{0,\text{cat}}^{(k)} = 0 $ and therefore no deviation would ever exist until one reaches the perturbation of order $ \lambda^{k\geqslant L/n_{\text{op}}} $.

	Let us summarize the results we obtain so far. Analytically derived scaling relations in Eqs.~\eqref{eq:scaling1}, \eqref{eq:scaling2}, and \eqref{eq:scaling3} (see Appendix~\ref{smsec:scaling} for rigorous proof), are verified by the corresponding scaling of IPR (Fig.~\ref{fig:iprscaling1} and \ref{fig:ipr_simplemodel}), Fock space localization (Fig.~\ref{fig:fock_loc}) and spectral gap deviation (Fig.~\ref{fig:sp}) respectively. They constitute a scheme to characterize the robustness of scars and predict the corresponding DTC behaviors as in Eq.~\eqref{eq:dtc}. These scalings all originate from the selection rule for perturbations in Eq.~\eqref{eq:selection}, which is enforced by the strong Ising interaction. Under this condition, we could start from a Fock state with spin patterns prescribed by the symmetry indicator in Eq.~\eqref{eq:symm} or its generalization (i.e. Eq.~\eqref{eq:symm_gen} for systems with sublattices, and Eq.~\eqref{eq:Jsign} \eqref{eq:morecats} for arbitrary patterns). Then, these scaling relations follow and capture the resulting DTC dynamics.

	\section{Two ways to distinguish interaction versus single-spin effects \label{sec:diffusive}}

	In the previous sections, we have investigated the cat scar structures and the corresponding DTC dynamics resulting from strong Ising interactions. When studying specific models, nevertheless, we may observe certain features that could possibly result from rather different reasons. As such, this section is devoted to distinguishing the cat scar DTCs from a class of rather subtle systems. It typically involves a weak interaction and relatively simple models, showing phenomena that very much resemble DTCs with robust SP and localization at early time. However, they are more related to single-spin effects instead. Specifically, we would discuss two possible reasons of spin echos and non-interacting integrability where such single-spin term dominated oscillations could arise, and offer the practical ways to distinguish them from the many-body DTCs enforced by cat scars based on early time dynamics.

	For our purposes here, let us start from the model introduced in Eq.~\eqref{eq:ufh}, which we reproduce here for convenience,
	\begin{align}\label{eq:simplemodel}
		U_F = U_2 U_1 = e^{-i\sum_{j=1}^L (J \tau^z_j \tau^z_{j+1} + h_z \tau^z_j) } e^{-i\left( \frac{\pi}{2} - \lambda \right) \sum_{j=1}^L \tau^x_j}.
	\end{align}
	(Note that here all dimensionless parameters $J, h_z, (\pi/2 - \lambda)$ are already compared with the Floquet driving frequency, i.e. the dimensionless parameter $J$ here is related to the interaction strength carrying energy unit $H_2 \sim \tilde{J} \tau^z_j \tau^z_{j+1}$ in Hamiltonians by $J = \tilde{J}T/2\hbar $).
	We would like to benchmark the roles of interaction and many-body effects, and therefore distinguish the two cases with strong (i.e. $ J =1$ discussed in Eq.~\eqref{eq:ufh}) and weak ($J=0.1$) interactions. The following analysis shows that although in certain cases, weakly interacting systems $J=0.1$ could mimic the many-body DTC behaviors, these oscillations in weakly interacting systems are more consistent with a fine-tuned single-particle physics description.

	\subsection{Single spin echos}
	
	\begin{figure}[h]
		\parbox[b]{4cm}{
			\includegraphics[width=4.2cm]{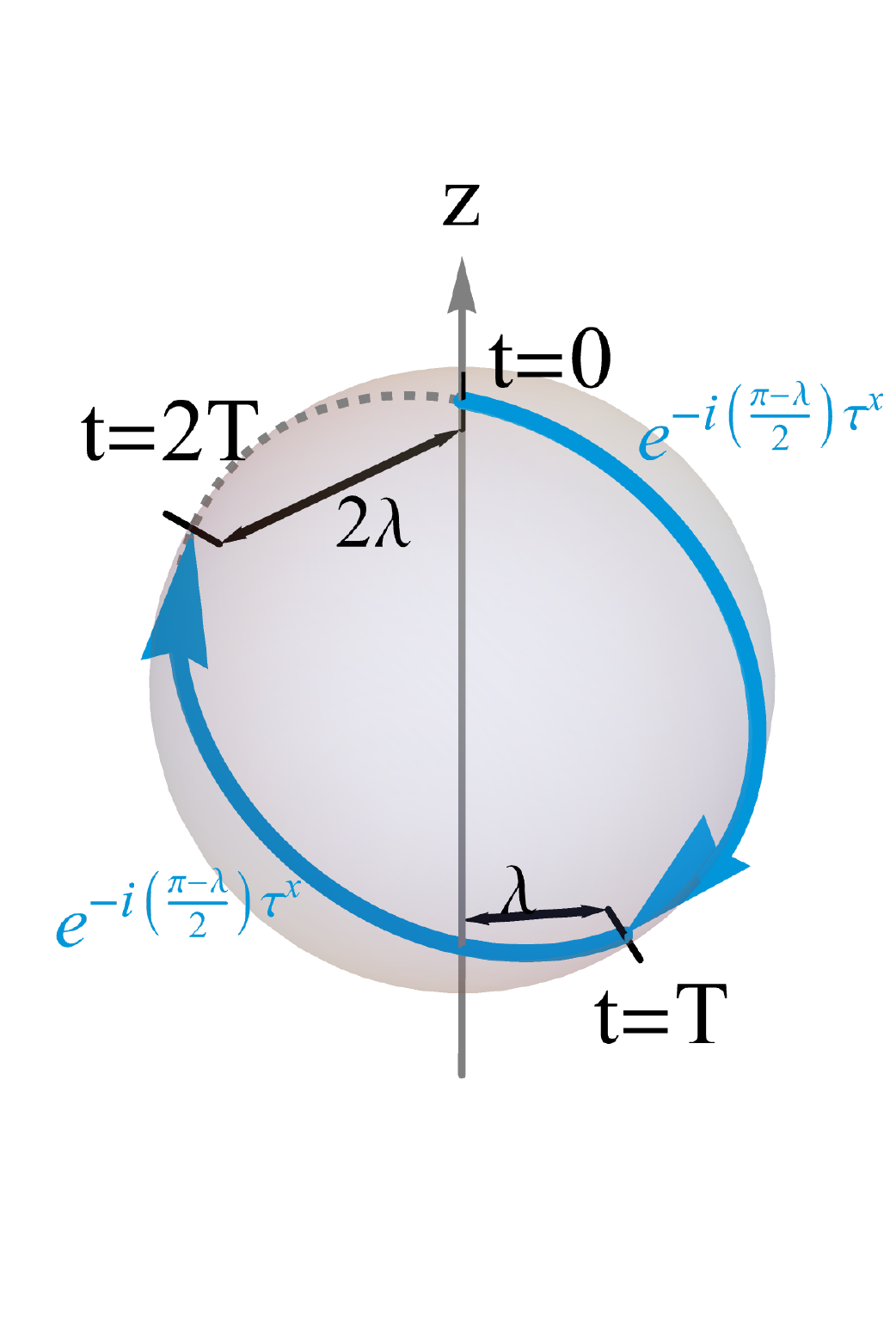}
			\\ (a) $h_z\rightarrow0$
		}
		\parbox[b]{4cm}{
			\includegraphics[width=4.2cm]{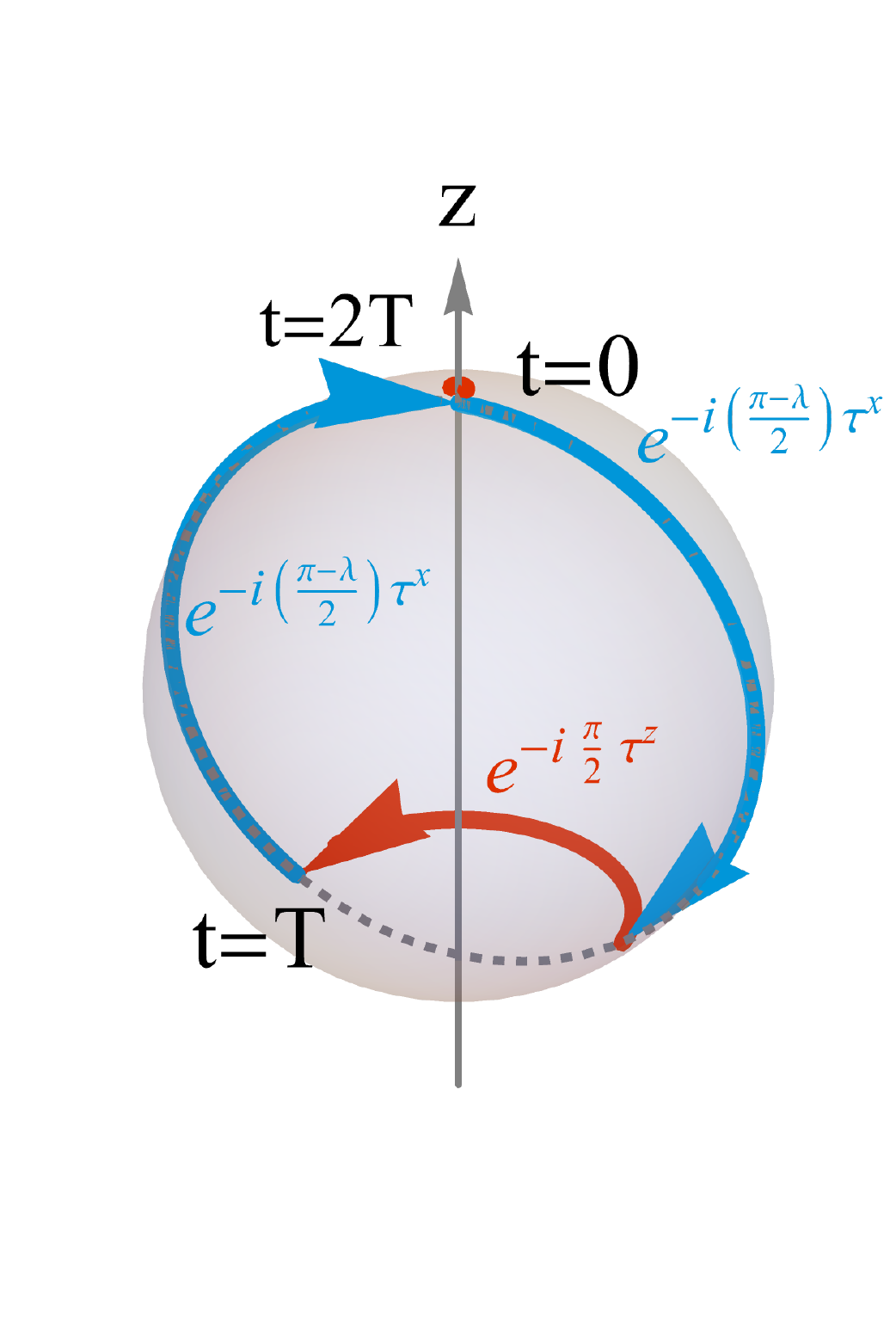}
			\\ (a) $h_z\rightarrow\pi/2$
		}
		\caption{\label{fig:echo}Schematic plot for the single-particle echo that generate early-time DTC-like oscillations in the weakly interacting case. (a) When $h_z\rightarrow0$, spin flip errors accumulate. (b) With a field strength $h_z\rightarrow\pi/2$, a single-particle spin echo is recovered every two periods. To distinguish cat scar DTCs from such single-particle physics, it is suggested to tune $h_z$ far away from the echo limit, like in Fig.~\ref{fig:dynamics}. }
	\end{figure}
	
	The first effect we would like to check is the {\bf approximate single-spin echo}, which is schematically illustrated in Fig.~\ref{fig:echo}. DTC phenomenon in many cases is observed in the following fashion: {\bf (1)} With $U_1$ in Eq.~\eqref{eq:simplemodel} alone, the spin oscillation frequency is fine-tuned. For instance, the perturbation $\lambda  \sum_{j=1}^L \tau^x_j $ in Eq.~\eqref{eq:simplemodel} will result in a frequency deviation from $\pi$, because the errors in spin flips, as shown in Fig.~\ref{fig:echo} (a), will accumulate during oscillations. {\bf (2)} Then, once we turn on $U_2$ in Eq.~\eqref{eq:simplemodel}, the $2T$-periodic oscillations may be recovered, indicating DTC physics due to the effects of interactions $J$. However, there is a subtlety here: in addition to the interactions $J$, there is also a single-spin rotation $\sim h_z$ in $U_2$ of Eq.~\eqref{eq:simplemodel}. If the values of $h_z$ is fine-tuned around $\pi/2$ --- corresponding to a spin $\pi$-pulse rotation around the $z$-axis --- a single-spin echo could be achieved in every two periods, as shown in Fig.~\ref{fig:echo} (b). Mathematically, the echo corresponds to the identity describing single-spin rotations over two periods,
	\begin{align}\nonumber
		&\quad e^{-i\frac{\pi}{2} \sum_{j=1}^L \tau^z_j} e^{-ig\sum_{j=1}^L \tau^x_j} e^{-i\frac{\pi}{2} \sum_{j=1}^L \tau^z_j} e^{-ig\sum_{j=1}^L \tau^x_j} 
		\\ \nonumber
		&= (-1)^L \left( \prod_{j=1}^L \tau^z_j \right) e^{-ig\sum_{j=1}^L \tau^x_j} \left( \prod_{j=1}^L \tau^z_j \right)  e^{-ig\sum_{j=1}^L \tau^x_j} 
		\\
		&= (-1)^L e^{+ig\sum_{j=1}^L \tau^x_j} e^{-ig\sum_{j=1}^L \tau^x_j}  = (-1)^L,
	\end{align}
	where we used $\tau^z_j \tau^x_j \tau^z_j = -\tau^x_j$. That means even if $J=0$ in Eq.~\eqref{eq:simplemodel} (in non-interacting cases), for {\em arbitrary} perturbation strength $e^{i \sum_{j=1}^L \tau^x_j} $, a sinlge-spin echo can always be achieved every two periods if $h_z \rightarrow \pi/2$.

	\begin{figure}[h]
		\parbox[b]{3.8cm}{
			\includegraphics[width=3.8cm]{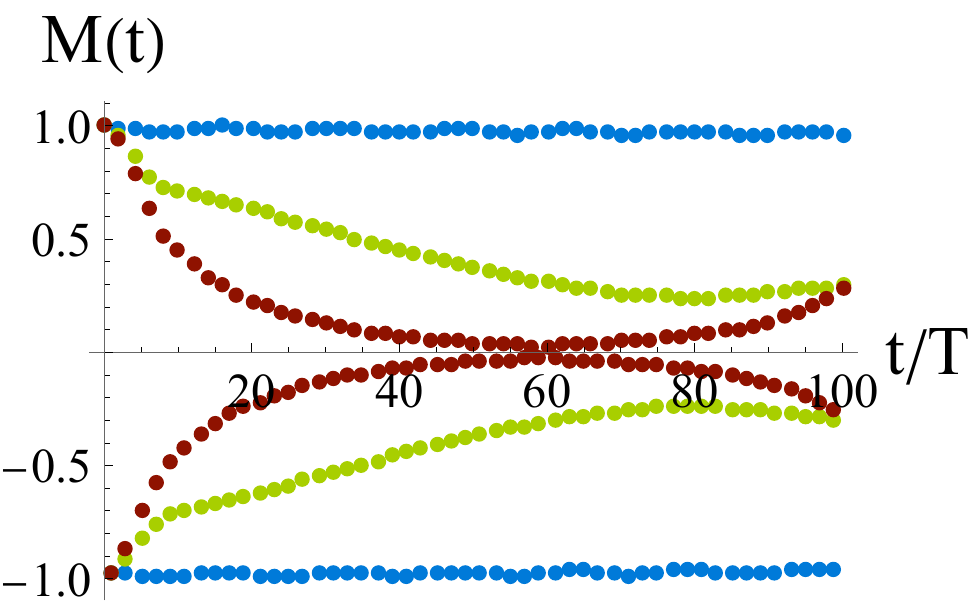}
			\\
			(a) $J=0.1$
		}
		\parbox[b]{4.5cm}{
			\includegraphics[width=4.8cm]{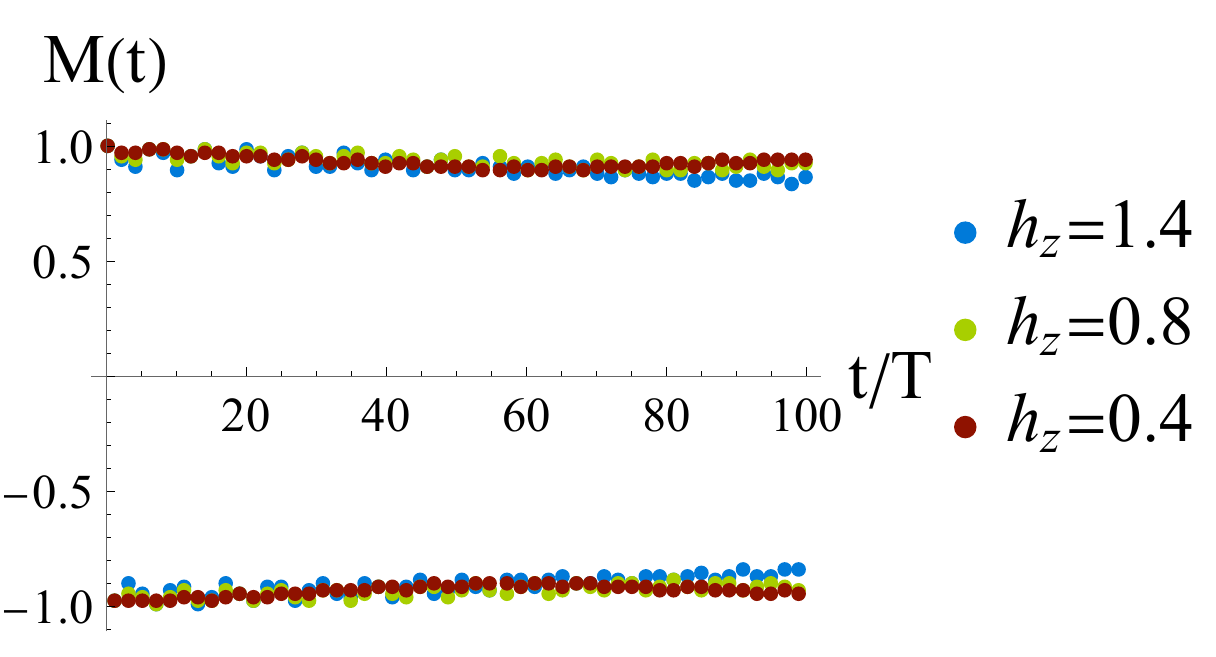}
			\\
			(b) $J=1$
		}
		\caption{\label{fig:preth_hz} Effects of changing single-particle terms $\sum_{j=1}^L h_z\tau^z_j$ in Eq.~\eqref{eq:simplemodel}. (a) In the weakly interacting case, the DTC-like oscillation at early time is sensitive to single-spin dynamics, where $h_z \rightarrow \pi/2\approx 1.57$ generates an almost perfect spin echo for every two periods. Consequently, tuning $h_z$ away from $\pi/2$ strongly supresses the $2T$ oscillations. (b) In contrast, the strongly interacting cat-scar DTC is insensitive to such single-particle parameter fine-tunings.   }
	\end{figure}
	
	Thus, we are prompted to double-check whether certain observed DTC-like oscillations are dominated by such single-spin echos, rather than many-body effects. An example is illustrated in Fig.~\ref{fig:preth_hz}, where we compare the evolutions for weakly ($J=0.1$ in Fig.~\ref{fig:preth_hz} (a)) versus strongly ($J=1$ in Fig.~\ref{fig:preth_hz} (b)) interacting cases concerning the model in Eq.~\eqref{eq:simplemodel}. In order to check whether local oscillations without diffusions occur, we start with the AFM initial state, and compute $M(t)$ as in Eq.~\eqref{eq:corr}. It is found numerically that for the case with weak interactions in Fig.~\ref{fig:preth_hz} (a), DTC like oscillations at early time sensitively depend on the longitudinal field strength $h_z$. Only in the single-particle echo limit, i.e. $ h_z = 1.4 \rightarrow\pi/2 \approx 1.57$, the system may exhibit a local DTC type of oscillation. When $h_z$ deviates from such a limit, like the cases with $  h_z = 0.8$ and $0.4$, local DTC oscillation is notably suppressed even at early time $t/T<100$. We have also verified that an analogous scenario occurs for disordered magnetic fields $h^z_j \in [-h_z/2, h_z/2]$ under uniform weak interactions $J=0.1$ as in Fig.~\ref{fig:preth_hz}. This is to be sharply contrasted against the strongly interacting case in Fig.~\ref{fig:preth_hz} (b) where the cat scar enforced DTCs are insensitive to such single-particle echos.

	To understand such a difference, let us compare the different energy scales in two cases more carefully. For weakly interacting situations with strong longitudinal fields, the dominant energy scales in Eq.~\eqref{eq:simplemodel} are single-spin magnetic fields $h_z, (\pi/2 - \lambda) \sim \pi/2$, both of which are comparable with the Floquet driving frequency. Interactions $J=0.1$ are then relatively small parameters that could be treated as perturbations. Thus, the mechanism underlying DTC-like oscillations observed for $h_z = 1.4$ in Fig.~\ref{fig:preth_hz} (a) is an almost perfect single-spin echo, perturbed by deviations from the echos $\lambda =0.1, |h_z-\pi/2| \approx 0.17$, and also interactions $J=0.1$ ensuring long-time thermalization. All of these perturbations are one order of magnitude weaker than driving frequency. In fact, in the limit $h_z\rightarrow\pi/2$, since spin echos occur for each individual spin, local DTC-like oscillations could occur for rather generic initial states at early time. Nevertheless, due to the single-particle nature of such oscillations, when the deviations from single-spin echos are enhanced for $h_z=0.8, 0.4$ (so that $|h_z-\pi/2| \approx 0.77, 1.17 $), DTC type of oscillations are destroyed. In contrast, for strongly interacting cases in Fig.~\ref{fig:preth_hz} (b), the dominant role is played instead by interactions $J=1$. That results in the selection rules for perturbations and the domain wall structures observed in Fig.~\ref{fig:dw}, prescribing the localization of eigenstates to different domain wall sectors of the many-body Fock space. In such cases, whether single spin echo occurs or not is no longer important for the essential mechanism.

	\subsection{Single-particle integrability}
	
	While the effects of single-spin echos are relatively intuitive to understand, there is a second effect due to {\bf approximate single-particle integrability} that could be slightly more subtle. Specifically, we could consider a gauge transformation of Eq.~\eqref{eq:simplemodel} to factor out a perfect $\pi$-pulse, and define an effective Hamiltonian for the remaining terms
	\begin{align}\nonumber
		\tilde{U}_F &=
		(-i)^L V U_F V^\dagger
		\qquad\qquad
		\left( V = e^{i \sum_{j=1}^L (h^z_j/2) \tau^z_j } \right) \\ \nonumber
		&=  \left( 
		e^{-i\sum_{j=1}^L J\tau^z_j \tau^z_{j+1}}  e^{i\lambda \sum_{j=1}^L ( \tau^x_j \cos(h_z) - \tau^y_j \sin(h_z)) }
		\right)  P
		\\ \label{eq:gaugetrans}
		&\equiv
		e^{-iH_{\text{eff}}}  P, 
		\qquad\qquad
		P = \prod_{j=1}^L \tau^x_j 
	\end{align}
	Such an operation shares certain similarity with the way prethermal Hamiltonians were defined in Ref.~\cite{Else2017}, but with the crucial difference that the interaction $J$ could be fairly large. Thus, if we were to treat $H_{\text{eff}}$ as a prethermal Hamiltonian, the associated prethermal time $\tau_* \sim e^{1/J} $ could be rather short if $J$ is strong. Instead, we define $H_{\text{eff}}$ here chiefly to capture the early time dynamics due to approximate integrability, for both weak and strong $J$ in a unified way. That only involves dominant lowest-order terms in $H_{\text{eff}}$. Late time dynamics, in contrast, requires incorporations of all higher-order smaller terms in Eq.~\eqref{eq:gaugetrans} and would not be considered here.

	{\bf (1)} For the {\bf weakly interacting} case $J\sim0.1$ in Eq.~\eqref{eq:gaugetrans},  the lowest-order effective Hamiltonian are simply the summation of all terms in the exponential, 
	\begin{align}\nonumber
		H_{\text{eff}} &= H^{(1)}|_{J\ll1} + O(\lambda^2, J^2, \lambda J), 
		\\
		 \label{eq:heffweak}
		H^{(1)}|_{J\ll 1} &= \sum_{j=1}^L \left( J \tau^z_j \tau^z_{j+1} - \lambda \left(\cos(h_z) \tau^x_j - \sin(h_z) \tau^y_j \right) \right) 
	\end{align}
	where the higher-order terms $ \sim \lambda^2, J^2, \lambda J \sim 10^{-2}$ would only be relevant on the time scale of several hundreds of periods. 
	%(Effective Hamiltonians of this form have been discussed in several previous works~\cite{Khemani2019b,Mi2022,Kyprianidis2021}, defined in slightly different ways by considering double-period ends $U_F^2 = e^{-2iH_{\text{eff}}'}$. $H'_{\text{eff}}$ generally becomes more integrable than $H_{\text{eff}}$ defined in our case, because $U_F^2$ or higher $U_F^{n>2}$ involve Floquet spectral folding, where several irrelevant parts of quasienergy spectrum may be layered on top of each other due to power-raising $e^{i\omega_m} \rightarrow e^{in\omega_m}$. The corresponding level spacing ratio $\langle r\rangle$ may approach Poissonian limit when consecutive gaps are considered for $U_F^{n\geq2}$ with increasing $n$). 
	We could observe that Eq.~\eqref{eq:heffweak} is a transverse-field Ising model, which can be mapped to non-interacting free fermions $\{f_j^\dagger, f_k\} = \delta_{jk}$ in one-dimension via Jordan-Wigner transformation~\cite{Sachdev2011}
	\begin{align} 
		\nonumber
		\tau^z_j 
		&\rightarrow (f_j^\dagger + f_j) (-1)^{\sum_{k=1}^{j-1} f_k^\dagger f_k}, 
		\\ \nonumber
		- \sin(h_z) \tau^x_j - \cos(h_z) \tau^y_j 
		&\rightarrow -i(f_j^\dagger - f_j) (-1)^{\sum_{k=1}^{j-1} f_k^\dagger f_k}
		\\
		\cos(h_z)\tau^x_j - \sin(h_z) \tau^y_j 
		&\rightarrow (2f_j^\dagger f_j - 1),
	\end{align}
	resulting in
	\begin{align} \label{eq:jordanweakh2}
		&
		H^{(1)}|_{J\ll1} = \sum_{j=1}^L \left(
		J(f_j^\dagger f_{j+1} + f_j^\dagger f_{j+1}^\dagger + h.c.) - 2 \lambda f_j^\dagger f_j
		\right)
	\end{align}
	where constant terms are neglected. Then, we see that Eq.~\eqref{eq:jordanweakh2} describes the Bloch band for mean-field spinless $p$-wave superconductors. The single-particle crystal momenta $k$, i.e. in the Fourier transformation $f_k = (1/\sqrt{L}) \sum_{j=1}^L f_j e^{ikx_j}$, are conserved quantities and therefore Eq.~\eqref{eq:jordanweakh2} is integrable.
	%For disordered magnetic fields, i.e. $h^z_j \in[-\delta h_z, \delta h_z]$, Eq.~\eqref{eq:jordanweakh2} corresponds to an Anderson insulator, as arbitrarily weak disorders $\delta h_z$ (in terms of random onsite chemical potentials $\lambda \cos(h^z_j)$ in Eq.~\eqref{eq:jordanweakh2}) would cause single-particle localization in one dimension. 
	
	Thus, for weakly interacting situation $J\sim 0.1$ in Eq.~\eqref{eq:simplemodel}, its early time dynamics is dominated by non-interacting single-particle integrable models. Such an approximate integrability causes a slow relaxation and delays the thermalization. Consequently, we observe that even for relatively weak $h_z =0.8, 0.4$ in Fig.~\ref{fig:preth_hz} (a), there still appears certain DTC-like oscillations with suppressed amplitudes at early time $t/T<100$, rather than showing thermalizing behaviors without dynamics (i.e. $M(t)$ decays to zero with small fluctuations). 
	
\begin{figure}[h]
		\parbox[b]{3.8cm}{
			\includegraphics[width=3.8cm]{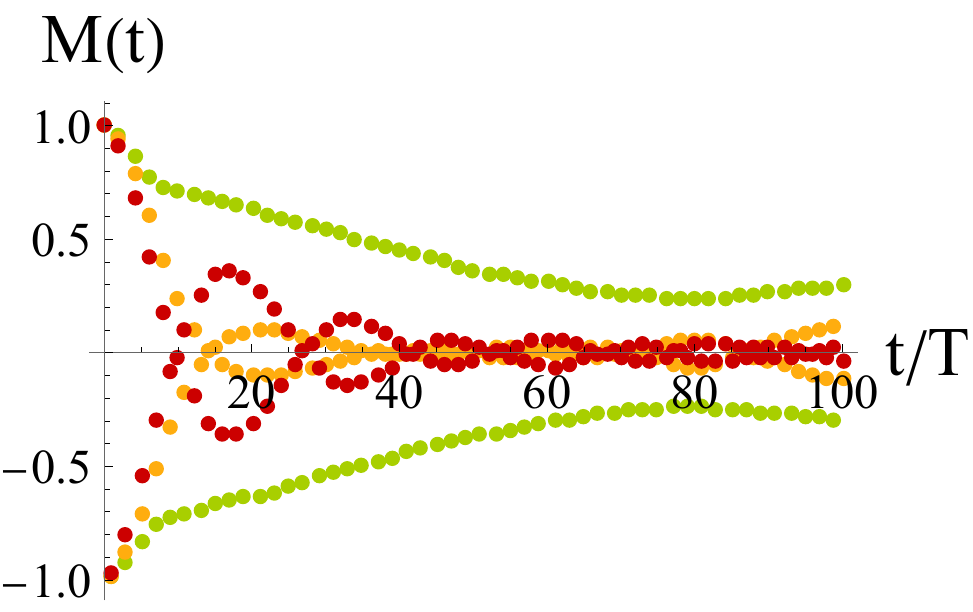}
			\\
			(a) $J=0.1, h_z=0.8$
		}
		\parbox[b]{4.5cm}{
			\includegraphics[width=4.9cm]{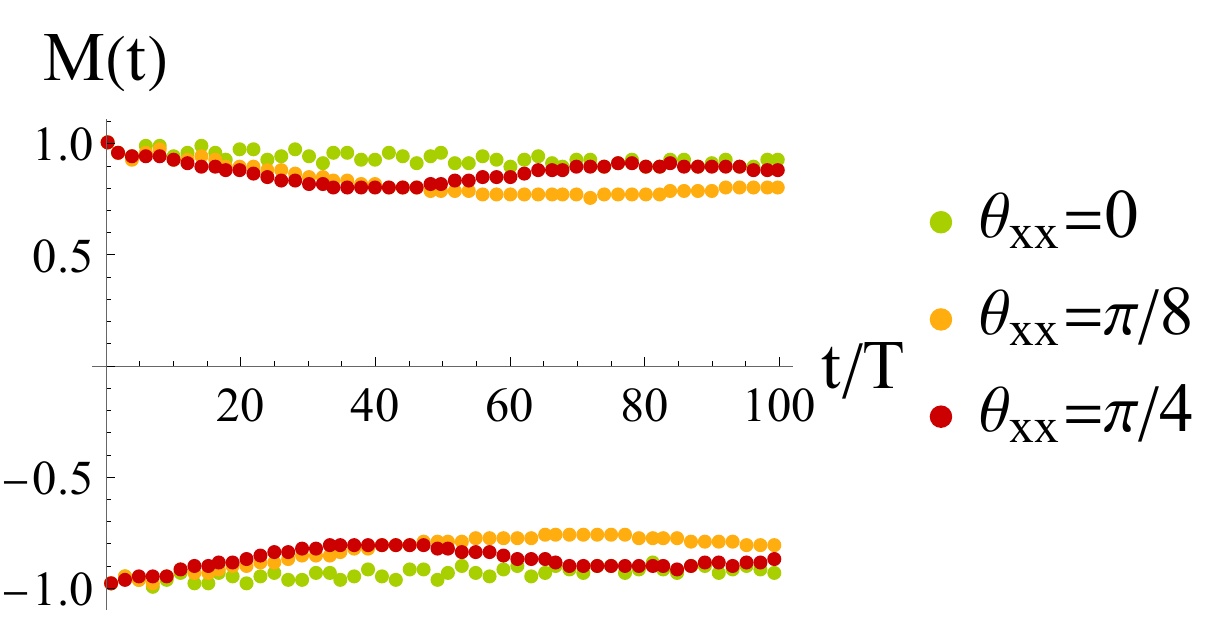}
			\\
			(b) $J=1, h_z=0.8$
		}
		\caption{\label{fig:preth} Effects of including more generic perturbation $e^{i\lambda \sum_{j=1}^L \tau^x_j} \rightarrow e^{i\lambda \sum_{j=1}^L \left( \cos(\theta_{xx}) \tau^x_j + \sin(\theta_{xx}) \tau^x_{j} \tau^x_{j+1} \right) } $, with total strength $\lambda$ unchanged. (a) The weakly interacting case can experience a slow relaxation, if the model exhibits fine-tuned integrability for the lowest-order effective Hamiltonians ($\theta_{xx}=0$ case). That delays the destruction of DTC-like oscillations at early time. Then, as we include more generic perturbation ($\theta_{xx}\ne0$), thermalization is significantly accelerated.  (b) In contrast, the strongly interacting system is already non-integrable for the effective Hamiltonian. DTCs induced by the cat scars do not rely on model fine-tuned integrability, and are robust against generic perturbation. Same results are observed in Fig.~\ref{fig:dynamics} where we sample over different perturbation types.  }
	\end{figure}

	In contrast, if we explicitly break the integrability by including more generic perturbations, thermalization processes would be significantly accelerated. For instance, in Fig.~\ref{fig:preth} (a), we change the form of perturbation in Eq.~\eqref{eq:simplemodel} into
	\begin{align}\nonumber
		U_1 &= e^{-i\left( \frac{\pi}{2} - \lambda \right) \sum_{j=1}^L \tau^x_j}
		\\
		&\rightarrow
		e^{-i \left( \frac{\pi}{2}  \sum_{j=1}^L \tau^x_j  
		+ \lambda \left( \cos(\theta_{xx}) \tau^x_j + \sin(\theta_{xx}) \tau^x_j \tau^x_{j+1}
		\right) \right)
		},
	\end{align}
	where $\theta_{xx} = 0$ returns the perturbation to its original form $ \sim\lambda \tau^x_j$, while for non-zero $\theta_{xx}$ additional two-spin perturbations $ \sim \tau^x_j \tau^x_{j+1}$ are included. We see in Fig.~\ref{fig:preth} (a) that the generic perturbation $\theta_{xx}\neq0$, with equal perturbation strength $\lambda$, indeed triggers a full thermalization behavior of the weakly interacting cases at early time. In contrast, the cat scar enforced DTC behaviors in strongly interacting cases [Fig.~\ref{fig:preth} (b)] is robust against the changes of perturbation forms, as we have already seen in Fig.~\ref{fig:dynamics}.

	Next, for completeness, let us also derive the effective Hamiltonian for early time dynamics in strongly interacting cases as well. 
	
\begin{figure}[h]
		\parbox{5cm}{
			\includegraphics[width=5cm]{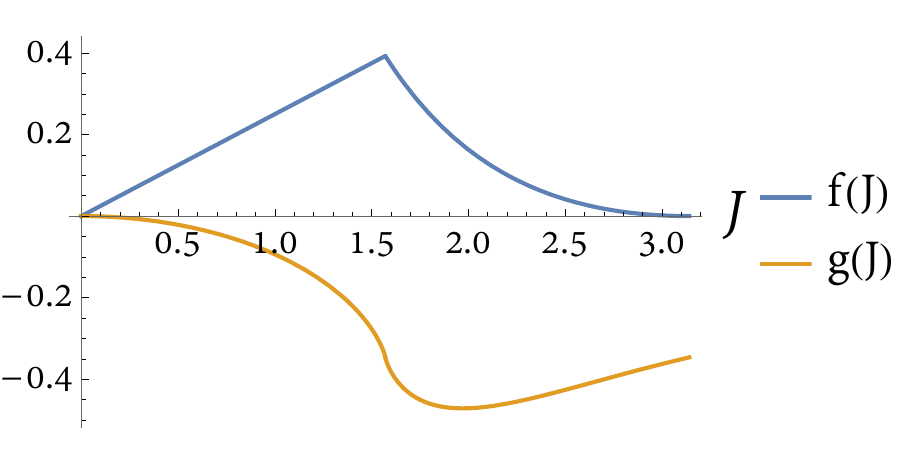}
		}
		\caption{\label{fig:fg} Illustration of the function magnitudes in Eq.~\eqref{eq:h2strong}.}
	\end{figure}

	{\bf (2)} For the {\bf strongly interacting} case $J\sim 1$ in Eq.~\eqref{eq:simplemodel}, however, the Hamiltonian $H^{(1)}$ cannot be obtained by simply summing over terms in the exponential parts, because a relatively strong interaction term $J\tau^z_j \tau^z_{j+1}$ with $J\sim1$ does not commute with other magnetic field terms. Nevertheless, we could still obtain a closed-form via BCHD formula in Appendix~\ref{smsec:hefflaregJ}, and arrive at the rigorous result
	\begin{align}\nonumber
		H_{\text{eff}} &= H^{(1)}|_{J\sim1} 
		+ O(\lambda^2), 
		\\ \nonumber
		H^{(1)}|_{J\sim 1} &= \sum_j \left(
		J\tau^z_j \tau^z_{j+1} - \lambda \left(\cos(h_z) \tau^x_j - \sin(h_z) \tau^y_j \right)  \right. \\ \nonumber
		& -
		\lambda f(J) (\cos(h_z) \tau^y_j + \sin(h_z) \tau^x_j) (\tau^z_{j-1} + \tau^z_{j+1})
		\\ \label{eq:h2strong}
		& \left. -
		\lambda g(J)
		(\cos(h_z) \tau^x_j - \sin(h_z) \tau^y_j) (1+\tau^z_{j-1}\tau^z_{j+1})  \right)
	\end{align}
	where the functions (which are plotted in Fig.~\ref{fig:fg})
	\begin{align}\nonumber
		f(J) &= \frac{1}{4J} \left( 
		\frac{\pi^2}{12} + \frac{\text{Li}_2(-e^{-2iJ}) + \text{Li}_2( -e^{2iJ})}{2}
		\right),
		\\
		g(J) &= \frac{1}{4J} \left( 
		-2J\ln(2) + \frac{\text{Li}_2(-e^{-2iJ}) - \text{Li}_2( -e^{2iJ})}{2i}
		\right),
	\end{align}
	and $ \text{Li}_s(z) $ is the polylogarithm function
	\begin{align}
		\text{Li}_s(z) = \sum_{k=1}^\infty \frac{z^k}{k^s}.
	\end{align}
	The terms $\sim f(J), g(J)$ in Eq.~\eqref{eq:h2strong} rigorously obtained here are new additions compared with results in previous literature for weakly interacting systems~\cite{Khemani2019b,Mi2022,Kyprianidis2021}. Due to the effects of these additional terms, the strongly interacting system already shows non-integrable nature at the lowest-order expansions. Thus, further adding two-spin terms in the perturbation, as in Fig.~\ref{fig:preth} (b), would not affect the dynamics because the cat scar enforced DTCs do not rely on fine-tuned integrability. 
	%We could also observe that these additional terms also obey the selection rules of flipping up to 1 spin. In particular, since $(\tau^z_{j-1} + \tau^z_{j+1})$ and $(1+\tau^z_{j-1} \tau^z_{j+1})$ are only nonzero if spins at sites $(j\pm1)$ are the same, these additional terms would take a Fock state to its nearby domain wall sector differing by $\delta w = \pm2$ total wall numbers.

	Finally, we should remark that while the effective Hamiltonian for weakly interacting system (Eq.~\eqref{eq:heffweak}) can be regarded as a prethermal static Hamiltonian, the strongly interacting one in Eq.~\eqref{eq:h2strong} cannot be viewed in the same way. To show it, let us consider the evolution at double-period ends for the transformed Floquet operator in Eq.~\eqref{eq:gaugetrans},
	\begin{align}\label{eq:hpreth}
		\tilde{U}_F^2 = e^{-iH^{(1)} (\tau^x_j, \tau^y_j, \tau^z_j) } 
		e^{-iH^{(1)} (\tau^x_j, - \tau^y_j, -\tau^z_j) } ,
	\end{align}
	where we used $P \tau^{y,z}_j P = \tau^x_j \tau^{y,z}_j \tau^x_j = -\tau^{y,z}_j$. For the weakly interacting case Eq.~\eqref{eq:heffweak}, all parameters in $H^{(2)}$ are small, and therefore we can simply sum up different terms and obtain
	\begin{align} \nonumber
		\tilde{U}_F^2 
		&= e^{-2i \sum_{j=1}^L \left( J\tau^z_j \tau^z_{j+1} - \lambda \tau^x_j \cos(h_z) \right) }  + O(\lambda^2, J^2, \lambda J) 
		\\
		&
		\equiv e^{-2iH_{\text{preth}}} + O(\lambda^2, J^2, \lambda J).
	\end{align}
	Thus, we can indeed arrive at a static Hamiltonian description with $H_{\text{preth}} = \sum_{j=1}^L \left( J\tau^z_j \tau^z_{j+1} - \lambda \tau^x_j \cos(h_z) \right) $, as further evolutions would be controlled by the same $e^{-2iH_{\text{preth}}}$ without drivings, until the time scale where higher-order terms $O(\lambda^2, J^2, \lambda J)$ show significant effects. 
	
	In contrast, for strongly interacting cases, the two terms $H^{(1)}(\tau^x_j, \tau^y_j, \tau^z_j), H^{(1)}(\tau^x_j, -\tau^y_j, -\tau^z_j) $ in the exponential of Eq.~\eqref{eq:hpreth}, where $H^{(1)}$'s are given by Eq.~\eqref{eq:h2strong}, both involve a large parameter $J$ and several two-spin terms that do not commute with each other. Then, it is no longer possible to obtain a closed-form local Hamiltonian to describe prethermal physics. Rather, in this case, we have a genuine strongly interacting system under periodic drivings, which typically absorbes energy and heats up quickly. This is consistent with the description in Fig.~\ref{fig:dw} (a), where the GOE type of level spacing statistics indicates the thermalizing behavior for non-scar eigenstates in each domain wall sector. Thus, in the strongly interacting case, we can no longer rely on a prethermal static description of the system with or without Landau's symmetry breaking. Instead, it is the unconventional scar localization enforced by strong interactions that leads to cat scar-induced DTC dynamics.

	\section{Conclusion\label{sec:con}}
	
	An analytical framework is constructed for SP of cat scars in clean Floquet systems, enabling systematic enumeration of their DTC oscillation patterns and  prediction of the scaling behaviors. For future works, symmetry indicators may yield more sophisticated phenomena in higher dimensions with richer space groups.  Also, our work paves the way to analytically bridging the two anchor points of clean DTCs illuminated here and the strongly disordered cases studied before. To extend the analysis in this work towards more disordered scenarios, one probably needs to carefully take into account possible resonances that may destabilize the localization of majority eigenstates, as revealed by recent studies on the avalanche mechanism. Finally, this work may help distinguish cat scar enforced DTC dynamics from other DTC-type of phenomena in translation-invariant systems, including the prethermal systems with Landau symmetry breaking similarly host robust SP, and also certain systems dominated by non-interacting effects instead.

	\begin{acknowledgments}	
		This work is supported by the National Natural Science	Foundation of China Grant No. 12174389.
	\end{acknowledgments}

	\appendix

	\section{Algebraic proof for selection rules regarding factored  perturbations of different orders}
	\label{smsec:selection}
	
	\subsection{Main proofs}
	Consider a Floquet operator made of two driving steps, 
	\begin{align}
		U_F(\lambda) = U_2(\lambda) U_1(\lambda) = e^{-i(H_2+\lambda H'_2)} e^{-i(H_1 + \lambda H'_1)}
	\end{align}
	Here we take generic perturbations up to two-spin terms,
	\begin{align}
		&
		H_{1}' = \sum_{j=1}^L \sum_{\mu=x,y,z} \theta_{\mu} \tau^\mu_j 
		+
		\sum_{j=1}^L \sum_{\mu,\nu = x,y,z} \phi_{\mu\nu} \tau^\mu_j \tau^\nu_{j+1}
		,
		\\ \label{smeq:hp12}
		&
		H_{2}' = \sum_{j=1}^L \sum_{\mu=x,y,z} \theta'_\mu \tau^\mu_j 
		+
		\sum_{j=1}^L \sum_{\mu,\nu = x,y,z} \phi'_{\mu\nu} \tau^\mu_j \tau^\nu_{j+1}
	\end{align}
	The model in Eq.~\eqref{eq:model} serves as a specific example with $ H_1'=H_2' $ and $ \phi_{\mu\nu}=\phi\delta_{\mu x} \delta_{\nu x} $. We define the operator product order $ n_{\text{op}} $ by counting at most how many spin operators are multiplied in individual terms of perturbation Hamiltonians, i.e.
	\begin{align}\nonumber
		n_{\text{op}} = 2:
		\quad & \quad
		\theta_\mu, \theta'_\mu, \phi_{\mu\nu}, \phi'_{\mu\nu} \ne 0, \quad 
		\sim \tau^\mu_j, \tau^\mu_j \tau^\nu_{j+1}
		\\
		\label{smeq:nop}
		n_{\text{op}} = 1:
		\quad & \quad
		\theta_\mu, \theta'_\mu \ne 0, \,\, \phi_{\mu\nu} = \phi'_{\mu\nu} = 0, 
		\quad
		\text{only } \sim \tau^\mu_j
	\end{align}
	Namely, with both one-spin and two-spin terms $ n_{\text{op}} = 2 $, and if only one-spin terms are present $ n_{\text{op}}=1 $. The selection rules would be intimately related to the quantity $ n_{\text{op}} $.
	
	To facilitate further analysis, we would like to factor the perturbation into the form $ U_F(\lambda) = U_0 U'(\lambda) $, where $ U_0 = e^{-iH_2} e^{-iH_1} $ does not involve perturbations and therefore can be solved exactly. We first formally factor out the perturbation by inserting an identity matrix,
	\begin{align}\nonumber
		&U_F(\lambda) = \left( e^{-iH_2} e^{-iH_1} \right)
		\left(
		e^{iH_1} e^{iH_2} U_F 
		\right) \equiv U_0 U', 
		\\
		& U' = e^{iH_1} e^{iH_2} e^{-i(H_2 + \lambda H'_2)} e^{-i(H_1 + \lambda H'_1)}.
	\end{align}
	Now, recall that the unperturbed Hamiltonians $ H_0(t) $ (Eq.~\eqref{eq:model}) at two driving steps read
	\begin{align} \nonumber
		H_1 = \frac{\pi}{2} \sum_j \tau^x_j, &\qquad 
		e^{-iH_1} = (-i)^L \prod_j \tau^x_j,\\
		\label{smeq:h12}
		H_2 = \sum_j \tau^z_j \tau^z_{j+1} , & \qquad
		e^{-iH_2} = \prod_j e^{-i\tau^z_j \tau^z_{j+1}}.
	\end{align}
	Then, $ e^{iH_1} (\tau^x_j, \tau^y_j, \tau^z_j) e^{-iH_1} = (\tau^x_j, -\tau^y_j, -\tau^z_j) $ gives
	\begin{align}\label{smeq:hpp}
		U' = \left( 
		e^{iH_2} e^{-i(H_2 + \lambda H_2'(\tau^x_j, -\tau^y_j, -\tau^z_j))} 
		\right)
		\left(
		e^{iH_1} e^{-i(H_1 + \lambda H_1')} 
		\right) 
		\equiv U_2' U_1'
	\end{align}
	Since $ H_{1,2}' $ are already taken as generic perturbations, we would neglect the sign flips for $ H_2'(\tau^x_j, -\tau^y_j, -\tau^z_j) $, and denote
	\begin{align}\label{smeq:ualpha}
		U'=U_2'U_1', 
		\qquad
		U_\alpha' = e^{iH_\alpha} e^{-i(H_\alpha + \lambda H'_\alpha)},
		\qquad
		\alpha = 1,2,
	\end{align}
	where $ H_{1,2} $ and $ H_{1,2}' $ are given by Eqs.~\eqref{smeq:h12} and \eqref{smeq:hp12} respectively.
	
	For the formal decomposition to be useful, we would like to sort out the structures of $ U'_\alpha $, and to prove a selection rule that paves the way for showing scaling relations later on. To do so, let us apply the Baker-Campbell-Hausdorff-Dynkin (BCHD) formula~\cite{Serre1992} to Eq.~\eqref{smeq:ualpha},
	\begin{widetext}

		\begin{align}\label{smeq:bchd}
			U_\alpha' &= \exp\left(
			\sum_{n=1}^\infty \frac{(-1)^{n+1}}{n}
			\sum_{p_j + q_j >0,\,\, p_j, q_j\ge0}
			\frac{
				[(iH_\alpha)^{(p_1)},[ (-i(H_\alpha + \lambda H_\alpha')^{(q_1)}), [\dots, [(iH_\alpha)^{(p_n)}, (-i(H_\alpha + \lambda H_\alpha')^{(q_n)}]\dots]
			}{\left(\sum_{j=1}^n (p_j + q_j)\right) \prod_{j=1}^n p_j! q_j!}
			\right),
		\end{align}
	\end{widetext}
	where the iterative brackets means, i.e. $ [A^{(2)}, B^{(3)}] = [A, [A, [B,[B,B ]]]] $, and in our case the non-vanishing terms in the exponential should satisfy $ q_n \le 1 $ and that not all $ q_j=0 $. 
	
	As a preliminary step, we confirm below that based on the form of Eq.~\eqref{smeq:bchd}, $ U_\alpha' $ can be written as
	\begin{align}\label{smeq:ualphavalpha}
		U'_\alpha(\lambda) = \exp\left(i\sum_{k=1}^\infty \lambda^k V_{\alpha,k} \right),
	\end{align}
	where $ V_{\alpha,k}^\dagger = V_{\alpha,k} $ are Hermitian operators. 
	
	First, note that the lowest order $ n=1 $ terms in the exponential
	\begin{align}\nonumber
		&iH_\alpha - i(H_\alpha + \lambda H'_\alpha) + \sum_{r_1\ge 1} 
		\frac{i^{p_1-1}}{(p_1+1) p_1!}
		\lambda[H_\alpha^{(p_1)}, H'_\alpha ]
		\\ &=
		-i\lambda (H_\alpha' + \sum_{p_1\ge1} \frac{i^{p_1}}{(p_1+1) p_1!} [H_\alpha^{(p_1)}, H'_\alpha]),
	\end{align}
	is of the order $ \lambda^1 $. Other higher-order terms would necessarily involve at least one commutator of $ [H_\alpha, \lambda H_\alpha'] $ in order to be non-vanishing. Thus, the perturbation $ U'(\lambda) $ indeed start from the $ \lambda^1 $ term. 
	
	Second, note that for a set of arbitrary Hermitian operators $ A_j^\dagger = A_j $, by repeatedly using $ [A_1, A_2]^\dagger = [A_2^\dagger, A_1^\dagger] = [A_2,A_1] = -[A_1,A_2] $, we have
	\begin{align}\nonumber
		&
		[A_1,[A_2,\dots [A_{n-1},A_n]]]^\dagger 
		\\ \nonumber
		&= [[[A_n, A_{n-1}], A_{n-2}], \dots,A_2], A_1] 
		\\ \nonumber
		&= 
		(-1)^{n-1} [A_1,[A_2,\dots,[A_{n-1},A_n]]]
		\\ \nonumber
		\Rightarrow
		\quad
		&
		[iA_1, [iA_2,\dots [iA_{n-1},iA_n]]]^\dagger 
		\\ \nonumber
		&= (i^n)^\dagger [A_1,[A_2,\dots [A_{n-1},A_n]]]^\dagger
		\\ &= (-1) [iA_1,[iA_2,\dots [iA_{n-1},iA_n]]].
	\end{align}
	Replacing $ A_j $'s with $ H_\alpha, H_\alpha' $ in Eq.~\eqref{smeq:bchd}, we see that each term in the exponential with given $ p_j, q_j $ is anti-Hermitian. Thus, operators $ V_{\alpha,k} $ in Eq.~\eqref{smeq:ualphavalpha} are Hermitian ones ($ (iV_{\alpha,k}) $'s are anti-Hermitian).

	Thus, the form of perturbations in Eq.~\eqref{smeq:ualphavalpha} is confirmed. Next we further prove the selection rule for Hermitian operators $ V_{1,k} $ and $ V_{2,k} $ separately using the form in Eqs.~\eqref{smeq:bchd}. To do so, let us define the Fock space distance $ \delta s $ by counting how many spins are flipped between two Fock configurations $ |\{s_j\}\rangle $ and $ |\{\tilde{s}_j\}'\rangle $,
	\begin{align}\label{smeq:deltas}
		\delta s(\{s_j\}, \{\tilde{s}_j\}') = \frac{1}{2}\sum_j |s_j - \tilde{s}_j|
	\end{align}
	The selection rule relates the perturbation orders $ \lambda^kV_{\alpha,k} $ with the maximal Fock space distance $ \delta s(\{s_j\}, \{\tilde{s}_j\}') $ for non-vanishing matrix elements $ \langle \{s_j\}| V_{\alpha, k} |\{\tilde{s}_j\}' \rangle $.
	
	{\bf\color{blue}(1) For $ U_1' $}, recall that $ H_1 = \frac{\pi}{2}\sum_j \tau^x_j $ only involves single spin terms. That means the commutation of $ H_1 $ with any product $ \tau^{\mu_1}_{j_1} \tau^{\mu_2}_{j_2} \dots \tau^{\mu_n}_{j_n} $ preserve the total number of spin operators being multiplied, i.e.
	\begin{align}\nonumber
		&[H_1^{(n)}, \tau^{\nu_1}_{j_1} \tau^{\nu_2}_{j_2} \dots \tau^{\nu_n}_{j_n}] = \sum_{\mu_1\dots \mu_n} A^{(n)}_{\mu_1\dots \mu_n} \tau^{\mu_1}_{j_1} \dots \tau^{\mu_n}_{j_n},
		\\
		&A^{(n)}_{\mu_1\dots\mu_n} \in \mathbb{C},\quad,
		\nu_k, \mu_k = x,y,z
	\end{align}
	where the RHS also only involves products of $ n $ operators. Thus, it solely depends on the number of $ H'_1 $, i.e. $ q_j $ in Eq.~\eqref{smeq:bchd} to determine the how many operators are multiplied in each term. Specifically, for the $ \lambda^k $-th order terms, there are $ k $ pieces of $ H_1' $ in the commutators of Eq.~\eqref{smeq:bchd}. Since $ H_1' $ involves at most multiplications of $ n_{\text{op}} $ spin operators in Eq.~\eqref{smeq:hp12}, there are at most $ n_{\text{op}}k $ spin operators being multiplied together for any terms in $ V_{1,k} $. Therefore, the selection rule for $ V_{1,k} $ is
	\begin{align}\label{smeq:selection1}
		\langle \{s_j\}_1 | V_{1,k} | \{s_j\}_2 \rangle \ne 0 
		\quad
		\Rightarrow
		\quad 
		\delta s(\{s_j\}_1, \{s_j\}_2) \leqslant n_{\text{op}}k
	\end{align}

	{\bf\color{blue} (2) For $ U_2' $}, note that the commutation of $ H_2 = \sum_j J_j \tau^z_j \tau^z_{j+1} $ with any operator product only exchanges $ \tau^x_j \leftrightarrow \tau^y_j $, and/or attached additional factors of $ \tau^z_j $'s in an operator product. Therefore, it cannot increase or decrease the number of spin-flipping operators $ \tau^{x,y}_j $ being multiplied in a term. For instance, $ H_2 $ commuting with one and two spin terms gives
	\begin{align}\nonumber
		&
		[H_2, \tau^x_{\color{red}j}] = iJ \tau^y_{\color{red}j}(\tau^z_{j+1} + \tau^z_{j-1}), 
		\quad
		[H_2, \tau^y_{\color{red}j}] = -iJ\tau^x_{\color{red}j} (\tau^z_{j+1} + \tau^z_{j-1})
		\\
		\label{smeq:u2p}
		&
		[H_2, \tau^x_{\color{red} j} \tau^x_{\color{red}j+1}] = 
		J
		(i\tau^z_{j-1} \tau^y_{\color{red}j} \tau^x_{\color{red}j+1}
		-
		\tau^y_{\color{red}j} \tau^y_{\color{red}j+1} 
		+ i\tau^x_{\color{red}j} \tau^y_{\color{red}j+1} \tau^z_{j+2}
		)
	\end{align}
	More generally, for an arbitrary term with $ n $ spin flipping operators, we have 
	\begin{align}\nonumber
		& \,
		[H_2^{(n)}, (\tau^{\nu_1}_{j_1} \tau^{\nu_2}_{j_2}\dots \tau^{\nu_n}_{j_n}) (\tau^z_{k_1} \tau^z_{k_2} \dots \tau^z_{k_m})] 
		\\ \nonumber
		&= 
		[H_2^{(n)}, \tau^{\nu_1}_{j_1} \tau^{\nu_2}_{j_2}\dots \tau^{\nu_n}_{j_n} ](\tau^z_{k_1} \tau^z_{k_2} \dots \tau^z_{k_m})
		\\ \label{smeq:temp1}
		=&
		\sum_{\mu_1\mu_2\dots\mu_n = x,y} 
		\tau^{\mu_1}_{j_1} \tau^{\mu_2}_{j_2}\dots \tau^{\mu_n}_{j_n}
		\\
		&
		\times \left(
		\sum_{m_{n+1} \dots m_L = 0,1} B^{(n)}_{\mu_1\dots\mu_n; m_{n+1},\dots, m_L} 
		(\tau^z_{j_{n+1}})^{m_{n+1}}
		(\tau^z_{j_{n+2}})^{m_{n+2}}
		\dots
		(\tau^z_{j_{L}})^{m_{L}}
		\right),
	\end{align}
	Here transverse components are labeled by $ \nu_1,\dots,\nu_n, \mu_1,\dots,\mu_n = x,y $, while longitudinal ones $ \tau^z_j $ are explicitly denoted. Coefficients are generally denoted as $ B^{(n)}_{\mu_1\dots \mu_n; m_{n+1},m_{n+2}, m_L} \in\mathbb{C} $. Now, although the total number of operators changes, the number of {\em spin-flipping} operators, i.e. $ \tau^{\nu_1}_{j_1}\dots \tau^{\nu_n}_{j_n} $ and $ \tau^{\mu_1}_{j_1}\dots \tau^{\mu_n}_{j_n} $, remain the same on both sides of Eq.~\eqref{smeq:temp1}.
	That means, again, $ V_{2,k} $ can flip as many spins as the $ k $ pieces of $ H_2' $ in the commutators of Eq.~\eqref{smeq:bchd}, rendering the selection rule 
	\begin{align}\label{smeq:selection2}
		\langle \{s_j\}_1 | V_{2,k} | \{s_j\}_2 \rangle \ne 0 
		\quad\Rightarrow\quad 
		\delta s(\{s_j\}_1, \{s_j\}_2) \leqslant n_{\text{op}}k.
	\end{align}

	Thus, the selection rules for $ V_{1,k} $ and $ V_{2,k} $ are the same as given by Eqs.~\eqref{smeq:selection1} and \eqref{smeq:selection2}.

	Finally, let us consider the total perturbations by using the BCHD formula again,
	\begin{widetext}

		\begin{align}\nonumber
			& U'(\lambda) = U_2' U_1' =
			e^{i\sum_{k=1}^\infty \lambda^k V_{1,k}} e^{i\sum_{k'=1}^\infty \lambda^{k'} V_{1,k'}}
			\\ \nonumber
			=\,&
			\exp\left(
			\sum_{n=1}^\infty \frac{(-1)^{n+1}}{n}
			\sum_{p_j + q_j >0,\,\, p_j, q_j\ge0}
			\frac{
				[(i\sum_{k_1}\lambda^{k_1} V_{1,k_1})^{(p_1)},
				[(i\sum_{k_1'}\lambda^{k_1'} V_{2,k_1})^{(q_1)}, 
				[\dots, 
				[(i\sum_{k_n}\lambda^{k_n} V_{1,k_n})^{(p_n)}, 
				(i\sum_{k_n'}\lambda^{k_n'} V_{2,k_n'})^{(p_1)}]\dots]
			}{\left(\sum_{j=1}^n (p_j + q_j)\right) \prod_{j=1}^n p_j! q_j!}
			\right),
			\\
			=\,& e^{i \sum_{k=1}^\infty \lambda^k V_k},
		\end{align}
	\end{widetext}
	where in the last step we use the same analysis as that for Eq.~\eqref{smeq:ualphavalpha} to obtain the exponential form and the Hermitian condition $ V_k^\dagger = V_k $. Then, a simple power counting gives that the $ \lambda^k $ terms involve commutations of
	\begin{align}
		&
		V_k \sim [V_{1,k_1}^{(p_1)}, [V_{2,k'_1}^{(q_1)}, [\dots [V_{1,k_n}^{(p_n)}, V_{2,k'_n}^{(q_n)}]\dots ],\quad
		\sum_{j=1}^n (k_j p_j + k'_jq_j) = k
	\end{align}
	Due to the selection rules Eqs.~\eqref{smeq:selection1} and \eqref{smeq:selection2}, $ V_{1,k_j} $ and $ V_{2,k_j'} $ at most involve multiplications of $ n_{\text{op}}k_j, n_{\text{op}}k'_j $ spin-flipping operators respectively, so $ V_k $ at most involves multiplications of $ \sum_{j=1}^n n_{\text{op}}(k_jp_j + k'_j q_j) = n_{\text{op}} k $ spin-flipping operators. Thus, we have the final form and selection rules for the full perturbation operator, 
	\begin{align}\nonumber
		&U_F(\lambda) = e^{-i(H_1 + \lambda H_1')} e^{-i(H_2+\lambda H_2')} = U_0U'(\lambda), 
		\\ \nonumber 
		&U_0\equiv U_F(\lambda=0) = e^{-iH_2} e^{-iH_1},
		\quad 
		U'(\lambda) = \exp\left(
		i\sum_{k=1}^\infty \lambda^k V_k
		\right),
		\\ \label{smeq:selection full}
		&
		\langle \{s_j\}_1 | V_{k} | \{s_j\}_2 \rangle \ne 0 
		\quad\Rightarrow\quad
		\delta s(\{s_j\}_1, \{s_j\}_2) \leqslant n_{\text{op}}k.
	\end{align}
	This is satisfied by $ H_1, H_2 $ in Eq.~\eqref{smeq:h12} under generic perturbations in Eq.~\eqref{smeq:hp12}.

	To give an example, let us write down the generic form for the first-order perturbation $ V_1 \sim \lambda^1 $ for the model in Eq.~\eqref{eq:model} $ H_1'=H_2'=H'=\sum_j (\phi \tau^x_j \tau^x_{j+1} + \sum_{\mu=x,y,z} \theta_\mu \tau^\mu_j) $. From Eq.~\eqref{smeq:bchd}, we see that $ V_1 $ is given by the commutation of $ H' $ with multiple $ H_1 $ and $ H_2 $, namely,
	\begin{align}
		V_1 = f_1(\{[H_1^{(n)}, H']| 
		0\le n\in \mathbb{Z}\}) +  f_2(\{[H_2^{(n)},H']| 
		0\le n\in \mathbb{Z}\}).
	\end{align}
	For the $ H_1 $ part, only single particle terms are involved, and we could easily obtain
	\begin{align}\nonumber
		[H_1^{(n)}, H'] &= (\frac{\pi}{2})^n [(\sum_j \tau^x_j)^{(n)}, \sum_k (\phi \tau^x_k \tau^x_{k+1} + \sum_{\mu=x,y,z}\theta_\mu \tau^\mu_{k}) ] 
		\\ \nonumber &=
		(\frac{\pi}{2})^n \sum_k [( \tau^x_k)^{(n)}, (\theta_y \tau^y_k + \theta_z \tau^z_k)]
		\\ \label{smeq:h1commute}
		&=
		\begin{cases}
			(\frac{\pi}{2})^{n} \sum_{k=1}^L i(\theta_y\tau^z_k - \theta_z \tau^x_k ), & \text{odd $ n $}
			\\
			(\frac{\pi}{2})^{n} \sum_{k=1}^L (\theta_y \tau^y_k + \theta_z \tau^z_k), 
			& \text{even $ n $}
		\end{cases},
	\end{align}
	On the other hand, the iterative commutation of $ H_2 $ with $ H' $ is a bit more complicated. Full details of computations are left to Sec.~\ref{smsec:h2commute}. From Eqs.~\eqref{smeq:h2commute1} and \eqref{smeq:h2commute2} there, combined with Eqs.~\eqref{smeq:h1commute}, we have the generic form
	\begin{align}\nonumber
		V_1 &= \sum_{j=1}^L \sum_{m_1,m_2=0,1} \left(
		\sum_{\mu_1 = x,y} \alpha_{m_1m_2,\mu_1} (\tau^z_{j-1})^{m_1} \tau^{\mu_1}_{j} (\tau^z_{j+1})^{m_2} \right. 
		\\  \label{smeq:v1form}
		&
		\qquad\qquad+
		\left. \sum_{\mu_1\mu_2=x,y}
		\beta_{m_1m_2,\mu_1\mu_2} (\tau^z_{j-1})^{m_1} \tau^{\mu_1}_{j} \tau^{\mu_2}_{j+1} (\tau^z_{j+2})^{m_2}
		\right),
	\end{align}
	where $ \alpha_{m_1m_2,\mu_1}^{(n)}, \beta^{(n)}_{m_1m_2,\mu_1\mu_2} \in \mathbb{C} $ are coefficients. We see that the first order factored perturbation $ V_1 $ takes a similar form as the bare perturbation $ H' $, with the only difference that certain $ \tau^z_j $'s are attached to nearby sites and the coefficients are modified. Most importantly, the number of spin-flipping operators, $ \tau^{x,y}_j $'s, are limited to 2 in each term, and therefore fulfilling the selection rules Eqs.~\eqref{smeq:selection full}.

	For later use, it is also helpful to note that a gauge transformation by the unperturbed Floquet operator $ U_0 $ in Eq.~\eqref{eq:u0} leaves the selection rules for $ V_k $ in Eq.~\eqref{smeq:selection full} unchanged, namely,
	\begin{align}\nonumber
		&\text{if } \langle \{s_j\}| V_k |\{\tilde{s}_j\}\rangle \ne0 
		\quad \Rightarrow\quad
		\delta s(\{s_j\}, \{\tilde{s}_j\}') \leqslant n_{\text{op}} k.
		\\ \nonumber
		&
		\text{then for }
		\tilde{V}_k \equiv U_0^\dagger V_k U_0, 
		\\  \label{smeq:tildevk}
		&
		\langle \{s_j\} | \tilde{V}_k | \{\tilde{s}_j\}' \rangle \ne 0 
		\quad \Rightarrow\quad
		\delta s(\{s_j\}, \{\tilde{s}_j\}') \leqslant n_{\text{op}} k.
	\end{align}
	Specifically, for individual spin operators, the gauge transformation performs a local linear mapping,
	\begin{widetext}

		\begin{align}\nonumber
			&U_0^\dagger  
			\begin{pmatrix}
				\tau^x_j \\ \tau^y_j \\ \tau^z_j
			\end{pmatrix}
			U_0 =
			(\cos J + i\sin J \tau^z_j \tau^z_{j+1})
			(\cos J + i\sin J \tau^z_j \tau^z_{j-1})
			\begin{pmatrix}
				\tau^x_j \\ -\tau^y_j \\ -\tau^z_j
			\end{pmatrix} 
			(\cos J - i\sin J \tau^z_j \tau^z_{j-1})
			(\cos J - i\sin J \tau^z_j \tau^z_{j+1})
			\\
			\nonumber
			&=
			\begin{pmatrix}
				\tau^x_j (\cos^2 2J - \tau^z_{j-1} \tau^z_{j+1}\sin^2 2J )
				- 
				\tau^y_j \cos 2J \sin 2J (\tau^z_{j-1} + \tau^z_{j+1})
				\\
				-\tau^x_j \cos 2J \sin 2J (\tau^z_{j-1} + \tau^z_{j+1}) 
				-
				\tau^y_j  (\cos^2 2J - \tau^z_{j-1} \tau^z_{j+1}\sin^2 2J )
				\\
				- \tau^z_j
			\end{pmatrix}
			\equiv K_j
			\begin{pmatrix}
				\tau^x_j \\ \tau^y_j \\ \tau^z_j
			\end{pmatrix}
			\\
			K_j &=
			\begin{pmatrix}
				\cos^22J - \sin^2 2J \tau^z_{j-1} \tau^z_{j+1} 
				&
				-\sin2J \cos2J (\tau^z_{j-1} + \tau^z_{j+1}) 
				& 
				0
				\\
				-\sin2J \cos2J (\tau^z_{j-1} + \tau^z_{j+1}) 
				&
				-(\cos^22J - \sin^2 2J \tau^z_{j-1} \tau^z_{j+1} )
				& 0
				\\
				0 & 0 & -1
			\end{pmatrix}.
		\end{align}
	\end{widetext}
	The matrix $ K_j $ is block-diagonalized as $ \tau^{x,y}_j $ and $ \tau^z_j $ are decoupled. Thus, the gauge transformation at a certain site $ j $ is to exchange $ \tau^x_j \leftrightarrow \tau^y_j $, and attach additional $ \tau^z_{j\pm1} $ at nearby sites. Further note that multiplications like $ \tau^z_{j-1} (\tau^x_{j-1}, \tau^y_{j-1}, \tau^z_{j-1}) = (i\tau^y_{j-1}, -i\tau^x_{j-1}, 1) $ again only exchanges $ \tau^x_{j-1}\leftrightarrow \tau^y_{j-1} $. We see that for a generic term, the gauge transformation cannot increase or decrease the number of spin-flipping operators $ \tau^{x,y}_j $, namely,
	\begin{align}\nonumber
		& \,
		U_0^\dagger (\tau^{\nu_1}_{j_1} \tau^{\nu_2}_{j_2}\dots \tau^{\nu_n}_{j_n}) (\tau^z_{k_1} \tau^z_{k_2} \dots \tau^z_{k_m})  
		U_0
		\\ \nonumber
		&=
		\sum_{\mu_1\mu_2\dots\mu_n = x,y} 
		\tau^{\mu_1}_{j_1} \tau^{\mu_2}_{j_2}\dots \tau^{\mu_n}_{j_n}
		\sum_{m_{n+1} \dots m_L = 0,1} 
		B_{\mu_1\dots\mu_n}^{m_{n+1},\dots, m_L} 
		\\
		&\qquad\times 
		(\tau^z_{j_{n+1}})^{m_{n+1}}
		(\tau^z_{j_{n+2}})^{m_{n+2}}
		\dots
		(\tau^z_{j_{L}})^{m_{L}},
	\end{align}
	where the number of spin flipping operator $ \nu_1\dots\nu_n, \mu_1\dots\mu_n = x,y $ is preserved on both sides as $ n $. Combined with the selection rules for the bare $ V_k $ in Eq.~\eqref{smeq:selection full}, we have proved the relations in Eq.~\eqref{smeq:tildevk}.

	\subsection{Algebras for proving Eq.~\eqref{smeq:v1form}: Iterative commutation of $ H_2 $ with generic one-spin and two-spin terms \label{smsec:h2commute}}
	
	Here the zeroth-order Hamiltonian reads
	\begin{align}\nonumber
		H_2 = J\sum_k \tau^z_k \tau^z_{k+1}.
	\end{align}
	We would consider repeated commutations between $ H_2 $ with one-spin and two-spin terms $ \tau^\mu_j, \tau^\mu_j \tau^\nu_j $. Independent ones are written in the following form
	\begin{align}
		A_j = \begin{pmatrix}
			\tau^x_j\\ \tau^y_j 
		\end{pmatrix},
		\qquad
		B_{j,j+1} = 
		\begin{pmatrix}
			\tau^x_j \tau^x_{j+1}\\
			\tau^x_j \tau^y_{j+1}\\
			\tau^y_j \tau^x_{j+1}\\
			\tau^y_j \tau^y_{j+1}
		\end{pmatrix},
	\end{align}
	while others like $ \tau^z_j\tau^x_{j+1} $ can be obtained from results for $ A_j $, as $ \tau^z_j $ commute with $ H_2 $. 
	In the following, commutations like $ [H_2,A_j] $ mean to commute $ H_2 $ with each component of $ A_j $. Since $ H_2 $ only exchanges $ \tau^x_j \leftrightarrow \tau^y_j $, its commutation with $ A_j, B_j $ amounts to a linear transformation within the subspace of $ A_j, B_{j,j+1} $, namely,
	\begin{align}
		[H_2^{(n)}, A_j] = K_j^{n} A_j,
		\qquad
		[H_2^{(n)}, B_{j,j+1}] = L^n_j B_{j,j+1},
	\end{align}
	where $ K^n_j, L^n_j $ are $ 2\times2 $ and $ 4\times 4 $ matrices respectively. We shall obtain the explicit formula for all $ K^n_j, L_j^n $. Below, we would use $ \tau^{x,y,z}_j $ to denote the spin operators. Instead, $ \sigma_{x,y,z,0}, s_{x,y,z,0} $ are just Pauli matrices to denote coefficients.
	
	{\bf (1) One-spin terms}
	\begin{align}\nonumber
		&
		[H_2, A_j] = 2J
		\begin{pmatrix}
			i\tau^y_{j} (\tau^z_{j-1} + \tau^z_{j+1})\\
			-i\tau^x_j (\tau^z_{j-1} + \tau^z_{j+1}) 
		\end{pmatrix}
		\\ \nonumber
		&=2J
		\begin{pmatrix}
			0 & i(\tau^z_{j-1} + \tau^z_{j+1}) \\
			-i(\tau^z_{j-1} + \tau^z_{j+1}) & 0
		\end{pmatrix}
		\begin{pmatrix}
			\tau^x_j \\ \tau^y_j
		\end{pmatrix}
		\equiv K_j^1A_j,
		\\
		&
		\Rightarrow
		\qquad
		K_j^1 = -2J(\tau^z_{j-1} + \tau^z_{j+1}) \sigma_y.
	\end{align}
	Here spin operators $ \tau^z_{j\pm1} $'s commute with all other operators. Further, note that all operators in $ K^1_j $ commute with $ H_2 $, repeated commutations of $ H_2 $ with $ A_j $ amounts to repeated linear transformation
	\begin{align}\nonumber
		&
		[H_2^{(n)}, A_j] = (K_j^1)^n A_j = K^n_j A_j,
		\\
		&
		K^n_j = (K_j^1)^n =
		\begin{cases}
			(-2J)^n 2^{n-1} (1+\tau^z_{j-1} \tau^z_{j+1}) \sigma_0 ,
			&
			\text{even $ n\ge 2 $}
			\\
			(-2J)^n 2^{n-1} (\tau^z_{j-1} + \tau^z_{j+1}) \sigma_y, & \text{odd $ n $}
		\end{cases}.
	\end{align}
	Thus, explicitly, for $ 1\le m \in \mathbb{Z} $,
	\begin{align}\nonumber
		&
		[H_2^{(2m)}, \tau^x_j] = (1,0)K_j^{2m}A_j =  J^{2m}2^{4m-1} (1+\tau^z_{j-1}\tau^z_{j+1}) \tau^x_j,
		\\ \nonumber
		&
		[H_2^{(2m)}, \tau^y_j] = (0,1)K_j^{2m}A_j =  J^{2m}2^{4m-1} (1+\tau^z_{j-1}\tau^z_{j+1}) \tau^y_j
		\\ \nonumber
		&[H_2^{(2m-1)}, \tau^x_j] = (1,0)K_j^{2m-1}A_j=  -J^{2m-1}2^{4m-2} (\tau^z_{j-1} + \tau^z_{j+1}) (-i\tau^y_j),
		\\
		\label{smeq:onespinterms}
		&[H_2^{(2m-1)}, \tau^y_j] = (0,1)K_j^{2m-1}A_j =  -J^{2m-1}2^{4m-2} (\tau^z_{j-1} + \tau^z_{j+1}) (i\tau^x_j)
	\end{align}
	Thus, commutation of $ H_2 $ with one-spin terms can be summarized into the generic form
	\begin{align}\label{smeq:h2commute1}
		[H_2^{(n)}, \tau^{\mu}_j] = \sum_{m_1,m_2=0,1} \sum_{\mu_1 = x,y} \alpha_{m_1m_2,\mu_1}^{(n)} (\tau^z_{j-1})^{m_1} \tau^{\mu_1}_{j} (\tau^z_{j+1})^{m_2}
	\end{align}
	with coefficients $ \alpha_{m_1m_2,\mu_1}^{(n)} $ as in Eq.~\eqref{smeq:onespinterms}.
	
	{\bf(2) Two-spin terms}
	\begin{align}\nonumber
		&
		[H_2, B_{j,j+1}] = 2J
		\begin{pmatrix}
			\tau^z_{j-1} i\tau^y_j \tau^x_{j+1} - \tau^y_j\tau^y_{j+1} + \tau^x_j i\tau^y_{j+1} \tau^z_{j+2} 
			\\
			\tau^z_{j-1} i\tau^y_j \tau^y_{j+1} - \tau^y_j\tau^y_{j+1} + \tau^x_j i\tau^y_{j+1} \tau^z_{j+2} 
			\\
			-\tau^z_{j-1} i\tau^x_j \tau^x_{j+1} + \tau^x_j\tau^y_{j+1} + \tau^y_j i\tau^y_{j+1} \tau^z_{j+2} 
			\\
			-\tau^z_{j-1} i\tau^x_j \tau^y_{j+1} - \tau^x_j\tau^x_{j+1} - \tau^y_j i\tau^x_{j+1} \tau^z_{j+2} 
		\end{pmatrix}
		\\
		\nonumber
		& \qquad\qquad
		\equiv L_j^1 B_{j,j+1},
		\\ \nonumber
		&
		L_j^1 = 2J
		\begin{pmatrix}
			0 & i\tau^z_{j+2} & i\tau^z_{j-1} & -1
			\\
			-i\tau^z_{j+2} & 0 & 1 & i\tau^z_{j-1}
			\\
			-i\tau^z_{j-1} & 1 & 0 & i\tau^z_{j+2}
			\\
			-1 & -i\tau^z_{j-1} & -i\tau^z_{j+2} & 0
		\end{pmatrix}
		\\
		&\quad = 2J(\sigma_y s_y - \tau^z_{j-1} \sigma_y s_0 - \tau^z_{j+2} \sigma_0 s_y).
	\end{align}
	where we verified again that commutation of $ H_2 $ with $ B_{j,j+1} $ only performs linear transformation within the 4-dimensional subspace. Since $ L_j^1 $ commute with $ H_2 $, repeated commutation of $ H_2 $ with $ B_{j,j+1} $ reduces to a multiplication
	\begin{align}
		[H_2^{(n)}, B_{j,j+1}] = (L_j^1)^n B_{j,j+1}  = L_j^n B_{j,j+1},
		\qquad
		L_j^n = (L_j^1)^n.
	\end{align}
	Computing $ L_j^n $ is slightly more involved than that for $ K_j^n $, which we will derive using deductions. Specifically, for $x_2 = 2$, we have
	\begin{align}\nonumber
		&
		(L_j^1)^2 = (2J)^2 (3 - 2\tau^z_{j-1} \sigma_0 s_y - 2\tau^z_{j+2} \sigma_y s_0 + 2\tau^z_{j-1} \tau^z_{j+2} \sigma_y s_0), 
	\end{align} 
	Now suppose
	\begin{align}\nonumber
		&(L_j^1)^n = (2J)^n ((x_n+1) - x_n \tau^z_{j-1} \sigma_0 s_y \\
		&\qquad - x_n \tau^z_{j+2} \sigma_y s_0 + x_n \tau^z_{j-1}\tau^z_{j+2} \sigma_y s_0).
	\end{align} 
	Then,
	\begin{align}\nonumber
		&
		(L_j^1)^{n+1} = (2J)^{n+1} ( 3x_n \tau^z_{j-1} \tau^z_{j+2} - (3x_n+1) \tau^z_{j+2} \sigma_0 s_y 
		\\ \nonumber
		&- (3x_n + 1) \tau^z_{j-1} \sigma_y s_0 + (3x_n + 1) \sigma_y s_y)
		\\ \nonumber
		&
		(L_j^1)^{n+2} = (2J)^{n+2} ((9x_n+3) - (9x_n+2) \tau^z_{j-1}\sigma_0 s_y 
		\\
		& 
		- (9x_n+2) \tau^z_{j+2} \sigma_y s_0 + (9x_n+2) \tau^z_{j-1} \tau^z_{j+2} \sigma_y s_y.
	\end{align}
	giving
	\begin{align}\nonumber
		&x_{n+2} = 9x_n + 2
		\quad \Rightarrow\quad
		(x_{n+2} + \frac{1}{4}) = 3^2(x_n + \frac{1}{4}) 
		\\
		\Rightarrow
		&x_n + \frac{1}{4} = 3^{n-2} (x_2 + \frac{1}{4})
		\quad \Rightarrow\quad
		x_n = \frac{3^n - 1}{4}.
	\end{align}
	Thus, for $ 1\le m \in \mathbb{Z} $,
	\begin{align}\nonumber
		&
		L_j^{2m} = \frac{(2J)^{2m}}{4}\left(
		(3^{2m}+3)\sigma_0s_0 
		\right. 
		\\ \label{smeq:temp2}
		& \left. + (3^{2m}-1) (-\tau^z_{j-1} \sigma_0 s_y - \tau^z_{j+2} \sigma_y s_0 + \tau^z_{j-1}\tau^z_{j+2} \sigma_y s_y )
		\right),
		\\ \nonumber
		&
		L_j^{2m-1} = \frac{(2J)^{2m-1}}{4} \left(
		(3^{2m-1}-3) \sigma_0 s_0 \tau^z_{j-1}\tau^z_{j+2} 
		\right. \\ \label{smeq:temp3}
		&  \left. 
		+ (3^{2m-1} + 1) ( - \tau^z_{j+2} \sigma_0 s_y - \tau^z_{j-1} \sigma_y s_0 + \sigma_y s_y )
		\right). 
	\end{align}
	From these results, one could obtain arbitrary iterated commutation between $ H_2 $ and the two-spin terms $ B_{j,j+1} $. For instance,
	\begin{align} \nonumber
		&[H_2^{(2m)}, \tau^x_j \tau^x_{j+1}] = (1,0,0,0) L_j^{2m} B_{j,j+1} 
		\\ \nonumber
		&=
		\frac{(2J)^{2m}}{4} \left(
		(3^{2m}+3) \tau^x_j \tau^x_{j+1} + (3^{2m}-1) 
		\right. 
		\\  \label{smeq:2spinterms1}
		&
		\quad\times 
		\left. ( \tau^z_{j-1} i\tau^x_j \tau^y_{j+1} + i\tau^y_j \tau^x_{j+1} \tau^z_{j+2} - \tau^z_{j-1} \tau^y_{j} \tau^y_{j+1} \tau^z_{j+2} )
		\right)
		\\ \nonumber
		& [H_2^{(2m-1)}, \tau^x_j \tau^x_{j+1}] =
		(1,0,0,0)L_j^{2m-1}B_{j,j+1}
		\\ \nonumber
		&  = \frac{(2J)^{2m-1}}{4} \left(
		(3^{2m-1}-3) \tau^z_{j-1} \tau^x_j \tau^x_{j+1} \tau^z_{j+2} + \right.
		\\  \label{smeq:2spinterms2}
		& 
		\left.
		(3^{2m-1}+1) (i\tau^x_j \tau^y_{j+1} \tau^z_{j+2} + \tau^z_{j-1} i\tau^y_j \tau^x_{j+1} - \tau^y_j \tau^y_{j+1})
		\right)
	\end{align}
	where the vector $ (1,0,0,0) $ is acting on the $ 4\times4 $ matrices $ \sigma_0 s_y $ etc. in Eqs.~\eqref{smeq:temp2} and \eqref{smeq:temp3}.
	In sum, the commutation of $ H_2 $ with two-spin terms would result in the generic form
	\begin{align}\label{smeq:h2commute2}
		[H_2^{(n)}, \tau^{\nu_1}_j \tau^{\nu_2}_{j+1}] =
		\sum_{
			\scriptsize
			\begin{array}{l}
				m_1m_2=0,1 \\
				\mu_1\mu_2=x,y
			\end{array}
		}
		\beta^{(n)}_{m_1m_2,\mu_1\mu_2} (\tau^z_{j-1})^{m_1} \tau^{\mu_1}_{j} \tau^{\mu_2}_{j+1} (\tau^z_{j+2})^{m_2},
	\end{align}
	where $ \beta^{(n)}_{m_1m_2,\mu_1\mu_2} $'s are given in Eqs.~\eqref{smeq:2spinterms1} and \eqref{smeq:2spinterms2}.

	\section{Algebraic proof for scaling relations\label{smsec:scaling}}
	
	\subsection{IPR scaling: amplitudes of original cat scar components and DTC amplitudes}
	\label{smsec:ipr}
	
	Here we would focus on the first-order correction to the wave function and obtain the leading order deviation of IPR from the unperturbed values. This is intimately related to the DTC amplitudes. Specifically, for the fine-tuned solutions in  Eq.~\eqref{eq:finetunesol}, the only non-degenerate scars take either the FM or AFM configuration. We would denote both of them by $ \{s_j^{(\text{cat})}\} $ as analysis for the two configurations are the same,
	\begin{align}\nonumber
		|\ell,\{s_j^{(\text{cat})}\}\rangle &= |\ell, \text{FM}\rangle = |\ell, \{s_j = (+1)^j\}\rangle 
		\\
		\quad \text{or}\quad &
		= |\ell, \text{AFM}\rangle = |\ell, \{s_j=(-1)^j\}\rangle.
	\end{align}
	Under perturbation, the first-order perturbation gives
	\begin{widetext}
		
		\begin{align}\label{smeq:wf1scar}
			|\tilde{\omega}_{\ell,\text{cat}} \rangle = 
			\alpha_0 \left(
			|\ell, \{s_j^{(\text{cat})}\}\rangle
			+
			i \lambda \sum_{\ell', \{s_j\}' \ne \ell, \{s_j^{(\text{cat})}\})}
			\frac{ \langle \ell', \{s_j\}'| V_1 |\ell, \{s_j^{(\text{cat})}\} \rangle  }{e^{i\left( E(\ell,\{s_j^{(\text{cat})}\}) - E(\ell', \{s_j\}') \right)} - 1 }
			|\ell', \{s_j\}'\rangle 
			\right)
			+ O(\lambda^2).
		\end{align}
	\end{widetext}
	Now, the generic form for $ V_1 $ in Eq.~\eqref{smeq:v1form} means that only consecutive one or two spins can be flipped. That means all $ |\ell',\{s_j\}'\rangle $ in Eq.~\eqref{smeq:wf1scar} must differ from the scar configurations $ |\ell,\{s_j^{(\text{cat})}\}\rangle $ by two domain walls. Recall $ E(\ell,\{s_j\}) = E(\ell,w) = \pi\ell - J(L-2w) $, with $ w=0,L $ for FM and AFM configurations respectively. All quasienergy differences in the denominator of Eq.~\eqref{smeq:wf1scar} then give identical contributions
	\begin{align}
		E(\ell, \{s_j^{\text{cat}}\}) - E(\ell',\{s_j\}') = \pi(\ell-\ell') \mp 4J,
	\end{align}
	where $ \mp $ signs correspond to FM or AFM scars. Thus, the denominator can be factored out from the summation. Meanwhile, for the generic form of $ V_1 $,
	\begin{align} \nonumber
		&
		V_1 = \sum_{j=1}^L V_{1j},
		\\ \nonumber
		&
		V_{1j} = \sum_{m_1,m_2=0,1} \left( 
		\sum_{\mu_1 = x,y} \alpha_{m_1m_2,\mu_1} (\tau^z_{j-1})^{m_1} \tau^{\mu_1}_{j} (\tau^z_{j+1})^{m_2}
		\right. 
		\\ 
		&+ \left.
		\sum_{\mu_1\mu_2=x,y}
		\beta_{m_1m_2,\mu_1\mu_2} (\tau^z_{j-1})^{m_1} \tau^{\mu_1}_{j} \tau^{\mu_2}_{j+1} (\tau^z_{j+2})^{m_2}
		\right),
	\end{align}
	translation invariance implies that
	\begin{align}
		\mathbb{T}_x V_1 \mathbb{T}_x^{-1} = V_1 
		\qquad
		\Rightarrow
		\qquad
		\mathbb{T}_x V_{1j} \mathbb{T}_x^{-1}
		= V_{1,j+1}.
	\end{align}
	Also, recall that cat scars satisfy the projective translation symmetry ($ \pm $ signs below are for FM and AFM configurations respectively)
	\begin{align}
		\mathbb{T}_x|\ell,\{s_j^{(\text{cat})}\} \rangle = (\pm1)^\ell |\ell,\{s_j^{(\text{cat})}\} \rangle.
	\end{align}
	Then, the matrix elements for each site $ V_{1j} $ are identical to the same terms two sites away,
	\begin{align}\nonumber
		&
		\sum_{\ell', \{s_j\}'} \langle \ell', \{s_j\}'| V_{1,j+2} |\ell, \{s_j^{(\text{cat})}\} \rangle |\ell', \{s_j\}'\rangle
		\\ \nonumber
		&=
		\sum_{\ell',\{s_j\}'} 
		\langle \ell',\{s_j\}'| \mathbb{T}_x^2 V_{1j} (\mathbb{T}_x^{-1})^2 |\ell,\{s_j^{(\text{cat})}\}\rangle |\ell',\{s_j\}'\rangle
		\\ \nonumber
		&=
		\sum_{\ell',\{s_j\}'} 
		\langle \ell',\{s_j\}'|  \mathbb{T}_x^2 V_{1j} |\ell, \{s_j^{(\text{cat})}\}\rangle |\ell',\{s_j\}'\rangle 
		\\ \nonumber
		&=
		\sum_{\ell', \{s_j\}'} \langle \ell',\{s_j\}'| V_{1,j}|\ell, \{s_j^{(\text{cat})}\}\rangle \left(
		\mathbb{T}_x^2 |\ell', \{s_j\}'\rangle
		\right),
	\end{align}
	where in the last step we shift the dummy configuration $ \sum_{\{s_j\}'} |\ell', \{s_j\}'\rangle \langle \ell', \{s_j\}' | = \sum_{\{s_j\}'} \mathbb{T}_x^2 |\ell', \{s_j\}'\rangle \langle \ell', \{s_j\}' | (\mathbb{T}_x^{-1})^{2} $.
	That means Eq.~\eqref{smeq:wf1scar} can be written as
	\begin{widetext} 
		
	\begin{align}\label{smeq:omega1storder}
		|\tilde{\omega}_{\ell,\text{cat}}\rangle = 
		\alpha_0 \left(
		|\ell,\{s_j^{(\text{cat})}\}\rangle 
		+ \lambda
		\sum_{\ell', \{s_j\}'}
		\frac{
			\langle \ell', \{s_j\}'| (V_{1,j=1} + V_{1,j=2}) | \ell, \{s_j^{(\text{cat})}\} \rangle
		}{e^{i(\pi(\ell - \ell') \mp 4J} -1}
		\sum_{m=0}^{L/2-1} \mathbb{T}_x^{2m}|\ell', \{s_j\}'\rangle 
		\right)
		+ O(\lambda^2).
	\end{align}
	\end{widetext}
	Denote the averaged strength of the first order perturbation as
	\begin{align}\nonumber
		\bar{V}_1^2 
		&= 
		\frac{1}{2}\sum_{\ell',\{s_j\}'\ne \ell,\{s_j^{(\text{cat})}\}} 
		\left|
		\frac{\langle \ell', \{s_j\}'|(V_{1,j=1} + V_{1,j=2})| \ell,\{s_j^{(\text{cat})}\}\rangle }{ e^{i(\pi(\ell-\ell') \mp 4J)} - 1 }
		\right|^2 
		\\ \nonumber
		&=
		\frac{1}{8} \sum_{\ell',\{s_j\}'\ne \ell,\{s_j^{(\text{cat})}\}} 
		\left|
		\langle \ell', \{s_j\}'|(V_{1,j=1} + V_{1,j=2})| \ell,\{s_j^{(\text{cat})}\}\rangle 
		\right|^2 
		\\ \label{smeq:v1bar}
		&\qquad\qquad \times
		\csc^2\left(
		\frac{\pi(\ell-\ell') \mp 4J}{2}
		\right)
		,
	\end{align}
	the normalization constant $ \alpha_0 $, related to amplitudes for the original cat scar components $ |\ell, \{s_j^{(\text{cat})}\} \rangle $ in the unperturbed cat scars $ |\tilde{\omega}_{\ell, \text{cat}} \rangle $, is rescaled to
	\begin{align}\nonumber
		&
		1 = |\langle \tilde{\omega}_{\ell, \text{cat}} | \tilde{\omega}_{\ell, \text{cat}}\rangle |^2  = 
		\alpha_0^2 (1+ \lambda^2 \bar{V}_1^2 L)
		\\ \label{smeq:amplitudescaling}
		\Rightarrow
		\qquad  &
		\alpha_0^2 = \frac{1}{1+ \bar{V}_1^2 \lambda^2 L}
		= |\langle \ell, \{s_j^{(\text{cat})}\} | \tilde{\omega}_{\ell, \text{cat}}\rangle|^2.
	\end{align}

	\subsection{Scaling for amplitudes of other spin components: Fock space localization}
	\label{smsec:fock}
	
	IPR scaling in the previous section characterizes the amplitude rescaling for FM or AFM components $ |\ell, \{s_j^{(\text{cat})}\}\rangle \sim |\pm \{s_j^{(\text{cat})}\}\rangle $ in perturbed cat scars eigenstates $ |\tilde{\omega}_{\ell, \text{cat}}\rangle $. It is chiefly contributed by the first-order DW fluctuations on top of the scar configurations $ \pm\{s_j^{(\text{cat})}\} $. Here, we would further consider all higher-order corrections to cat scar eigenstates and observe the amplitudes for other spin configurations in $ |\tilde{\omega}_{\ell,\text{cat}} \rangle $. To quantitatively describe the deviations of spin configurations away from the cat scar patterns, we recall the Fock space distance $ \delta s $ in Eq.~\eqref{smeq:deltas} and further define the pairwise Fock space distance $ \Delta s $,
	\begin{align}\nonumber
		\delta s(\{s_j\}, \{s_j^{(\text{cat})}\}) &= 
		\frac{1}{2}\sum_{j=1}^L |s_j - s_j^{(\text{cat})}|,
		\\
		\label{smeq:Deltas}
		\Delta s(\{s_j\}, \pm \{s_j^{(\text{cat})}\}) &= 
		\frac{1}{2} \min\left( \sum_{j=1}^L |s_j - s_j^{(\text{cat})}|, 
		\sum_{j=1}^L |s_j + s_j^{(\text{cat})}|
		\right).
	\end{align}
	Here $ \Delta s(\{s_j\}, \pm \{s_j^{(\text{cat})}\}) $ is abbreviated in the main text as $ \Delta s_{\text{cat}}(\{s_j\}) $.
	An example for $ L=4 $ is given in Fig.~\ref{fig:fock_loc} (a) and (d). Intuitively, $ \delta s(\{s_j\}, \{s_j^{(\text{cat})}\}) $ characterizes that staring from $ +\{s_j^{(\text{cat})}\} $, how many spins are flipped in order to end up with $ \{s_j\} $. Similarly, $ \Delta s(\{s_j\}, \pm \{s_j^{(\text{cat})}\}) $ counts the minimal numbers of spin flips to go from either of the cat scar configuration pair $ \pm\{s_j^{(\text{cat})}\} $ to $ \{s_j\} $. Correspondingly, Fock space localization means that certain eigenstates, i.e. the perturbed cat scars $ |\tilde{\omega}_{\ell, \text{cat}}\rangle $, exhibit exponential decay for the Fock basis coefficients $ |\langle \{s_j\}| \tilde{\omega}_{\ell, \text{cat}}\rangle| $ with the increase of $ \Delta s $.

	In the following, let us quantitatively analyze the scaling of Fock basis amplitudes for cat scars including all perturbation orders. General perturbation series for the corrected cat scar eigenstates reads
	\begin{align}
		|\tilde{\omega}_{\ell, \text{cat}}\rangle = \cdots e^{i\lambda^k S_k} \cdots e^{i\lambda^2 S_2} e^{i\lambda S_1} | \ell, \{s_j^{(\text{cat})}\} \rangle,
	\end{align}
	where the Hermitian matrices $ S_1,S_2,\cdots, S_k $ would diagonalize the perturbed Floquet operator up to the $ \lambda^k $-th order,
	\begin{align}\nonumber
		&
		\langle \ell_1, \{s_j\}_1 | e^{-i\lambda S_1} e^{-i\lambda^2 S_2} \cdots e^{-i\lambda^k S_k} \left(U_0 e^{i\sum_{k=1}^\infty \lambda^k V_k} \right) e^{i\lambda^k S_k}
		\cdots \\ \label{smeq:skpert}
		&\qquad
		\times
		e^{i\lambda^2 S_2} e^{i\lambda S_1} | \ell_2, \{s_j\}_2 \rangle 
		\propto 
		\delta_{\ell_1, \ell_2} \delta_{\{s_j\}_1, \{s_j\}_2} + O(\lambda^{k+1}) .
	\end{align}
	For our purposes here, it will be most useful and sufficient to obtain the formal operator solutions for $ S_k $'s, and especially to prove their associated selection rules. To gain some intuition for general forms, we check the first order results, where off-diagonal elements for $ S_1 $ satisfy (i.e. expand Eq.~\eqref{smeq:skpert} up to $ \lambda^1 $ orders)
	\begin{align}\nonumber
		&
		1-(1-i\lambda S_1) (U_0 (1+i\lambda V_1)) (1+ i\lambda S_1) = 0 \\ \label{smeq:temp6}
		\Rightarrow
		\qquad &
		[S_1,U_0] = U_0 V_1 \quad
		\Rightarrow
		S_1 - U_0^\dagger S_1 U_0 = -V_1.
	\end{align}
	Then, formal solutions for Eq.~\eqref{smeq:temp6} can be written as
	\begin{align}\label{smeq:s1sol}
		S_1 = -\sum_{p=0}^\infty (U_0^\dagger)^p V_1 U_0^p.
	\end{align}
	\iffalse
	To verify, one could sandwich it with eigenstates ($ \ell_1, \{s_j\}_1 \ne \ell_2, \{s_j\}_2 $)
	\begin{align}\nonumber
		&
		\langle \ell_1, \{s_j\}_1 |S_1| \ell_2, \{s_j\}_2 \rangle =  - \langle \ell_1, \{s_j\}_1 |V_1| \ell_2, \{s_j\}_2 \rangle \\ \label{smeq:temp5}
		&
		\times \sum_p e^{i(E(\ell_2, \{s_j\}_2 - E(\ell_1, \{s_j\}_1)))p} 
		= 
		\frac{\langle \ell_1, \{s_j\}_1 |V_1| \ell_2, \{s_j\}_2 \rangle }{e^{i(E(\ell_2, \{s_j\}_2 - E(\ell_1, \{s_j\}_1)))}-1},
	\end{align}
	where the last step uses the expansion $ 1/(1-x) = \sum_{p=0}^\infty x^p $. Then, Eq.~\eqref{smeq:temp5} recovers the previous results Eq.~\eqref{smeq:s1} for non-diagonal elements. 
	\fi
	Using the solution forms in Eq.~\eqref{smeq:s1sol}, we see that $ S_1 $ corresponds to repeated gauge transformation by $ U_0 $ for the first-order perturbation $ V_1 $. So according to the generalized selection rules in Eq.~\eqref{smeq:tildevk}, $ S_1 $ inherits the selection rules of $ V_1 $ as
	\begin{align}\label{smeq:selections1}
		\langle \{s_j\}| S_1 | \{\tilde{s}_j\}'\rangle \ne 0
		\qquad
		\Rightarrow
		\qquad
		\delta s(\{s_j\}, \pm \{\tilde{s}_j\}) \leqslant n_{\text{op}},
	\end{align}
	where $ n_{\text{op}} $ for operator product powers in perturbation Hamiltonians is given in Eq.~\eqref{smeq:nop}. 
	
	Now, we obtain the generic form for $ S_k $ using deductions. Suppose $ S_{k-1} $ already diagonalize Eq.~\eqref{smeq:skpert} up to the $ \lambda^{k-1} $-th order, and $ \{S_q| q=1,2,\dots,k-1\}  $ satisfy the selection rule that $ \langle \{s_j\}_1| S_q| \{s_j\}_2\rangle \ne 0 \Rightarrow \delta s(\{s_j\}_1, \{s_j\}_2) \leqslant n_{\text{op}}q $. Note the expansion 
	\begin{align}\nonumber
		\nonumber
		e^{i\sum_{k=1}^\infty \lambda^k V_k} 
		&= \sum_{\alpha=0}^\infty \frac{i^\alpha}{\alpha!} (\sum_{k=1}^\infty \lambda^k V_k)^\alpha
		\\
		&=
		\sum_{\alpha=0}^\infty \frac{i^\alpha}{\alpha!} 
		\sum_{\{k_1,k_2,\dots, k_\alpha  \geqslant 1 \} } \lambda^{\sum_{j=1}^\alpha k_j}V_{k_1} V_{k_2} \dots V_{k_\alpha}
	\end{align}
	Then, for the $ \lambda^k $-th order, we have for non-diagonal elements
	\begin{align}\nonumber
		&0 = i[U_0, S_k] + i U_0 V_k 
		\\ \nonumber
		&+ \sum_{\alpha=0}^{k-1}
		\sum_{\scriptsize
			\begin{array}{c}
				\{m_j, n_j, =0,1,\dots,k-1,
				\\
				k_j = 1,2,\dots, k-1|
				\\
				\sum_{j=1}^{k-1} j(m_j+n_j) + \sum_{j=1}^\alpha k_j = k \}
			\end{array}
		}
		\frac{i^{\alpha + \sum_{j=1}^{k-1} (m_j-n_j) }}{\alpha! \prod_{j=1}^k m_j! n_j! } 
		\\
		\nonumber
		&
		\times \left(
		S_1^{n_1} S_2^{n_2} \cdots S_{k-1}^{n_{k-1}} 
		\right)
		U_0 
		\left(
		V_{k_1} V_{k_2} \dots V_{k_\alpha}
		\right)
		\left(
		S_{k-1}^{m_{k-1}}  \cdots S_2^{m_2}S_1^{m_1} 
		\right)
		\\ \nonumber
		&\Rightarrow
		\\ \nonumber
		& S_k - U_0^\dagger S_k U_0 
		\\
		\nonumber
		&=
		-V_k
		+i
		\sum_{\alpha=1}^{k-1}
		\sum_{\scriptsize
			\begin{array}{c}
				\{m_j, n_j =0,1,\dots,k-1,
				\\
				k_j = 1,2,\dots, k-1|
				\\
				\sum_{j=1}^{k-1} j(m_j+n_j) + \sum_{j=1}^\alpha k_j = k \}
			\end{array}
		}
		\frac{i^{\alpha + \sum_{j=1}^{k-1} (m_j-n_j) }}{\alpha! \prod_{j=1}^k m_j! n_j!} 
		\\ \nonumber
		& \times 
		U_0^\dagger \left(
		S_1^{n_1} S_2^{n_2} \cdots S_{k-1}^{n_{k-1}} 
		\right)
		U_0 
		\left(
		V_{k_1} V_{k_2} \dots V_{k_\alpha}
		\right)
		\left(
		S_{k-1}^{m_{k-1}}  \cdots S_2^{m_2}S_1^{m_1} 
		\right).
	\end{align}
	Thus, the formal solutions can be similarly written as
	\begin{align}\nonumber
		&S_k = -\sum_{p=0}^\infty (U_0^\dagger)^p V_k U_0^p \\
		\nonumber
		&
		+i \sum_{p=0}^\infty (U_0^\dagger)^p (
		\sum_{\alpha=1}^{k-1}
		\sum_{\scriptsize
			\begin{array}{c}
				\{m_j, n_j =0,1,\dots,k-1,
				\\
				k_j = 1,2,\dots, k-1|
				\\
				\sum_{j=1}^{k-1} j(m_j+n_j) + \sum_{j=1}^\alpha k_j = k \}
			\end{array}
		}
		\frac{i^{\alpha + \sum_{j=1}^{k-1} (m_j-n_j) }}{\alpha! \prod_{j=1}^k m_j! n_j!} 
		\\ \nonumber
		&
		\times 
		U_0^\dagger \left(
		S_1^{n_1} S_2^{n_2} \cdots S_{k-1}^{n_{k-1}} 
		\right)
        U_0 
		\left(
		V_{k_1} V_{k_2} \dots V_{k_\alpha}
		\right)
		\\ 
		\label{smeq:temp8}
		& \times 
		\left(
		S_{k-1}^{m_{k-1}}  \cdots S_2^{m_2}S_1^{m_1} 
		\right)
		) U_0^p.
	\end{align}
	The first line in Eq.~\eqref{smeq:temp8} takes the same form as Eq.~\eqref{smeq:s1sol}, which is a gauge transformation of $ V_k $ with $ U_0 $. Thus, it inherits the selection rules of $ V_k $ of relating configurations at most $ \delta s\leqslant n_{\text{op}}k $ spins apart. Also, operators $ V_{1},\dots, V_{k-1} $ and $ S_1, \dots, S_{k-1} $, already satisfies the selection rules according to the assumptions of deduction. The second line corresponds to gauge transformations of these operators. Thus, they also satisfy the selection rules of flipping at most $ \delta s \leqslant  \sum_{j=1}^{k-1} (n_{\text{op}}j) (m_j + n_j) + \sum_{j=1}^\alpha (n_{\text{op}} p_j) q_j = n_{\text{op}}k $ spins. In sum, the operator $ S_k $ satisfy the selection rule
	\begin{align}\label{smeq:selectionsk}
		\langle \{s_j\}| S_k | \{\tilde{s}_j\}' \rangle \ne 0
		\quad \Rightarrow\quad
		\delta s(\{s_j\}, \{\tilde{s}_j\}) \leqslant n_{\text{op}}k.
	\end{align}
	That completes the deductive proof for arbitrary orders $ k $.

	Finally, the perturbed cat scar eigenstate can be sorted by orders of $ \lambda^k $ as
	\begin{align}\nonumber
		|\tilde{\omega}_{\ell, \text{cat}}\rangle &= 
		\cdots e^{i\lambda^k S_k}\cdots e^{i\lambda^2 S_2} e^{i\lambda S_1} | \ell, \{s_j^{(\text{cat})}\} \rangle 
		\\ \nonumber
		&=
		|\ell, \{s_j^{(\text{cat})}\}\rangle 
		+
		\sum_{k=1}^\infty \lambda^k 
		\sum_{
			\scriptsize
			\begin{array}{l}
				\{m_j=0,1,\dots, k |  \\
				\sum_{j=1}^k jm_j = k\} 
			\end{array}
		}
		\frac{i^{\sum_{j=1}^k m_j}}{\prod_{j=1}^k m_j! }
		\\
		&\qquad \times 
		S_{k}^{m_k} \cdots S_{2}^{m_2} S_{1}^{m_1}
		|\ell, \{s_j^{(\text{cat})}\}\rangle.
	\end{align}
	Using the selection rules for $ S_k $ in Eq.~\eqref{smeq:selectionsk}, the maximal number of spins that can be flipped by the $ \lambda^k $ terms $ 
	\leqslant \sum_{j=1}^k n_{\text{op}}j\times m_j = n_{\text{op}}k $. Now, note that in the unperturbed eigenstate $ |\ell, \{s_j^{(\text{cat})}\} \rangle = (1/\sqrt2) (|\{s_j^{(\text{cat})}\}\rangle + (-1)^\ell |-\{s_j^{(\text{cat})}\}\rangle) $, there are pairwise Fock product states $ |\pm\{s_j^{(\text{cat})}\} \rangle $ to start with. Thus, Fock states $ |\{s_j\} \rangle $ with  $ \Delta s (\{s_j\}, \pm \{s_j^{(\text{cat})}\}) $ spins flipped with respect to cat configurations $ |\pm \{s_j^{(\text{cat})}\} \rangle $ will only show up above perturbation orders $ \lambda^{k \geqslant \Delta s(\{s_j\}, \pm \{s_j^{(\text{cat})}\})/n_{\text{op}} } $, with pairwise Fock space distance $ \Delta s $ in Eq.~\eqref{smeq:Deltas}.
	Then, we have the scaling relation characterizing Fock space localization
	\begin{align}\label{smeq:fock_loc_scaling}
		|\langle \{s_j\}| \tilde{\omega}_{\ell, \text{cat}}\rangle|^2 = O(\lambda^{\Delta s (\{s_j\}, \pm\{s_j^{(\text{cat})}\})/\xi}),
		\qquad
		\xi \leqslant n_{\text{op}}.
	\end{align}
	Recall that $ \lambda\lesssim 0.1 $ in the DTC regime, so indeed the cat scar eigenstates $ |\tilde{\omega}_{\ell, \text{cat}}\rangle $ is exponentially centered at the Fock space around $ |\pm \{s_j^{(\text{cat})}\}\rangle $ as the coefficients for other spin configurations decay exponentially $ \lambda^{\Delta s (\{s_j\}, \pm\{s_j^{(\text{cat})}\})/\xi} = e^{-\Delta s (\{s_j\}, \pm\{s_j^{(\text{cat})}\})|\ln(1/\lambda)| /\xi} $ with the increase of spin differences between $ \{s_j\} $ and $ \pm\{s_j^{(\text{cat})}\} $.

	\subsection{Spectral gap scaling: exponential growth of DTC lifetime with the increase of $ L $}
	\label{smsec:spectral}
	
	In the previous two sections, we have proved that for the perturbed cat scars $ |\tilde{\omega}_{\ell, \text{cat}=\text{FM/AFM}} \rangle $, the amplitudes for original FM or AFM components experience a minor reduction to $ \alpha_0^2 = |\langle \ell, \{s_j^{(\text{cat})}\} | \tilde{\omega}_{\ell, \text{cat}}\rangle |^2 = 1/(1+\bar{V}_1^2 \lambda^2 L) $ by the first-order domain wall fluctuations. Other spin configurations show exponential Fock space localization. Thus, in intermediate scale systems $ L\lesssim 1/\lambda^2 $, an FM (or AFM) initial state will chiefly overlap with two cat scars $ |\tilde{\omega}_{\ell = 0, \text{FM}} \rangle, |\tilde{\omega}_{\ell = 1, \text{FM}} \rangle $ (or $ |\tilde{\omega}_{\ell = 0, \text{AFM}} \rangle $, $ |\tilde{\omega}_{\ell = 1, \text{AFM}} \rangle $). In this section, we would further prove that the spectral gap for the scars of a certain configuration, i.e. $ \tilde{\omega}_{1,\text{FM}} - \tilde{\omega}_{0,\text{FM}} $ or $ \tilde{\omega}_{1,\text{AFM}} - \tilde{\omega}_{0,\text{AFM}} $, approaches the unperturbed value $ \pi $ with exponential accuracy as the system size $ L $ increases. In particular, we would explicitly demonstrate the origin of exponentially small gap deviation from $ \pi $ in finite-size systems, and give a bound on the scaling exponents.

	Let us first gain some intuitions for the general proof by checking the first-order quasienergy correction,
	\begin{align}
		\omega_{\ell,\text{cat}}^{(1)} &= \langle \ell, \{s_j^{(\text{cat})}\}| V_1 | \ell, \{s_j^{(\text{cat})}\}\rangle.
	\end{align}
	Note that $ V_1 $ can flip up to $ n_{\text{op}} $ spins, i.e. for perturbations up to two-spin terms like $ \phi \tau^x_j \tau^x_{j+1} $ we have $ n_{\text{op}}=2 $. Then, for system sizes $ L>4 $, matrix elements like
	\begin{align}\nonumber
		&
		\langle \ell, \text{FM}| V_{1} | \ell, \text{FM}\rangle 
		\\ \nonumber
		&= 
		{\color{Green}\frac{1}{2}(\langle \uparrow \uparrow \uparrow \uparrow \cdots |V_{1}|\uparrow \uparrow \uparrow \uparrow \cdots \rangle + \langle \downarrow \downarrow \downarrow \downarrow \cdots |V_{1}|\downarrow \downarrow \downarrow \downarrow \cdots \rangle )}
		\\ \nonumber
		& + 
		{\color{red}\frac{(-1)^\ell}{2}(\langle \uparrow \uparrow \uparrow \uparrow \dots| V_{1} |\downarrow \downarrow \downarrow \downarrow \dots\rangle + \langle \downarrow \downarrow \downarrow \downarrow \cdots | V_{1} |\uparrow \uparrow \uparrow \uparrow \cdots\rangle  )},
		\\ 
		\nonumber
		&
		\langle \ell, \text{AFM}| V_{1} | \ell, \text{AFM}\rangle 
		\\ \nonumber
		&= 
		{\color{Green}\frac{1}{2}(\langle \uparrow \downarrow \uparrow \downarrow \dots |V_{1}|\uparrow \downarrow \uparrow \downarrow \dots \rangle + \langle \downarrow \uparrow \downarrow \uparrow \dots |V_{1}|\downarrow \uparrow \downarrow \uparrow \dots \rangle )}
		\\ 
		\label{smeq:temp7}
		& + 
		{\color{red}\frac{(-1)^\ell}{2}(\langle \uparrow \downarrow \uparrow \downarrow \dots| V_{1} |\downarrow \uparrow \downarrow \uparrow \dots\rangle + \langle \downarrow \uparrow \downarrow \uparrow \dots | V_{1} |\uparrow \downarrow \uparrow \downarrow \dots\rangle  )},
	\end{align}
	shows vanishing cross terms (denoted by red), because those red terms require {\em simultaneously} flipping $ L $ spins, which violates the selection rules. Thus, these matrix elements are completely independent of spectral pairing quantum numbers $ \ell $, and we see that it is again the selection rules enforcing the identical quasienergy correction for pairwise scars,
	\begin{align}\nonumber
		&\omega_{0, \text{cat}} = \omega_{1, \text{cat}} 
		=
		\langle \ell, \{s_j^{(\text{cat})}\} | V_{1} |\ell, \{s_j^{(\text{cat})}\} \rangle 
		\\ \label{smeq:2orderele1}
		&= 
		{\color{Green}
			\frac{1}{2} \sum_{m=0,1} \langle (-1)^m \{s_j^{(\text{cat})}\} | V_{1} | (-1)^m \{s_j^{(\text{cat})}\} \rangle. 
		}
	\end{align}
	
	Based on the first-order solutions, we now prove the fixed spectral gap for higher-order quasienergy corrections $ \tilde{\omega}_{\ell, \text{cat}} $ via deductions. Assuming that $ \omega_{\ell, \text{cat}}^{1\leqslant q \leqslant k-1} $ up to the $ \lambda^{k-1} $-th order  are all independent of spectral pairing quantum numbers $ \ell $, let us consider the $ \lambda^k $-th order corrections. Generically, with the corrected eigenstates $ |\tilde{\omega}_{\ell, \text{cat}}\rangle = \cdots e^{i\lambda^k S_k}\dots e^{i\lambda S_1}|\ell, \{s_j^{(\text{cat})}\}\rangle $, the corresponding corrected quasienergy reads
	\begin{align}\nonumber
		e^{i\tilde{\omega}_{\ell, \text{cat}} } &= \langle \ell, \{s_j^{(\text{cat})}\} | e^{-i\lambda S_1} e^{-i\lambda^2S_2} \cdots e^{-i\lambda^k S_k} \cdots
		\left(
		U_0 e^{i\sum_{k=1}^\infty \lambda^k V_k}
		\right)
		\cdots 
		\\ \nonumber & 
		\times e^{i\lambda^k S_k} \cdots e^{i\lambda^2 S_2} e^{i\lambda S_1}| \ell, \{s_j^{(\text{cat})}\}\rangle
		\\
		&= e^{i\left( 
			E(\ell, \{s_j^{(\text{cat})}\}) + \delta\omega_{\ell, \text{cat}} \right)
		},
		\qquad
		\delta\omega_{\ell, \text{cat}} = \sum_{k=1}^\infty \lambda^k \omega_{\ell, \text{cat}}^{(k)},
	\end{align}
	where $ E(\ell, \{s_j^{(\text{cat})}\}) $ is the cat scar quasienergy at $ \lambda = 0 $. Using
	$ \langle\ell, \{s_j^{(\text{cat})}\} | U_0^\dagger  = e^{-iE(\ell, \{s_j^{(\text{cat})}\})} \langle \ell, \{s_j^{(\text{cat})}\}| $, we have
	\begin{align}\nonumber
		&
		e^{i\delta\omega_{\ell, \text{cat}} } = 
		e^{i\left( 
			\tilde{\omega}_{\ell, \text{cat}} {\color{red}- E(\ell, \{s_j^{(\text{cat})}\}) } \right)
		}
		\\ \nonumber
		&= 
		\langle \ell, \{s_j^{(\text{cat})}\} |
		{\color{red} U_0^\dagger} e^{-i\lambda S_1} e^{-i\lambda^2S_2} \cdots e^{-i\lambda^k S_k} \cdots
		\left(
		U_0 e^{i\sum_{k=1}^\infty \lambda^k V_k}
		\right)
		\cdots 
		\\ 
		&
		\qquad
		\times e^{i\lambda^k S_k} \cdots e^{i\lambda^2 S_2} e^{i\lambda S_1}| \ell, \{s_j^{(\text{cat})}\}\rangle,
	\end{align}
	and therefore a power counting gives that the $ \lambda^k $-th order terms read
	\begin{align}\nonumber
		i\omega_{\ell, \text{cat}}^{(k)}
		&=
		- \sum_{\alpha=2}^k \frac{i^\alpha}{\alpha!}
		\sum_{\{1 \leqslant k_j \leqslant k-1| \sum_{j=1}^\alpha k_j = k\}} 
		\omega_{\ell,\text{cat}}^{(k_1)}\dots \omega_{\ell,\text{cat}}^{(k_\alpha)}
		\\ \nonumber
		&+ 
		\sum_{\alpha=0}^k
		\sum_{\scriptsize
			\begin{array}{c}
				\{m_j, n_j =0,1,\dots,k, 
				\\
				k_j = 1,2,\dots, k|
				\\
				\sum_{j=1}^k j(m_j+n_j) + \sum_{j=1}^\alpha k_j = k \}
			\end{array}
		}
		\frac{i^{\alpha + \sum_{j=1}^k (m_j-n_j) }}{\alpha! \prod_{j=1}^k m_j! n_j!} 
		\\ \nonumber
		&
		\times
		\langle \ell, \{s_j^{(\text{cat})}\} |
		F_{\{n_j, m_j, k_j\}}^{(k,\alpha)}
		|\ell, \{s_j^{(\text{cat})}\} \rangle, 
		\\ \nonumber
		F_{\{n_j, m_j, k_j\}}^{(k,\alpha)} &= U_0^\dagger
		\left(
		S_1^{n_1} S_2^{n_2} \cdots S_k^{n_k} 
		\right)
		U_0 
		\left(
		V_{k_1}  V_{k_2}  \dots V_{k_\alpha} 
		\right)
		\\ & 
		\qquad \times
		\left(
		S_k^{m_k}  \cdots S_2^{m_2}S_1^{m_1} 
		\right).
	\end{align}
	The lower-order corrections $ \omega_{\ell, \text{cat}}^{q\leqslant k-1} $ are already independent of $ \ell $ according to the assumptions of deductions. Further, for $ F_{\{n_j, m_j, k_j\}}^{(k,\alpha)} $, one could apply the selection rules Eq.~\eqref{smeq:selection full} for $ V_{k_j} $, Eq.~\eqref{smeq:selectionsk} for $ S_{1}\dots S_k $, as well as Eq.~\eqref{smeq:tildevk} for operators under gauge transformation by $ U_0^\dagger \dots U_0 $. Altogether, $ F_{\{n_j, m_j, k_j\}}^{(k,\alpha)} $ could flip as many as $ \sum_{j=1}^{n_k} (n_{\text{op}}j) (n_j + m_j) + \sum_{j=1}^\alpha n_{\text{op}}k_j = n_{\text{op}}k $ spins, namely,
	\begin{align}
		\langle \{s_j\}_1| F_{\{n_j, m_j, k_j\}}^{(k,\alpha)} | \{\tilde{s}_j\}_2\rangle \ne 0
		\quad\Rightarrow\quad
		\delta s(\{s_j\}_1, \{\tilde{s}_j\}_2) \leqslant n_{\text{op}}k.
	\end{align}
	Thus, for a system of size $ L $, up to perturbation order $ k< L/n_{\text{op}} $, the cross terms denoted by red color below  vanish,
	\begin{align}\nonumber
		&
		\langle \ell, \{s_j^{(\text{cat})}\} | F_{\{n_j, m_j, k_j\}}^{(k,\alpha)} | \ell, \{s_j^{(\text{cat})}\} \rangle 
		\\ \nonumber
		&=
		{\color{Green}
			\frac{1}{2} \left(
			\langle \{s_j^{(\text{cat})}\}| F_{\{n_j, m_j, k_j\}}^{(k,\alpha)} | \{s_j^{(\text{cat})}\}\rangle + \langle -\{s_j^{(\text{cat})}\}| F_{\{n_j, m_j, k_j\}}^{(k,\alpha)} |- \{s_j^{(\text{cat})}\}\rangle
			\right)
		}
		\\ \label{smeq:spectralgapgeneral}
		&
		+ 
		{\color{red}\frac{(-1)^\ell}{2}
			\left(
			\langle \{s_j^{(\text{cat})}\}| F_{\{n_j, m_j, k_j\}}^{(k,\alpha)} |- \{s_j^{(\text{cat})}\}\rangle 
			+
			\langle -\{s_j^{(\text{cat})}\}| F_{\{n_j, m_j, k_j\}}^{(k,\alpha)} | \{s_j^{(\text{cat})}\}\rangle 
			\right)
		}.
	\end{align}
	This is again because the cross terms require flipping all $ L $ spins in order to related $ \pm \{s_j^{(\text{cat})}\} $, which violates the selection rules for $ F_{\{n_j, m_j, k_j\}}^{(k,\alpha)} $ when $ k< L/n_{\text{op}} $. Then, we have proved that up to the order  $ \lambda^{k < L/n_{\text{op}}} $, quasienergy corrections $ \omega_{0, \text{cat}}^{(k)} = \omega_{1,\text{cat}}^{(k)} $ at each perturbation order. On the other hand, for $ k\geqslant L/n_{\text{op}} $, operators $ F_{\{n_j, m_j, k_j\}}^{(k,\alpha)} $ can indeed flip all $ L $ spins in the system, so the red terms in Eq.~\eqref{smeq:spectralgapgeneral} no longer vanish such that $ \omega_{\ell, \text{cat}}^{(k\geqslant L/n_{\text{op}})} $ start to depend on the quantum numbers $ \ell $. In sum, the quasienergy difference for pairwise cat scars $ \tilde{\omega}_{0, \text{cat}}, \tilde{\omega}_{1, \text{cat}} $ approaches the unperturbed value $ \pi $ with exponential accuracy
	\begin{align}\nonumber
		& 
		\exp\left( i\left(\tilde{\omega}_{1,\text{cat}} - \tilde{\omega}_{0,\text{cat}} \right)
		\right)
		\\ \nonumber
		&=
		\exp\left( 
		i\left(
		E(1, \{s_j^{(\text{cat})}\})
		-
		E(0, \{s_j^{(\text{cat})}\})
		\right)
		+ 
		i\left( 
		\delta \omega_{1,\text{cat}} 
		- 
		\delta \omega_{0, \text{cat}}
		\right) 
		\right)
		\\ \nonumber
		&=
		\exp\left( 
		i\pi
		+ i\sum_{k=1}^\infty \lambda^k \left(
		\omega_{1,\text{cat}}^{(k)} 
		-
		\omega_{0,\text{cat}}^{(k)} 
		\right) 
		\right)
		\\ \nonumber
		&=
		\exp\left( 
		i\pi
		+ i\sum_{k=L/n_{\text{op}}}^\infty \lambda^k \left(
		\omega_{1,\text{cat}}^{(k)} 
		-
		\omega_{0,\text{cat}}^{(k)} 
		\right) 
		\right)
		\\
		&= \exp\left(
		i(\pi + O(\lambda^{k\geq L/n_{\text{op}}}))
		\right)
	\end{align}
	Thus, we could introduce the exponent $ \nu $ and write the spectral gap deviations for finite size systems as
	\begin{align}\nonumber
		&
		\tilde{\omega}_{1,\text{cat}} - \tilde{\omega}_{0,\text{cat}} 
		= \pi + 
		(\delta\omega_{1,\text{cat}} - \delta\omega_{0,\text{cat}})
		= \pi + O(\lambda^{L/\nu}),
		\\ 
		\label{smeq:nu}
		&
		\nu\leqslant n_{\text{op}}.
	\end{align}

	\section{Derivation of effective Hamiltonian for strongly interacting cases}
	\label{smsec:hefflaregJ}
	
	For the model in Eq.~\eqref{eq:simplemodel}, with strong interactions $ J\sim1$, we need to include all higher-order terms $J^n$ using the BCHD formula
	\begin{align}
		e^A e^B = 
		e^{
		\sum_{n=1}^\infty \frac{(-1)^{n+1}}{n}
		\sum_{ 
			p_j+q_j>0, p_j, q_j\geqslant 0
			} 
		\frac{ 
			[A^{(p_1)}, [B^{(q_1)}, [\dots, [A^{(p_n)}, B^{(q_n)}]\dots]
		}{(\sum_{j=1}^n (p_j+q_j)) \prod_{j=1}^n p_j! q_j!}
		}.
	\end{align}
	Here we keep terms up to perturbation strength $ \lambda^1 $, while include all powers of $ J $. Then,
	\begin{align}\nonumber
		&e^{-i\sum_j J\tau^z_j \tau^z_{j+1}} e^{i\lambda \sum_j (\cos(h) \tau^x_j - \sin(h) \tau^y_j)} 
		\\ 
		&= e^{i\sum_j (-J\tau^z_j \tau^z_{j+1} + \lambda (\cos(h) \tau^x_j - \sin(h) \tau^y_j)) + i\lambda H'},
	\end{align}
	where
	\begin{align}\nonumber
		H' &= 
		\sum_{n=1}^\infty \frac{(-1)^{n+1}}{n} 
		\sum_{p_1\dots p_n = 0}^\infty 
		\frac{ (-iJ)^{p_1+p_2+\dots+p_n} }{(1+\sum_{j=1}^n p_j) \prod_{j=1}^n p_j! }
		\\ \nonumber
		& \quad \times
		[(\tau^z_j \tau^z_{j+1})^{(p_1+p_2+\dots + p_n)}, \cos(h)\tau^x_j - \sin(h) \tau^y_j]
		\\ \nonumber
		&=
		\sum_{n=1}^\infty \frac{(-1)^{n+1}}{n}
		\sum_{r=1}^\infty \frac{(-iJ)^r}{r+1}
		[(\tau^z_j\tau^z_{j+1})^{(r)}, \cos(h)\tau^x_j - \sin(h) \tau^y_j]
		\\ \nonumber
		& \quad \times 
		\sum_{p_1+\dots +p_n = r } \frac{1}{p_1!p_2! \dots p_n!}
		\\ \nonumber
		&=
		\sum_{n=1}^\infty \frac{(-1)^{n+1}}{n}
		\sum_{r=1}^\infty \frac{(-iJ)^r}{r+1}
		[(\tau^z_j\tau^z_{j+1})^{(r)}, \cos(h)\tau^x_j - \sin(h) \tau^y_j]
		\frac{n^r}{r!}
		\\
		&=
		\sum_{n=1}^\infty \frac{(-1)^{n+1}}{n}
		\sum_{r=1}^\infty \frac{(-iJn)^r}{(r+1)!}
		[(\tau^z_j\tau^z_{j+1})^{(r)}, \cos(h)\tau^x_j - \sin(h) \tau^y_j]
	\end{align}
	Now, use
	\begin{align}\nonumber
		&
		[(\sum_j\tau^z_j \tau^z_{j+1})^{(1)}, \sum_k \cos(h) \tau^x_k - \sin(h) \tau^y_k]
		\\ \nonumber
		&=
		\sum_j i(\cos(h) \tau^y_j + \sin(h) \tau^x_j) (\tau^z_{j-1} + \tau^z_{j+1})
		,
		\\ \nonumber
		&
		[(\sum_j\tau^z_j \tau^z_{j+1})^{(2)}, \sum_k \cos(h) \tau^x_k - \sin(h) \tau^y_k]
		\\
		\nonumber
		&=
		\sum_j (\cos(h) \tau^x_j - \sin(h) \tau^y_j) 2(1+\tau^z_{j-1}\tau^z_{j+1}),
		\\ \nonumber
		&
		[(\sum_j\tau^z_j \tau^z_{j+1})^{(3)}, \sum_k \cos(h) \tau^x_k - \sin(h) \tau^y_k]
		\\
		&=
		\sum_j i(\cos(h) \tau^y_j + \sin(h) \tau^x_j) 4(\tau^z_{j-1} + \tau^z_{j+1}),
	\end{align}
	which gives
	\begin{align} \nonumber
		&
		[(\sum_j \tau^z_j \tau^z_{j+1})^{(2m-1)}, \sum_k \cos(h) \tau^x - \sin(h) \tau^y_k] 
		\\ \nonumber
		&=
		\sum_j i(\cos(h) \tau^y_j + \sin(h) \tau^x_j) 2^{2(m-1)} (\tau^z_{j-1} + \tau^z_{j+1}),
		\\ \nonumber
		&
		[(\sum_j \tau^z_j \tau^z_{j+1})^{(2m)}, \sum_k \cos(h) \tau^x - \sin(h) \tau^y_k] 
		\\&=
		\sum_j (\cos(h) \tau^x_j - \sin(h) \tau^x_j) 2^{2m-1} (1 + \tau^z_{j-1}  \tau^z_{j+1}),
	\end{align}
	Thus,
	\begin{align}\nonumber
		H' &= \sum_{n=1}^\infty \frac{(-1)^{n+1}}{n}
		\\ \nonumber
		&\times \sum_{m=1}^\infty \left(
		\frac{(-iJn)^{2m-1}}{(2m)!} 2^{2m-2} i(\cos(h) \tau^y_j + \sin(h) \tau^x_j) (\tau^z_{j-1} + \tau^z_{j+1})
		\right.
		\\ \nonumber
		&\qquad
		+ \left.
		\frac{(-iJn)^{2m}}{(2m+1)!} 2^{2m-1}
		(\cos(h)\tau^x_j - \sin(h) \tau^y_j) (1+\tau^z_{j-1}\tau^z_{j+1})
		\right)
		\\
		\nonumber
		&=
		\sum_{n=1}^\infty \frac{(-1)^{n+1}}{n}
		\left(
		\frac{1}{-4iJn} (-1+\cos(2Jn)) \right. \\
		\nonumber
		&\qquad \times 
		\left. 
		i(\cos(h) \tau^y_j + \sin(h) \tau^x_j) (\tau^z_{j-1} + \tau^z_{j+1})
		\right.
		\\ \nonumber
		&\qquad
		+ \frac{1}{-4iJn}(2iJn -i\sin(2Jn)) 
		\\ \nonumber
		&\qquad 
		\times \left. (\cos(h) \tau^x_j - \sin(h) \tau^y_j) (1+\tau^z_{j-1} \tau^z_{j+1})
		\right)
		\\ \nonumber
		&=
		\sum_{n=1}^\infty 
		\frac{(-1)^{n}}{4Jn^2}
		\left( (-1+\cos(2Jn)) (\cos(h) \tau^y_j + \sin(h)\tau^x_j) (\tau^z_{j-1} + \tau^z_{j+1})  \right. 
		\\ \nonumber
		&\qquad  \left.
		 + 
		(2Jn - \sin(2Jn)) (\cos(h) \tau^x_j - \sin(h) \tau^y_j) (1+\tau^z_{j-1} \tau^z_{j+1}) \right)
		\\  \nonumber
		&=
		\frac{1}{4J} \left(
		f(J) (\cos(h) \tau^y_j + \sin(h) \tau^x_j) (\tau^z_{j-1} + \tau^z_{j+1})
		\right. 
		\\ &
		\quad \left. +
		g(J)
		(\cos(h) \tau^x_j - \sin(h) \tau^y_j) (1+\tau^z_{j-1}\tau^z_{j+1})
		\right) 
	\end{align}
	where
	\begin{align}\nonumber
		f(J) &= \frac{1}{4J}\left( 
		\frac{\pi^2}{12} + \frac{\text{Li}_2(-e^{-2iJ}) + \text{Li}_2( -e^{2iJ})}{2}
		\right),
		\\
		g(J) &= \frac{1}{4J} \left( 
		-2J\ln(2) + \frac{\text{Li}_2(-e^{-2iJ}) - \text{Li}_2( -e^{2iJ})}{2i}
		\right)
	\end{align}
	and $ \text{Li}_s(z) $ is the polylogarithm function
	\begin{align}
		\text{Li}_s(z) = \sum_{k=1}^\infty \frac{z^k}{k^s}.
	\end{align}
	Thus, we recover the results in Eq.~\eqref{eq:h2strong}.

	\bibliography{TC}
	
\end{document}